\documentclass[10pt,reqno]{amsart}

\usepackage{ifpdf}
\ifpdf
 \usepackage[pdftex]{graphicx}
\else
 \usepackage{graphicx}
\fi

\usepackage{fullpage}
\usepackage{color}
\usepackage{mathrsfs}

 \newtheorem{thm}{Theorem}
\newtheorem{cor}{Corollary}
\newtheorem{lem}{Lemma}

 \theoremstyle{definition}

 \theoremstyle{remark}

 \newcommand{\abs}[1]{\left\vert#1 \right\vert}

 \newcommand{\e}{\varepsilon}

\title{The Semiclassical Modified Nonlinear Schr\"odinger Equation I:  Modulation Theory and Spectral Analysis}
\author{Jeffery C. DiFranco and Peter D. Miller}
\date{\today}
\begin{document}
\begin{abstract}
  We study an integrable modification of the focusing nonlinear
  Schr\"odinger equation from the point of view of semiclassical
  asymptotics.  In particular, (i) we establish several important
  consequences of the mixed-type limiting quasilinear system including
  the existence of maps that embed the limiting forms of both the
  focusing and defocusing nonlinear Schr\"odinger equations into the
  framework of a single limiting system for the modified equation,
  (ii) we obtain bounds for the location of discrete spectrum for the
  associated spectral problem that are particularly suited to the
  semiclassical limit and that generalize known results for the
  spectrum of the nonselfadjoint Zakharov-Shabat spectral problem, and
  (iii) we present a multiparameter family of initial data for which
  we solve the associated spectral problem in terms of special
  functions for all values of the semiclassical scaling parameter.  We
  view our results as part of a broader project to analyze the
  semiclassical limit of the modified nonlinear Schr\"odinger equation
  via the noncommutative steepest descent procedure of Deift and Zhou,
  and we also present a self-contained development of a
  Riemann-Hilbert problem of inverse scattering that differs from
  those given in the literature and that is well-adapted to
  semiclassical asymptotics.
\end{abstract}
\maketitle
\section{Introduction}
\label{sec:intro}
This paper will be concerned with the modified nonlinear Schr\"odinger
(MNLS) equation for a complex-valued function $\phi(x,t)$:
\begin{equation}\label{MNLScauchy}
i\e\frac{\partial \phi}{\partial t} +
\frac{\e^2}{2}\frac{\partial^2 \phi}{\partial x^2} + |\phi|^2\phi + 
i\alpha\e\frac{\partial}{\partial x}\left(|\phi|^2\phi\right)=0\,,
\quad\quad \alpha\ge 0\,,\quad\e>0\,.
\end{equation}
The correct problem to pose for this equation when $x\in\mathbb{R}$ is
the Cauchy problem, where initial data $\phi(x,0)$ are given at $t=0$.
Here $\alpha$ and $\e$ are real parameters which we assume throughout
the paper satisfy the inequalities given in \eqref{MNLScauchy}.

The focusing nonlinear Schr\"odinger (NLS) equation is a special case
of \eqref{MNLScauchy} when $\alpha=0$.  The focusing NLS equation is
of course very well known, arising naturally in the modeling of any
nearly monochromatic, weakly nonlinear, dispersive wave
propagation. In particular, it is in many circumstances an adequate
model to describe the evolution of the electric field envelope of
picosecond pulses in monomode optical fibers \cite{HT}.  The so-called
``semiclassical'' or zero-dispersion limit of the focusing NLS Cauchy
problem consists of taking the initial condition in the form
$\phi(x,0)=A(x)e^{iS(x)/\e}$ for fixed real functions $A(\cdot)>0$ and
$S(\cdot)$ and considering the behavior of the corresponding solution
$\phi(x,t)$ of the focusing NLS equation in the limit $\e\downarrow
0$.  This asymptotic problem is important in the design of
increasingly prevalent dispersion-shifted fibers as well as being of
intrinsic mathematical interest.

The utility of the focusing NLS equation as a mathematical model stems
from the fact that this equation is integrable, and in principal the
Cauchy problem may be solved using the scattering/inverse scattering
transform \cite{ZS72} that is precisely adapted to this
equation. However, in applications if higher-order effects become
important then it is necessary to account for them in the model. To
consider sub-picosecond electromagnetic pulses in monomode optical
fibers, one should consider the more general evolution equation,
 \begin{equation}\label{MNLSgen}
i\e \frac{\partial \phi}{\partial t}+\frac{\e^2}{2}
\frac{\partial^2\phi}{\partial x^2}+\abs{\phi}^2\phi=
-i\alpha\e\frac{\partial}{\partial x}\left(\abs{\phi}^2\phi\right)+
\alpha'\e\frac{\partial}{\partial x}\left(\abs{\phi}^2\right)\cdot\phi+
i\alpha''\e^3\frac{\partial^3\phi}{\partial x^3}\,,
\end{equation}
where $\alpha$, $\alpha'$, and $\alpha''$ are real constants and
where the three additional terms on the right hand side account for
nonlinear dispersion, intrapulse Raman scattering, and higher-order
linear dispersion respectively \cite{A}. In general, (\ref{MNLSgen})
is not integrable; however it turns out that there are special cases
of \eqref{MNLSgen} besides $\alpha=\alpha'=\alpha''=0$ that are
integrable, although by different machinery than applies to the
focusing NLS equation.  In fact, the MNLS equation \eqref{MNLScauchy}
is such an integrable system \cite{WKI} (related to the so-called
``derivative'' NLS equation).  As the MNLS equation takes into account
at least the higher-order effect of nonlinear dispersion,
understanding the predictions of this equation can serve as a
mathematical ``stepping stone'' in the understanding of the more
general problem \eqref{MNLSgen} as well as any weakly nonlinear,
nearly monochromatic, dispersive wave propagation problem when
nonlinear dispersion is not negligible.

One of the well-known features of the focusing NLS equation is the
so-called \emph{modulational instability}.  This instability may be
manifested in several different ways, and so the term ``modulational
instability'' can mean different things in different contexts.
However, the simplest context of modulational instability is the
linearized perturbation theory of small disturbances of
exact ``plane-wave'' solutions of the focusing NLS equation.
Indeed, the NLS equation
\begin{equation}\label{NLS}
i\e\frac{\partial \phi}{\partial t}+\frac{\e^2}{2}\frac{\partial^2\phi}{\partial x^2}+\kappa\abs{\phi}^2\phi=0
\end{equation}
(the focusing case is $\kappa=1$ and the defocusing case is $\kappa=-1$)
has exact solutions (plane waves) of the form
$\phi_o(x,t)=Ae^{i(kx-\omega t)/\e}$ for real constants $A>0$, $k$ and
$\omega$ necessarily related by the \emph{nonlinear dispersion
  relation}
\begin{equation}
\omega=\frac{1}{2}k^2-\kappa
A^2\,. 
\end{equation}
Consider a small perturbation of this plane-wave solution by
substituting $\phi=\phi_o\cdot(1+p)$ into \eqref{NLS},
where $p$ is a new unknown.  In this way, one obtains an
(equivalent) equation for the ``in-phase'' perturbation
$p$:
\begin{equation}
  i\e \frac{\partial p}{\partial t} +
ik\e \frac{\partial p}{\partial x} +
\frac{\e^2}{2}\frac{\partial^2p}{\partial x^2}+
\kappa A^2(p+p^*+2|p|^2+p^2+
|p|^2p)=0\,,
\end{equation}
where the asterisk denotes complex conjugation.  Assuming $p$ to be
small, this equation may be linearized and written as a system for 
$a=\text{Re}\{p\}$ and $b=\text{Im}\{p\}$:
\begin{equation}
\begin{split}
\e\frac{\partial a}{\partial t}+k\e \frac{\partial a}{\partial x}
+\frac{\e^2}{2}\frac{\partial^2b}{\partial x^2}&=0\\
\e\frac{\partial b}{\partial t} +k\e \frac{\partial b}{\partial x}
-\frac{\e^2}{2}\frac{\partial^2a}{\partial x^2}-2\kappa A^2a&=0\,.
\end{split}
\label{linearized}
\end{equation}
This is a linear constant-coefficient system of partial differential
equations that may be solved using Fourier techniques.  We look for
elementary solutions of the form
\begin{equation}
\begin{bmatrix}a(x,t)\\b(x,t)\end{bmatrix}=\begin{bmatrix}\hat{a}(t)\\\hat{b}(t)
\end{bmatrix}e^{i\Delta x/\e}\,,
\end{equation}
where $\Delta$ is a ``relative wavenumber'' (to $k$), which reduces
\eqref{linearized} to a system of ordinary differential equations:
\begin{equation}
\e\frac{d}{d t}\begin{bmatrix}\hat{a}\\\hat{b}
\end{bmatrix}=\begin{bmatrix}-ik\Delta  &\frac{1}{2}\Delta^2\\
2\kappa A^2-\frac{1}{2}\Delta^2 & -ik\Delta\end{bmatrix}
\begin{bmatrix}\hat{a}\\\hat{b}\end{bmatrix}\,.
\end{equation}
There is a basis of solutions proportional to eigenvectors of the
coefficient matrix on the right-hand side, and the time dependence
enters through factors of the form $e^{\sigma t/\e}$ where $\sigma$ is
the corresponding eigenvalue of the coefficient matrix.  The 
eigenvalues are given by
\begin{equation}
\sigma = -ik\Delta \pm \frac{\Delta}{2}\sqrt{4\kappa A^2-\Delta^2}\,.
\end{equation}
If we are in the defocusing case ($\kappa=-1$) then $\sigma$ is always
purely imaginary, and all Fourier-type solutions of \eqref{linearized}
are oscillatory in $t$.  This behavior indicates the \emph{absence} of
modulational instability in the defocusing case.  However, if we are
in the focusing case ($\kappa=1$) then $\sigma$ has a nonzero real part
if $\Delta^2<4A^2$ and therefore there is a band of unstable relative
wavenumbers about $\Delta=0$ (and hence a \emph{sideband} of the
unperturbed wavenumber $k$).  This is the modulational instability in
its simplest form.  These calculations show that in the focusing case,
the only perturbations that have a chance of avoiding growth correspond to
waves of length $O(\e)$, and worse yet, the exponential growth rates
for the unstable modes scale as $1/\e$.  In this sense, the
modulational instability is enhanced when $\e$ is small.

The modulational instability may also be understood from a nonlinear
perspective, via the so-called \emph{modulation equations}.  Without
loss of generality, we write the solution $\phi(x,t)$ of \eqref{NLS}
in the form $\phi(x,t)=A(x,t)e^{i S(x,t)/\e}$ where $A(x,t)>0$ is a
real amplitude and $S(x,t)$ is a real phase, and then \eqref{NLS}
implies corresponding (real) nonlinear equations governing these two
functions.  If we think of $A(x,t)$ and $S(x,t)$ as functions independent of
$\e$ (this assumption is of course not exactly consistent with
\eqref{NLS}) then locally in the neighborhood of fixed $x$ and $t$,
$\phi(x,t)$ resembles a plane wave solution with amplitude $A=A(x,t)$,
wavenumber $k=\partial S/\partial x(x,t)$, and frequency
$\omega=-\partial S/\partial t(x,t)$.  This observation is the grounds
for expecting a connection between the preceding perturbative
calculation and what will now follow.  To proceed, it is more natural
to introduce $u(x,t):=\partial S/\partial x$ and $\rho(x,t):=A^2$ and obtain
equations for these unknowns equivalent to \eqref{NLS}:
\begin{equation}
\begin{split}
\frac{\partial \rho}{\partial t} +u\frac{\partial\rho}{\partial x}+\rho\frac{\partial u}{\partial x} &=0\\
\frac{\partial u}{\partial t} -\kappa\frac{\partial\rho}{\partial x}+u\frac{\partial u}{\partial x} &=\frac{\e^2}{2}\left(
\frac{1}{2\rho}\frac{\partial^2\rho}{\partial x^2}-\left(\frac{1}{2\rho}\frac{\partial\rho}{\partial x}\right)^2\right)\,.
\end{split}
\end{equation}
This time, the approximation we make (in lieu of the formal
linearization step in the preceding discussion) is to drop the
formally small terms in $\e$.  Thus, one arrives at the modulation equations
associated to (\ref{NLS}),
\begin{equation}\label{modfocNLS}
\frac{\partial}{\partial t}\begin{bmatrix} \rho \\ u \end{bmatrix}+
\begin{bmatrix} u& \rho\\ -\kappa & u
\end{bmatrix}\frac{\partial}{\partial x}
\begin{bmatrix} \rho \\ u\end{bmatrix}=0\,.
\end{equation}

A system of quasilinear partial differential equations of the form
\eqref{modfocNLS} is called \emph{hyperbolic} if the coefficient
matrix of the $x$-derivatives has distinct real eigenvalues and
\emph{elliptic} if the eigenvalues are complex.  Cauchy problems for
hyperbolic systems can be solved by the method of characteristics, one
implication of which is continuous dependence of the solution on
initial data for small time.  The hyperbolic Cauchy problem is said to
be \emph{well-posed}.  On the other hand, the Cauchy problem is
ill-posed for elliptic systems.  Worse yet, the Cauchy problem for an
elliptic system can only be solved at all (by the method of
Cauchy-Kovalevskaya) if the initial data $\rho(x,0)$ and $u(x,0)$ are
analytic functions of the variable $x$; this should be regarded as a
overly restrictive condition for a physically relevant mathematical
model.

It is easy to check that for $u\in\mathbb{R}$ and $\rho>0$, the system
\eqref{modfocNLS} is elliptic for $\kappa=1$ and hyperbolic for
$\kappa=-1$.  This calculation therefore reveals the same dichotomy
between the focusing and defocusing versions of the NLS equation as
did linearization about a plane wave.  The linearization calculation
may be viewed as a ``local in $x$'' version of the deduction of type
(hyperbolic or elliptic) of modulation equations.  In general,
hyperbolicity of plane-wave modulation equations is equivalent to the
absence of linear instabilities of plane waves of sideband type,
although it is possible for a system to have hyperbolic modulation
equations while admitting instabilities of relatively short waves
\cite{MillerB98}.  Interestingly, analyticity of initial data as
required by elliptic modulation equations plays a fundamental role in
the spectral and inverse-spectral analysis of the focusing NLS
equation in the semiclassical limit \cite{KMM}, even though the latter
analysis is not based on modulation equations at all.  (For an
approach based directly on the Cauchy-Kovalevskaya series solution of
the analytic data Cauchy problem for the elliptic modulation
equations, see G\'erard \cite{Gerard}.)

One might notice that in the linearization calculation the conclusion
of stability or instability was independent of the pair $(k,A)$ that
characterizes the underlying plane wave.  Indeed, in the focusing
case, there is a band of unstable relative wavenumbers no matter what
values $k$ and $A$ take.  Similarly, the modulation equations
\eqref{modfocNLS} are elliptic or hyperbolic independently of
$u\in\mathbb{R}$ and $\rho>0$.  In principle, it could have been
otherwise; some waves may be stable to all sideband perturbations
while others are not.  For such a system, the modulation equations may
be hyperbolic for some values of $u$ and $\rho$ and elliptic for
others, and the system is said to admit a \emph{change of type}.  It
will turn out that the system of modulation equations for the MNLS
equation admits a change of type.

To summarize, the focusing NLS equation exhibits modulational instability,
while the defocusing NLS equation is modulationally stable.  Now, the
MNLS equation \eqref{MNLScauchy} appears to be a perturbation of the focusing
NLS equation (through the limit $\alpha\rightarrow 0$).  One might expect
on these grounds that the MNLS problem should also experience modulational
instability.  On the other hand, 
consider making the following substitution in \eqref{MNLScauchy}:
\begin{equation}
\phi(x,t)=e^{i(c\xi+c^2\tau/2)/\e}\psi(\xi,\tau)\,,
\label{eq:Galilean}
\end{equation}
where $\xi=x-ct$ and $\tau=t$, and $c\in\mathbb{R}$ is an arbitrary
parameter.  Then, by direct calculation, $\psi$ satisfies
\begin{equation}
i\e\frac{\partial \psi}{\partial \tau} +\frac{\e^2}{2}
\frac{\partial^2 \psi}{\partial \xi^2} + (1-\alpha c)|\psi|^2\psi + i\alpha\e
\frac{\partial}{\partial \xi}(|\psi|^2\psi)=0\,.
\end{equation}
If $\alpha=0$, we recover the well-known result that the focusing NLS
equation is invariant under the one-parameter group of Galilean
transformations \eqref{eq:Galilean}.  More generally, the MNLS
equation is certainly not invariant under this group.  In particular,
one may note that by choosing $c$ so that $\alpha c>1$, the sign of
the $|\psi|^2\psi$ term becomes negative\footnote{If $c$ is chosen so that
  $\alpha c=1$ exactly, then under the Galilean boost the MNLS
  equation for $\phi$ reduces to the so-called derivative NLS equation
  for $\psi$.}, and thus the equation governing $\psi(\xi,\tau)$ takes the
form of a perturbed \emph{defocusing} NLS equation rather then a
perturbation of the \emph{focusing} NLS equation.

Given that when $\alpha=0$ one expects drastically different behavior
in the semiclassical limit depending on whether the sign of the cubic
term is positive (strong modulational instability) or negative
(modulational stability), the above calculation suggests two things:
(i) the term in the MNLS equation \eqref{MNLScauchy} proportional to
$\alpha$ certainly does not constitute a small perturbation uniformly
in the phase space of fields $\phi$ (that is, taking $\alpha\neq 0$
amounts to a singular perturbation of the NLS equation), and (ii) that
part of the phase space of fields $\phi$ may evolve under the MNLS
equation \eqref{MNLScauchy} in such a way as to {\em avoid all
  catastrophic modulational instability} associated with the
undifferentiated cubic term $|\phi|^2\phi$.  If this latter statement can be
placed on rigorous footing, there are immediate implications for the
design of optical fiber telecommunication links (for example).
Indeed, one of the difficulties arising in long-distance fiber-optic
systems is the so-called Gordon-Haus effect, in which pulses
experience a random ``jitter'' in their arrival times which leads to
detection errors.  This effect has been shown to be a consequence of
the modulational instability associated with the focusing nonlinearity
(and anomalous dispersion) in the focusing NLS model.  The possibility
of avoiding Gordon-Haus jitter in the context of the MNLS model has
already been explored \cite{DK} at the level of single soliton
solutions.  One of our goals is to understand this change in stability
for the MNLS equation at the level of more general initial data, and
in particular initial data that ``contain'' many solitons as is
characteristic of the semiclassical scaling.

Due to its integrable structure, solving the Cauchy problem for the
MNLS equation \eqref{MNLScauchy} requires three steps:  
\begin{itemize}
\item[] {\bf Step 1.}  Calculate certain ``spectral data'' associated
  with the given initial condition for the problem, viewed here as a
  coefficient in a linear system of differential equations with a
  spectral parameter.  This is a ``forward transform'' of the initial
  data.
\medskip
\item[] {\bf Step 2.} Determine how the spectral data change as
  the field $\phi$ evolves under the MNLS equation, and obtain the
  spectral data corresponding to the evolved field.
\medskip
\item[] {\bf Step 3.} Find the field $\phi(x,t)$ from the time-evolved
  spectral data, that is, invert the mapping from step 1.  This is an
  ``inverse transform'' step that in many cases can be written as a
  matrix-valued Riemann-Hilbert problem from analytic function theory.
\end{itemize}
The miracle of integrability is that for an appropriate linear system
of differential equations with spectral parameter, the inverse of the
spectral map of step 1 that is required in step 3 actually exists, and
as importantly the evolution of the spectral data in step 2 is
completely explicit.  For semiclassical problems, the small parameter
$\e$ enters significantly in the first and third steps.  Therefore, to
understand the stability transition in the MNLS equation and its role
in semiclassical asymptotics, we must first understand the forward
transform (step 1) in the semiclassical regime.  Once the the
semiclassical asymptotics of the forward transform have been
understood, it will remain to utilize the growing library of tools of
asymptotic analysis for Riemann-Hilbert problems to complete the
solution of the Cauchy problem for the MNLS equation in the limit
$\e\downarrow 0$.  In this paper, we will consider the spectral
problem associated with the MNLS equation (see \eqref{eq:Laxx})
paying particular attention to results that are meaningful in the
limit $\e\downarrow 0$.

This paper is organized as follows.  In Section \ref{sec:Modeqns} we
follow the procedure applicable to the NLS equation to obtain the
modulation equations associated to the MNLS equation.  Through the
analysis of these modulation equations we will obtain a condition for
modulational stability. Throughout the remainder of the paper we note
the importance of this condition and the role it plays in the spectral
problem. In Section \ref{sec:hyperbolae} we establish some bounds on
the discrete eigenvalues of the spectral problem \eqref{eq:Laxx}
associated to the MNLS equation. In Section \ref{sec:hypergeometric}
we use the theory of special functions to calculate the spectral data
explicitly for a certain multiparameter family of initial conditions;
this calculation is valid for all $\e>0$ and provides an explicit
avenue to the semiclassical analysis of the corresponding Cauchy
problem.  The appendix contains the general spectral and
inverse-spectral theory associated with the MNLS equation (that is,
all theoretical details of steps 1--3 above) and in particular
includes a derivation of the associated Riemann-Hilbert problem that
we will return to in a later publication.  The theory outlined in the
appendix frames all of our analysis of the MNLS equation.

\subsection*{Notation.}  Throughout the paper, we use boldface capital letters
to refer to square matrices, with the exception of the identity matrix which
we write as $\mathbb{I}$, the Pauli matrices
\begin{equation}
\sigma_1:=\begin{bmatrix}0 & 1\\1 & 0\end{bmatrix}\,,\quad
\sigma_2:=\begin{bmatrix}0 & -i\\i & 0\end{bmatrix}\,,\quad
\text{and}\quad
\sigma_3:=\begin{bmatrix}1 & 0\\0 & -1\end{bmatrix}\,,
\end{equation}
and certain diagonal matrices denoted $e^{a\sigma_3}$
for $a\in\mathbb{C}$ and defined by
\begin{equation}
e^{a\sigma_3}:=\begin{bmatrix}e^a & 0\\0 & e^{-a}\end{bmatrix}\,.
\end{equation}
Lowercase boldface letters refer to column vectors.  We use the
``dagger'' notation (superscript $\dagger$) to indicate the transpose
and elementwise complex conjugate of a matrix or vector, and we use
asterisk notation (superscript $*$) for complex conjugation.  We
sometimes use prime notation for derivatives of functions of just one
variable.

\section{Modulation Equations}
\label{sec:Modeqns}
In this section we derive the modulation equations associated to the
MNLS equation and find a condition on the local amplitude and
wavenumber equivalent to the hyperbolicity of these equations.  By
analogy with the connection between hyperbolicity of modulation
equations and stability of plane waves described in the context of the
NLS equation in Section \ref{sec:intro}, this calculation gives a
condition under which the MNLS equation is (locally) modulationally stable.
The fact that one can have local stability without global stability is
related to the fact the MNLS modulation equations admit change of type.

\subsection{Derivation of the modulation equations.}
Consider the MNLS equation \eqref{MNLScauchy}, and for some real-valued
functions $A(x,t)$ and $S(x,t)$ write the solution $\phi(x,t)$ as:
\begin{equation}
\phi(x,t)=A(x,t)e^{iS(x,t)/\e}\,.
\end{equation}
Substitution into (\ref{MNLScauchy}), canceling the phase factor
$e^{iS/\e}$, taking the imaginary part, multiplying by $A$ and
dividing by $\e$ gives:
\begin{equation}\label{ModA}
  \frac{\partial}{\partial t}A^2+
  \frac{\partial}{\partial x}\left(\frac{\partial S}{\partial x}A^2+
\frac{3\alpha}{2}A^4\right)=0\,.
\end{equation}
Proceeding similarly but taking the real part instead, one arrives at:
\begin{equation}\label{Modsx}
  \frac{\partial}{\partial t}\,\frac{\partial S}{\partial x}+
\frac{\partial}{\partial x}
\left(\frac{1}{2}\left(\frac{\partial S}{\partial x}\right)^2-A^2+
\alpha \frac{\partial S}{\partial x}A^2\right)=
\frac{\e^2}{2}\frac{\partial}{\partial x}
\left(\frac{1}{A}\frac{\partial^2 A}{\partial x^2}\right).
\end{equation}
Setting $\rho(x,t):=A(x,t)^2$ and $u(x,t):=\partial S(x,t)/\partial
x$, the system of {\em modulation equations} is obtained by dropping
the formally small terms in $\e$:
\begin{equation}\label{modeq}
\frac{\partial}{\partial t}
\begin{bmatrix} \rho\\ u\end{bmatrix}+
\begin{bmatrix} 3\alpha\rho+u & \rho\\ \alpha u-1& u+\alpha\rho\end{bmatrix}
\frac{\partial}{\partial x}\begin{bmatrix} \rho\\ u \end{bmatrix}=0.
\end{equation}

\subsection{Modulational stability criterion.}
The eigenvalues $\zeta$ of the coefficient matrix of the
$x$-derivatives in \eqref{modeq} are the roots of the equation:
\begin{equation}\label{eigcond}
\zeta^2-\left(4\alpha\rho+2u\right)\zeta+3\alpha u\rho+3\alpha^2\rho^2+u^2+\rho=0.
\end{equation}
The discriminant of this equation is $4\rho(\alpha^2\rho+\alpha u-1)$. Since $\rho>0$, the condition
\begin{equation}\label{stabcond}
\alpha^2\rho+\alpha u-1>0,
\end{equation}
is equivalent to having real distinct eigenvalues, which makes
\eqref{modeq} a hyperbolic system.  Note that for a given value of the
parameter $\alpha\in\mathbb{R}$, this condition involves both
dependent variables $\rho$ (square of local plane-wave amplitude) and
$u$ (local wavenumber) and this dependence is characteristic of a
quasilinear system that can change type.  The inequality
\eqref{stabcond} will appear at every stage of our analysis of the
spectral problem (see \eqref{eq:Laxx}) associated with the MNLS
equation.

\subsection{Implications of the stability criterion on Riemann-Hilbert analysis
of the inverse spectral problem.}
\label{sec:stabilityRHP}
While this paper is concerned mostly with the semiclassical limit of
the spectral transform associated with the MNLS equation and a
subsequent paper will analyze the corresponding inverse-spectral
transform in this limit, it is worthwhile pointing out how we might
expect the stability criterion \eqref{stabcond} to influence the
inverse-spectral part of the solution of the MNLS Cauchy problem for
rapidly decaying initial data of the form $\phi(x,0)=A(x)e^{iS(x)/\e}$,
$x\in\mathbb{R}$.  

How do modulated plane waves emerge from the asymptotic analysis of
the inverse-spectral problem?  As shown in the appendix, the
inverse-spectral problem for the MNLS Cauchy problem can be cast as a
Riemann-Hilbert problem in which one seeks a $2\times 2$ matrix-valued
unknown depending principally upon a complex variable $k$ and
parametrically upon $x$, $t$, $\alpha$, and $\e$.  The unknown matrix
is characterized by a normalization condition as $k\rightarrow 0$,
prescribed discontinuities along the real and imaginary $k$-axes,
simple poles with prescribed residues at certain other points in the
complex $k$-plane, and analyticity in $k$ for all other
$k\in\mathbb{C}$.  Such a Riemann-Hilbert problem involving a
parameter $\e$ in a singular way may be analyzed by a two-step
process.  First, one introduces an explicit transformation of the
matrix-valued unknown designed to remove the poles at the expense of
introducing new jump discontinuities.  Thus one arrives at an
equivalent Riemann-Hilbert problem for a piecewise-analytic
matrix-valued function of $k$ whose jump discontinuities lie along a
system, $\Sigma$, of contours including (generally) the real and
imaginary $k$-axes.  Second, one introduces a so-called $g$-function
(an innovation discovered in \cite{DVZ94}), a scalar function of $k$
analytic for $k\in\mathbb{C}\setminus\Sigma$, and by multiplying the
matrix-valued unknown on the right by the diagonal matrix
$e^{g(k)\sigma_3/\e}$ one arrives at a second equivalent
Riemann-Hilbert problem on the same system $\Sigma$ of contours but
for which the jump conditions involve the boundary values of $g$.  The
function $g$ is then chosen to cast the jump conditions for the matrix
unknown in as simple a form as possible.  By the phrase ``as simple a
form as possible,'' we really mean that $g$ is chosen so that the
steepest descent method of Deift and Zhou \cite{DZ93} applies,
implying a set of conditions that one expects to uniquely determine
$g$.  A matrix-valued Riemann-Hilbert problem to which the steepest
descent method applies is one for which the ratio of the boundary
values of the unknown is a matrix taking two alternate forms in
complementary systems of subintervals of $\Sigma$.  The number of
subintervals and their common endpoints depend on the remaining
parameters of the problem, including $x$ and $t$.  

The common endpoints $k=k_j(x,t)$ of these subintervals have special
significance as branching points of the Lax eigenfunctions (solutions
${\bf v}$ of the linear equation \eqref{eq:Laxx}; see the appendix)
corresponding to the local asymptotic solution of the MNLS equation in
a small neighborhood of $(x,t)$.  If there are just four of these
endpoints, then one expects the local solution to have plane-wave form
$\phi(x,t)=A(x,t)e^{iS(x,t)/\e}+O(\e)$, where one may consider $\rho=A^2$
and $u=\partial S/\partial x$ to be \emph{constants}.  To understand
the interpretation of the branching points $k_j$ in terms of $\rho$
and $u$, note that the relevant scattering problem (see
\eqref{system6} below) is written in terms of $\rho$ and $u$ as
\begin{equation}
2\alpha\e\frac{d{\bf w}}{dx}=i\begin{bmatrix}
-4k^2+1-\alpha u & 4\alpha \rho^{1/2}k \\
4\alpha \rho^{1/2} k & 4k^2-1+\alpha u\end{bmatrix}{\bf w}\,.
\end{equation}
With $\rho$ and $u$ constant, this is a constant-coefficient linear problem with
eigensolutions ${\bf w}(x)=e^{E_\pm x/\e}{\bf w}_\pm$, where
\begin{equation}
i\begin{bmatrix}
-4k^2+1-\alpha u & 4\alpha \rho^{1/2}k \\
4\alpha \rho^{1/2} k & 4k^2-1+\alpha u\end{bmatrix}
{\bf w}_\pm=2\alpha
E_\pm{\bf w}_\pm\,.
\end{equation}  
The eigenvalues are
\begin{equation}
2\alpha E_\pm=\pm i\sqrt{16\alpha^2\rho k^2+(4k^2+\alpha u-1)^2}\,,
\end{equation}
and branching points of the eigenfunctions in the complex
$k$-plane will occur where $E_+=E_-$ or equivalently where $E_\pm=0$.
There are therefore at most four branching points satisfying
\begin{equation}
4k_j^2=-2\alpha^2\rho-\alpha u+1\pm  
2\alpha\sqrt{(\alpha^2\rho+\alpha u-1)\rho}\,.
\end{equation}

It follows that if (i) steepest descent analysis of the inverse-spectral
Riemann-Hilbert problem in the limit $\e\downarrow 0$ leads to a
solution of the MNLS equation of the form $\phi=A(x,t)e^{iS(x,t)/\e} +
O(\e)$ where $\rho(x,t)=A^2$ and $u(x,t)=\partial S/\partial x$, and
(ii) the plane wave characterized by $\rho$ and $u$ is modulationally
stable according to the hyperbolicity criterion \eqref{stabcond}, then
it is necessary that the $g$-function of steepest descent theory
provides four endpoints $k_j(x,t)$ all lying on the imaginary axis.
On the other hand, if $\rho$ and $u$ characterize a modulationally
unstable plane wave, then the four endpoints $k_j(x,t)$ will need to
form a quartet invariant under the reflections $k\rightarrow k^*$ and
$k\rightarrow -k^*$.  

The unstable case of course requires that the contour system, $\Sigma$,
contain components that are disjoint from the real and imaginary
$k$-axes, and determining the correct boundary values of the function
$g$ on such a contour further requires comparing with certain spectral
functions computed on the real and imaginary $k$-axes (see the
appendix) that are analytically continued to the complex $k$-plane.
This process of analytic continuation, that we expect to only be
required in the unstable case, is the spectral analogue of analytic
continuation of the functions $\rho(x,0)$ and $u(x,0)$ to the complex
$x$-plane as would be necessary to solve the Cauchy initial-value
problem for the elliptic modulation equations.

\subsection{Other forms of the modulation equations.}
\subsubsection{Local conservation laws.}
\label{sec:Localconslaws}
With the dependent variable $v:=\rho u+\alpha \rho^2$ used instead of $u$,
the system \eqref{modeq} takes the form
 \begin{equation}\label{Modrho}
\begin{split}
\frac{\partial\rho}{\partial t}+\frac{\partial}{\partial x}
\left(v-\alpha\rho^2+\frac{3\alpha}{2}\rho^2\right)&=0\,,\\
\frac{\partial v}{\partial t}+\frac{\partial}{\partial x}
\left(\frac{v^2}{\rho}+\alpha\rho v-\frac{\rho^2}{2}\right)&=0\,.
\end{split}
\end{equation}
That is, \eqref{modeq} may be written as a system of local
conservation laws for densities $\rho$ and $v$.  Strictly speaking,
the system \eqref{modeq} is already in this form:
\begin{equation}
\begin{split}
\frac{\partial\rho}{\partial t} +\frac{\partial}{\partial x}
\left(\rho u + \frac{3\alpha}{2}\rho^2\right) &=0\,,\\
\frac{\partial u}{\partial t} +\frac{\partial}{\partial x}
\left(\alpha\rho u-\rho + \frac{1}{2}u^2\right)&=0\,;
\end{split}
\end{equation}
however since $u=\partial S/\partial x$, the local conservation law for
$u$ is trivial.  Consequently one should seek a second conserved local
density that is not a derivative with respect to $x$ of a local
expression in $A$ and $S$.  It is well-known and easy to verify that
the \emph{momentum} $v_o:=\rho u$ is a nontrivial conserved local density
for the modulation equations \eqref{modfocNLS} associated to the
focusing NLS equation, and the nontrivial conserved local density $v$
for \eqref{modeq} (or equivalently, \eqref{Modrho}) is clearly an
$\alpha$-perturbation of the momentum.  

In the paper \cite{DesjardinsLT00} the quantity $v$ is called a
\emph{noncanonical momentum}.  The authors of that paper give an
analysis of the (equivalent, by a Galilean boost with speed
$c=\alpha^{-1}$) derivative NLS equation in the semiclassical limit.
In the spirit of the paper of Grenier \cite{Grenier}, the authors use
the system \eqref{Modrho} (with $u$ replaced by $u+\alpha^{-1}$ as is
appropriate for the derivative NLS equation rather than the MNLS
equation) to model the behavior of $\psi(x,t)$ before singularities form
in the solution of the limiting system.  The development in the paper
\cite{DesjardinsLT00} appears to assume, however, that the system of
conservation laws \eqref{Modrho} is hyperbolic for arbitrary initial
data, and we have seen (see \eqref{stabcond} which becomes simply
$\alpha^2\rho+\alpha u>0$ when $u$ is replaced by $u+\alpha^{-1}$)
that this is not the case.

\subsubsection{Connection with focusing NLS in the semiclassical limit.}
\begin{figure}[h]
\begin{center}
\includegraphics{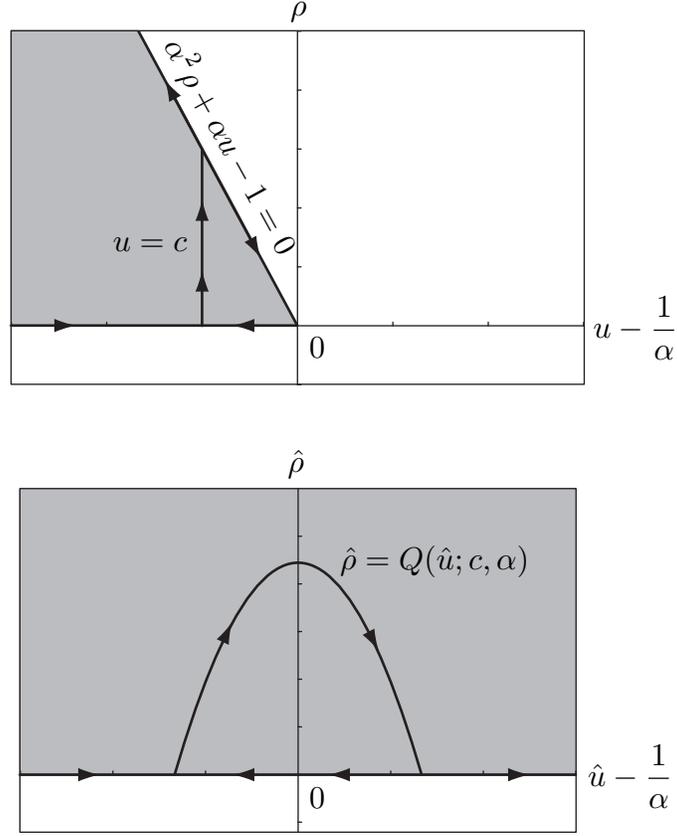}
\end{center}
\caption{\em Top:  the domain of $F_-$ is shaded.  Bottom:  the range of $F_-$ 
is shaded.  
Note that $Q(\hat{u};c,\alpha):=[(c-\alpha^{-1})^2-(\hat{u}-\alpha^{-1})^2]/4$.}
\label{fig:Fminus}
\end{figure}

Suppose that $\alpha^2\rho+\alpha u-1<0$ and that $\rho>0$, defining
an open domain $D_-(\alpha)\subset\mathbb{R}^2$ for each $\alpha>0$.
Then consider the map $F_-: (\rho,u)\in D_-(\alpha)\mapsto
(\hat{\rho},\hat{u})\in\mathbb{R}^2$ defined by the formulae
\begin{equation}
\hat{\rho}=-\rho\cdot (\alpha^2\rho+\alpha u-1)\,, \quad\quad
\hat{u}=u+2\alpha\rho\,.
\end{equation}
The mapping $F_-$ is one-to-one and maps $D_-(\alpha)$ onto the 
upper half-plane 
\begin{equation}
\text{$R_-=\{(\hat{\rho},\hat{u})\in\mathbb{R}^2$ such
that $\hat{\rho}>0\}$}\,.
\end{equation}
The inverse $F_-^{-1}:R_-\to D_-(\alpha)$ is given by the formulae
\begin{equation}
\rho=\frac{1}{2\alpha^2}\left(\alpha \hat{u}-1+\sqrt{(\alpha\hat{u}-1)^2+4\alpha^2\hat{\rho}}\right)\,,\quad\quad
u=\frac{1}{\alpha}\left(1-
\sqrt{(\alpha\hat{u}-1)^2+4\alpha^2\hat{\rho}}\right)\,,
\end{equation}
where the positive square root is meant.  The mapping $F_-$ is illustrated
in Figure~\ref{fig:Fminus}.

It is a direct calculation to
check that when the MNLS modulation equations are restricted to functions
$(\rho(x,t),u(x,t))\in D_-(\alpha)$, they take the form
\begin{equation}
\frac{\partial}{\partial t}\begin{bmatrix}\hat{\rho}\\\hat{u}\end{bmatrix}
+\begin{bmatrix}\hat{u} &\hat{\rho}\\-1 &\hat{u}\end{bmatrix}
\frac{\partial}{\partial x}\begin{bmatrix}\hat{\rho}\\\hat{u}\end{bmatrix}
=0
\label{eq:modfocNLSagain}
\end{equation}
in the new variables $(\hat{\rho},\hat{u})$.  Comparing with
\eqref{modfocNLS} for $\kappa=1$ we see that on the modulationally
unstable sector of the phase space the formal semiclassical limit for
the MNLS equation is, when viewed in the correct variables
$(\hat{\rho},\hat{u})$, the same\footnote{Note however, that the
  mapping $F_-$ does not appear to establish any equivalence between
  the MNLS equation and the focusing NLS equation at the level of
  finite nonzero $\e$.  We do not know whether the $\e$-independent
  mapping $F_-$ can be prolonged into a gauge transformation relating
  the MNLS equation to the focusing NLS equation for general $\e$.
  However, in \cite{Hayashi93} (see also \cite{Ozawa96}) Hayashi has
  shown that if $\psi(\xi,\tau)$ is a solution of the derivative NLS
  equation
\[
i\e\frac{\partial\psi}{\partial \tau} + 
\frac{\e^2}{2}\frac{\partial^2\psi}{\partial\xi^2} + i\alpha\e\frac{\partial}
{\partial \xi}\left(|\psi|^2\psi\right)=0
\]
(which is itself equivalent to the MNLS equation via a Galilean boost
with velocity $c=1/\alpha$ as discussed in the introduction) then
\[
f:=\sqrt{2\alpha}
\exp\left(\frac{2i\alpha}{\e}\int_{-\infty}^\xi|\psi(\xi',\tau)|^2\,d\xi'
\right)\psi\;\;\text{and}\;\;
g:=\sqrt{2\alpha}
\exp\left(\frac{2i\alpha}{\e}\int_{-\infty}^\xi|\psi(\xi',\tau)|^2\,d\xi'
\right)\left[\e\frac{\partial\psi}{\partial \xi}+i\alpha|\psi|^2\psi\right]
\]
satisfy the coupled system
\[
i\e\frac{\partial f}{\partial\tau} +
\frac{\e^2}{2}\frac{\partial^2 f}{\partial\xi^2}-if^2g^*=0\quad\quad
\text{and}\quad\quad
i\e\frac{\partial g}{\partial\tau} +
\frac{\e^2}{2}\frac{\partial^2 g}{\partial\xi^2}+i g^2f^*=0
\]
which reduces to the focusing NLS equation if $g=-if$ and to
the defocusing NLS equation if $g=if$.  However neither of these reductions
is consistent with the definition of $f$ and $g$ in terms of $\psi$.} as that
for the focusing NLS equation.

\begin{figure}[h]
\begin{center}
\includegraphics{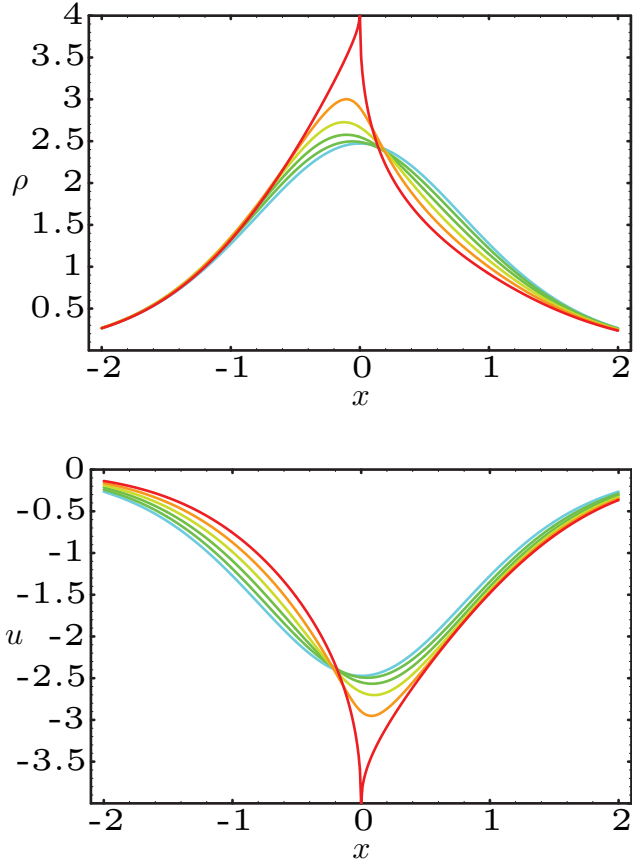}
\end{center}
\caption{\em Snapshots of the Akhmanov-Sukhorukov-Khokhlov solution of
  the focusing NLS modulation equations interpreted via $F_-^{-1}$ as a
  solution of the MNLS modulation equations for $\alpha=1/2$.  The initial
conditions are shown with blue curves and the snapshot corresponding to
the singularity formation time $t=1/4$ is drawn with red curves.}
\label{fig:ASKhalf}
\end{figure}

\begin{figure}[h]
\begin{center}
\includegraphics{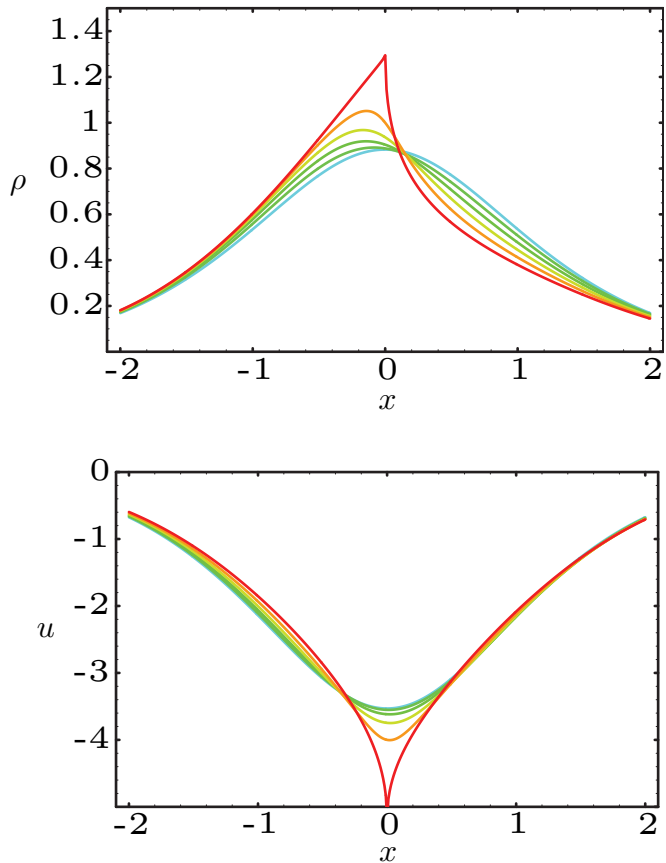}
\end{center}
\caption{\em Snapshots of the Akhmanov-Sukhorukov-Khokhlov solution of
  the focusing NLS modulation equations interpreted via $F_-^{-1}$ as
  a solution of the MNLS modulation equations for $\alpha=2$.  The color
scheme is the same as in Figure~\ref{fig:ASKhalf}.}
\label{fig:ASKtwo}
\end{figure}

This identification allows us to immediately apply certain known results for
the focusing NLS equation to the MNLS equation.  For example, a particular
solution to \eqref{eq:modfocNLSagain} was obtained in implicit algebraic
form by Akhmanov, Sukhorukov, and Khokhlov \cite{ASK66} in 1966:
\begin{equation}
\hat{\rho}=(A_o^2 + t^2\hat{\rho}^2)\,\text{sech}^2(x-\hat{u}t)\,,\quad\quad
\hat{u}=-2t\hat{\rho}\tanh(x-\hat{u}t)\,.
\label{eq:ASKimplicit}
\end{equation}
Here $A_o>0$ is an amplitude parameter, and of course $x$ and $t$ are the
independent variables in \eqref{eq:modfocNLSagain}.  Setting $t=0$ gives
the initial conditions
\begin{equation}
\hat{\rho}(x,0)=A_o^2\,\text{sech}^2(x)\,,\quad\quad\hat{u}(x,0)\equiv 0\,,
\label{eq:ASKIC}
\end{equation}
(which corresponds to a ``chirp-free, return-to-zero'' pulse in the
optical fiber context) and for small $t$ the Implicit Function Theorem
applies to \eqref{eq:ASKimplicit} and allows us to solve for
$\hat{\rho}(x,t)$ and $\hat{u}(x,t)$.  The earliest positive $t$ for
which the Implicit Function Theorem fails is $t=(2A_o)^{-1}$, and the
singularity occurs for $x=0$.  For the implicit solution of Akhmanov,
Sukhorukov, and Khokhlov, $\hat{\rho}(x,t)$ is an even function of $x$
and $\hat{u}(x,t)$ is an odd function of $x$ for each $t\in
(0,(2A_o)^{-1})$.  For any $\alpha>0$ it is possible to compose the
Akhmanov-Sukhorokov-Khokholov solution with the transformation
$F_-^{-1}$ and thus obtain a solution of the MNLS modulation equations
that remains restricted to the modulationally unstable sector of the
phase space.  The transformation $F_-^{-1}$ breaks the even/odd symmetry
of the functions $\hat{\rho}$ and $\hat{u}$.  Indeed, the initial conditions
\eqref{eq:ASKIC} both become even under $F_-^{-1}$:
\begin{equation}
\rho(x,0)=-\frac{1}{2\alpha^2}\left(1-\sqrt{1+4\alpha^2
A_o^2\,\text{sech}^2(x)}\right)\,,
\quad
u(x,0)=\frac{1}{\alpha}\left(1-\sqrt{1+4\alpha^2A_o^2\,
\text{sech}^2(x)}\right)\,.
\end{equation}
These have the following asymptotics in $\alpha$:
\begin{equation}
\rho(x,0)=A_o^2\,\text{sech}^2(x)+O(\alpha^2)\,,\quad u(x,0)=
-2\alpha A_o^2\,\text{sech}^2(x)+O(\alpha^3)\,,\quad
\text{as $\alpha\rightarrow 0$}
\end{equation}
and
\begin{equation}
\rho(x,0)=A_o\alpha^{-1}\,\text{sech}(x) + O(\alpha^{-3})\,,\quad
u(x,0)=-2A_o\,\text{sech}(x) + O(\alpha^{-2})\,,\quad
\text{as $\alpha\rightarrow\infty$}\,,
\end{equation}
which shows that the MNLS pulse corresponding to the
Akhmanov-Sukhorukov-Khokhlov solution has ``fatter tails'' when
$\alpha$ is larger, but is of lower amplitude.  The even symmetry of
$\rho$ and $u$ at $t=0$ is, however, irrelevant under evolution.  To
show the dynamics of the Akhmanov-Sukhorukov-Khokhlov solution
composed with $F_-^{-1}$ for different $\alpha$ we have numerically
inverted the implicit relations \eqref{eq:ASKimplicit} and composed
explicitly with $F_-^{-1}$, plotting the solution snapshots for six
equally spaced times $t_k\in [0,(2A_o)^{-1}]$ for $A_o=2$ and $\alpha=1/2$
(Figure~\ref{fig:ASKhalf}) and for $A_o=2$ and $\alpha=2$
(Figure~\ref{fig:ASKtwo}).  In both cases we see leftward-leaning
pulses evolving toward a singularity at $t=(2A_o)^{-1}=1/4$ and $x=0$.  

\subsubsection{Connection with defocusing NLS in the semiclassical limit.  
Riemann invariants.}
\begin{figure}[h]
\begin{center}
\includegraphics{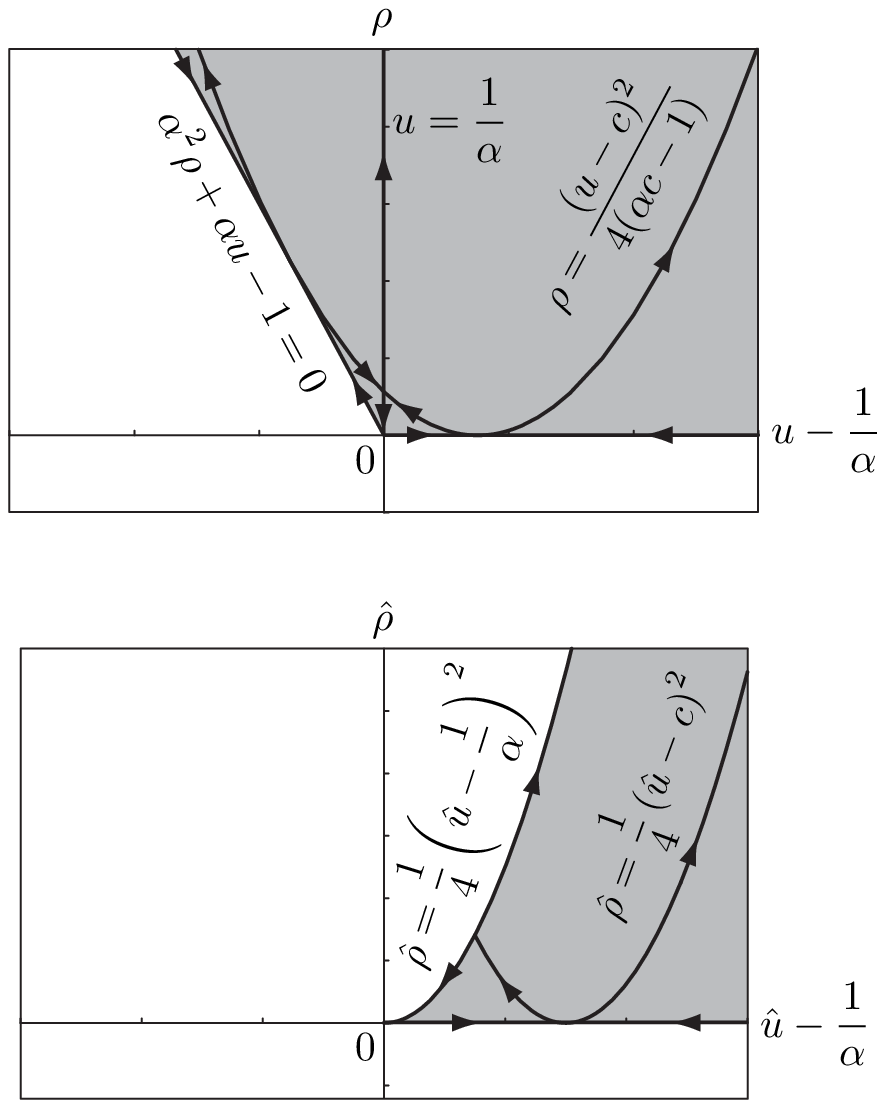}
\end{center}
\caption{\em Top:  the domain of $F_+$ is shaded.  Bottom:  the range of $F_+$ 
is shaded.}
\label{fig:Fplus}
\end{figure}

Now suppose that $\alpha^2\rho+\alpha u-1>0$ and $\rho>0$, defining an open
domain $D_+(\alpha)\subset\mathbb{R}^2$ for each $\alpha>0$, and consider the
map $F_+:(\rho,u)\in D_+(\alpha)\mapsto(\hat{\rho},\hat{u})\in\mathbb{R}^2$
defined by the formulae
\begin{equation}
\hat{\rho}=\rho\cdot(\alpha^2\rho+\alpha u-1)\,,\quad\quad
\hat{u}=u+2\alpha\rho\,.
\end{equation}
The range of the mapping $F_+$ is a strict subset of the upper half-plane:
\begin{equation}
\text{$R_+(\alpha)=\{(\hat{\rho},\hat{u})\in\mathbb{R}^2$ such that 
$0<\hat{\rho}\le (\hat{u}-\alpha^{-1})^2/4\}$}\,.
\end{equation}
Unlike $F_-$, the map $F_+$ is not one-to-one.  Each point in the
interior of $R_+(\alpha)$ has exactly two preimages in $D_+(\alpha)$,
one with $u<\alpha^{-1}$ and one with $u>\alpha^{-1}$.  On the other
hand, each point on the boundary curve
$\hat{\rho}=(\hat{u}-\alpha^{-1})^2/4$ has exactly one preimage in
$D_+(\alpha)$, satisfying $u=\alpha^{-1}$.  See Figure~\ref{fig:Fplus}.

Again, by direct calculation, the MNLS modulation equations \eqref{modeq}
for $\rho(x,t)$ and $u(x,t)$ are transformed under $F_+$ into the system
\begin{equation}
\frac{\partial}{\partial t}\begin{bmatrix}\hat{\rho}\\\hat{u}\end{bmatrix}
+\begin{bmatrix}\hat{u} &\hat{\rho}\\1 & \hat{u}\end{bmatrix}
\frac{\partial}{\partial x}\begin{bmatrix}\hat{\rho}\\\hat{u}\end{bmatrix}
=0\,,
\label{eq:moddefocNLS}
\end{equation}
which upon comparison with \eqref{modfocNLS} for $\kappa=-1$ we
recognize as the modulation equations for the defocusing NLS equation.
Once again, we are led to expect\footnote{with the same caveat as
  indicated in the previous footnote} that in the semiclassical limit
of $\e\downarrow 0$, the dynamics of the MNLS equation on the stable
sector of its phase space should be equivalent (under the mapping
$F_+$) to the dynamics of the defocusing NLS equation.

In principle, every hyperbolic quasilinear system with two dependent
variables can be written in diagonal form through the introduction of
new dependent variables called \emph{Riemann invariants}.  A suitable
definition (see, for example, \cite{I}) of Riemann invariants for the
system \eqref{eq:moddefocNLS} is
\begin{equation}
R_\pm:=\frac{\hat{u}}{2}\pm\sqrt{\hat{\rho}}\,,
\end{equation}
where the positive square root is meant.  Indeed, in terms of these
variables, the system \eqref{eq:moddefocNLS} takes diagonal form:
\begin{equation}\label{RIevolution}
\begin{split}
\frac{\partial R_+}{\partial t}+\frac{1}{2}\left[\vphantom{\hat{R}_+}3R_++R_-\right]\frac{\partial R_+}{\partial x}&=0\,,\\
\frac{\partial R_-}{\partial t}+\frac{1}{2}\left[\vphantom{\hat{R}_+}R_++3R_-\right]\frac{\partial R_-}{\partial x}&=0\,.
\end{split}
\end{equation}
The quantities
\begin{equation}
\sigma_\pm:=\frac{1}{2}\left[\vphantom{\hat{R}_+}3R_\pm +R_\mp\right]
\end{equation}
are the \emph{characteristic velocities}, that is, the eigenvalues of
the coefficient matrix of \eqref{eq:moddefocNLS} written in terms of
the Riemann invariants.  The parabolae in the range $R_+(\alpha)$ of
$F_+$, as shown in the bottom diagram of Figure~\ref{fig:Fplus}, are
coordinate curves for the Riemann invariants; indeed, the relation
$\hat{\rho}=(\hat{u}-c)^2/4$ corresponds to the ray $R_+=c/2$, $R_-<c/2$
if $u<c$ and the ray $R_-=c/2$, $R_+>c/2$ if $u>c$.

Diagonal form made possible by the introduction of Riemann invariants
is useful for many purposes.  For example, the way we discovered that
the transformation $F_+$ transforms the MNLS modulation equations
\eqref{modeq} on the modulationally stable sector into the defocusing
NLS modulation equations \eqref{eq:moddefocNLS} was by first writing
\eqref{modeq} in diagonal form and subsequently comparing with the
known diagonal form of \eqref{eq:moddefocNLS}.  (Later we generalized
this result to obtain the corresponding result that $F_-$ links
\eqref{modeq} on its modulationally unstable sector to
\eqref{eq:modfocNLSagain} although as both are elliptic systems real
Riemann invariants cannot exist as an intermediate step.)  Another
advantage of diagonal form is that one can easily see that there are
nontrivial solutions for which either $R_+$ or $R_-$ is identically
constant.  Such solutions are called \emph{simple waves} and their
dynamics are characterized by the scalar inviscid Burgers equation.
Indeed, if $R_\pm=c/2$, then it is easy to show that $\sigma_\mp$
evaluated for $R_\pm\equiv c/2$ and $R_\mp$ arbitrary satisfies
\begin{equation}
\frac{\partial\sigma_\pm}{\partial t}+\sigma_\pm\frac{\partial\sigma_\pm}{\partial x}=0\,,
\end{equation}
whose general solution with arbitrary initial condition
$\sigma_\pm(x,0)=f(x)$ can be written in implicit form as
\begin{equation}
\sigma_\pm = f(x-\sigma_\pm t)\,.
\end{equation}
As is well-known, if $f$ has critical points, then this solution only
exists in the classical sense for a finite time due to the formation
of shocks (derivative singularities).  The number of shocks that can
form is bounded above by the number of critical points of $f$ (whether
or not more than one shock forms is a sensitive question of
regularization that we do not address here; in particular we are
ultimately interested in dispersive regularization where an
early-occurring shock can ``spread'' into microscopic oscillations and
wipe out other shocks that might occur in nonconservative
regularizations).  A final advantage of diagonal form that is useful
in considering solutions more general than simple waves is that as
long as a solution remains classical it is bounded in terms of its
initial data by sharp inequalities of the form
\begin{equation}
\inf_{y\in\mathbb{R}}R_\pm(y,0)\le R_\pm(x,t)\le\sup_{y\in\mathbb{R}}R_\pm(y,0)\,.
\label{eq:RIinequalities}
\end{equation}
In other words, the smallest coordinate box containing the solution
at a given $t\ge 0$ is in fact independent of $t$.

Perhaps with the help of Riemann invariants, the mapping $F_+$ and its
double-valued inverse $F_+^{-1}$ can be used to obtain solutions of
\eqref{modeq} from known solutions of the hyperbolic system
\eqref{eq:moddefocNLS}, a problem that is better understood than the
mixed-type system \eqref{modeq}, having received much attention in the
literature (see, for example, \cite{BiondiniK06}).  As the mapping
$F_+$ is not one-to-one, some care must be taken in using given initial
data for \eqref{modeq} in its modulationally stable sector to obtain
corresponding initial data for \eqref{eq:moddefocNLS} and then porting
the solution thereof back to the domain $D_+(\alpha)$ via $F_+^{-1}$.

Consider, for example, smooth pulse-like initial data for
\eqref{modeq} for which $\rho(x,0)\rightarrow 0$ as $x\rightarrow
\pm\infty$ and for which the modulational stability condition
\eqref{stabcond} holds for each $x\in\mathbb{R}$.  If additionally
$u(x,0)>\alpha^{-1}$ holds for all $x\in\mathbb{R}$, then the image of
this initial data in the $(\hat{\rho},\hat{u})$ plane is a curve
parametrized by $x\in\mathbb{R}$ that is attached to the
$\hat{\rho}=0$ axis at one or more points with $\hat{u}>\alpha^{-1}$
and that avoids the boundary curve
$\hat{\rho}=(\hat{u}-\alpha^{-1})^2/4$ of the range $R_+(\alpha)$.  As
parabolic curves of the form $\hat{\rho}=(\hat{u}-c)^2/4$ for $c\ge
\alpha^{-1}$ are coordinate curves of the Riemann invariants for
\eqref{eq:moddefocNLS}, the evolution in time prior to shock formation
will not alter this basic picture due to the inequalities
\eqref{eq:RIinequalities}.  Thus, although when applying the inverse
mapping $F_+^{-1}$ pointwise in $x$ one has to make a choice between
two distinct preimages, there are only two distinct pairs of
continuous \emph{functions} $(\rho(x,t),u(x,t))$ for each $t$
corresponding to the evolved initial data.  Furthermore, exactly one
of these will satisfy the boundary condition that
$\rho(x,t)\rightarrow 0$ as $x\rightarrow\pm\infty$; the remaining
solution of \eqref{modeq} will satisfy $\alpha^2\rho(x,t)+\alpha
u(x,t)-1\rightarrow 0$ as $x\rightarrow\pm\infty$ instead, and taking
into account the initial condition this extraneous
solution\footnote{Of course this extraneous solution may be meaningful
  if different boundary conditions are imposed at $x=\pm\infty$.} is
not continuous in time $t$.  On the other hand, if for some finite
$x\in\mathbb{R}$ the initial data crosses the threshold into the
subregion of the domain $D_+(\alpha)$ for which $u<\alpha^{-1}$, then
even the image of the initial data as a curve in the
$(\hat{\rho},\hat{u})$-plane may have some unexpected properties.
Indeed, a direct calculation shows that all curves in the
$(\rho,u)$-plane passing through the vertical axis $u=\alpha^{-1}$
with slope $d\rho/du\neq -(2\alpha)^{-1}$ will have images in the
$(\hat{\rho},\hat{u})$-plane that are tangent to the boundary curve
$\hat{\rho}=(\hat{u}-\alpha^{-1})^2/4$ while this is not the case for
curves passing through with $d\rho/du=-(2\alpha)^{-1}$ (as holds for all
of the coordinate parabolae).  Therefore, if $d\rho/du=-(2\alpha)^{-1}$
for $u=\alpha^{-1}$ then the transformed initial data functions
$\hat{\rho}(x,0)$ and $\hat{u}(x,0)$ will necessarily have a common
critical point for some $x\in\mathbb{R}$.  Moreover, for initial data
crossing the $u=\alpha^{-1}$ threshold, there exist multiple preimage
function pairs $(\rho(x,t),u(x,t))$ that are continuous in
$x$ with the same boundary conditions as $x\rightarrow\pm\infty$.
Care must be taken to select the correct preimage to have continuity for $t\ge
0$.

\begin{figure}[h]
\begin{center}
\includegraphics{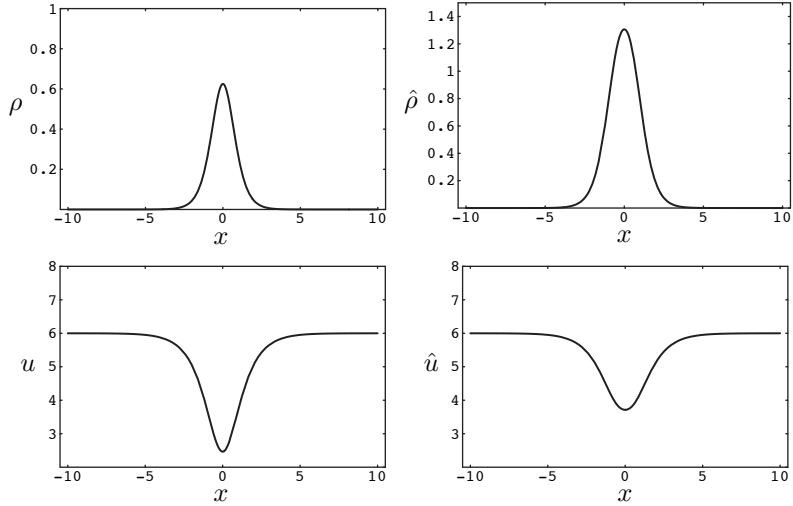}
\end{center}
\caption{\em The initial data and its image under $F_+$ for $\alpha=1$ and $c=6$
with $A_o^2=5/8$.}
\label{fig:MultiHump1}
\end{figure}

\begin{figure}[h]
\begin{center}
\includegraphics{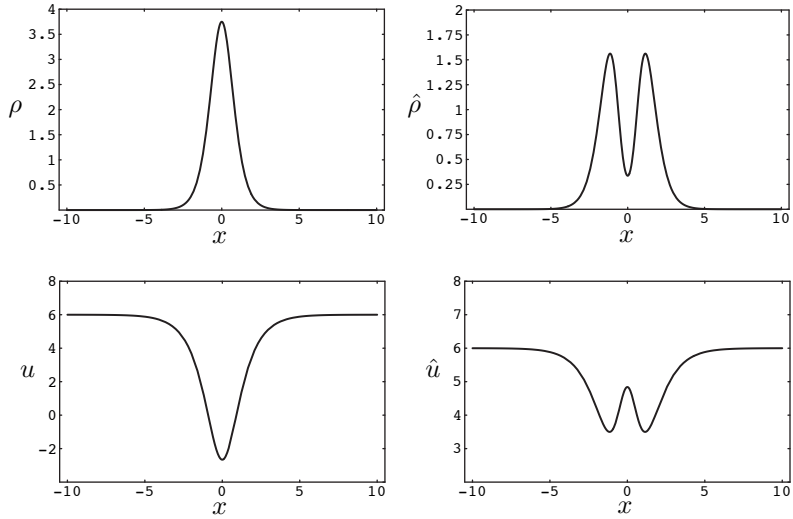}
\end{center}
\caption{\em The initial data and its image under $F_+$ for $\alpha=1$ and $c=6$
with $A_o^2=30/8$.}
\label{fig:MultiHump2}
\end{figure}

\begin{figure}[h]
\begin{center}
\includegraphics{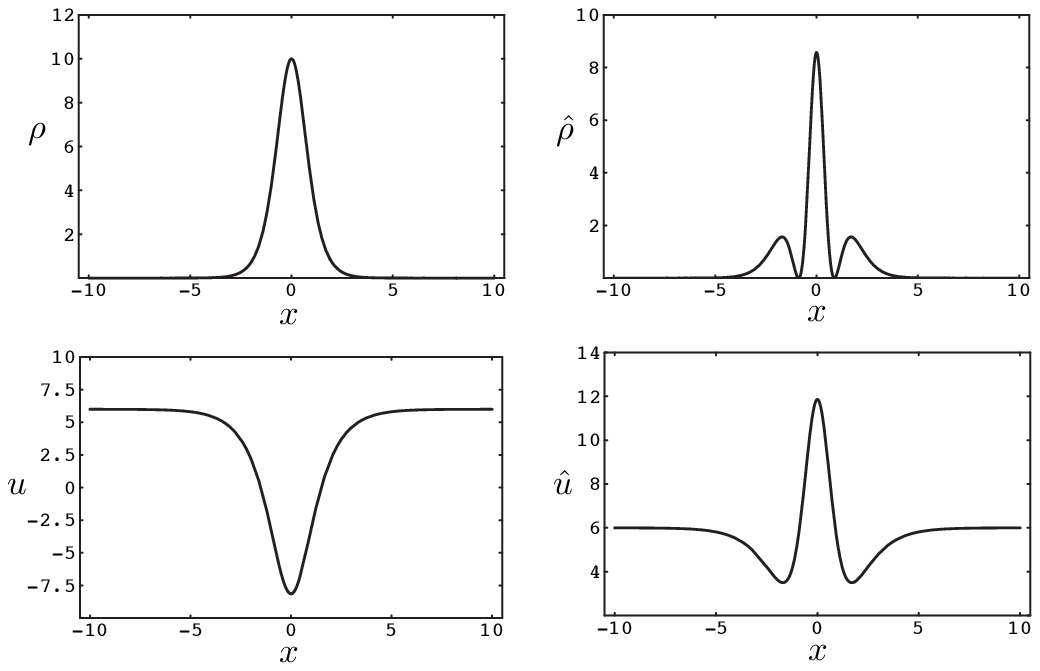}
\end{center}
\caption{\em The initial data and its image under $F_+$ for $\alpha=1$ and $c=6$
with $A_o^2=80/8$.}
\label{fig:MultiHump3}
\end{figure}

These facts suggest that while the modulationally stable sector of the
MNLS modulation equations \eqref{modeq} can be mapped to the
defocusing NLS modulation equations \eqref{eq:moddefocNLS} which we
may consider to be a problem for which some intuition is available,
new and important dynamical features of the MNLS modulation equations
can be introduced simply through the change of variables $F_+$.  For
example, let $c>\alpha^{-1}$ and $A_o>0$ be fixed, and consider initial data for
\eqref{modeq} of the form
\begin{equation}
\rho(x,0)=A_o^2\,\text{sech}^2(x)\,,\quad\quad u(x,0)=c-2A_o\sqrt{\alpha c-1}
\cdot\text{sech}(x)\,.
\end{equation}
The corresponding curve parametrized by $x\in\mathbb{R}$ lies along
one of the coordinate parabolae in the $(\rho,u)$-plane with $u\le c$
and $\rho\rightarrow 0$ as $x\rightarrow\pm\infty$.  If
$0<A_o^2<(c-\alpha^{-1})/(4\alpha)$, then the initial data satisfies
$u>1/\alpha$ for all $x$ and thus a single-peaked pulse is mapped to a
single-peaked pulse as shown in Figure~\ref{fig:MultiHump1}.
If $(c-\alpha^{-1})/(4\alpha)<A_o^2<(c-\alpha^{-1})/\alpha$, then the
initial data crosses the threshold of $u=\alpha^{-1}$ but does not yet
reach the point of tangency between the coordinate parabola and the
stability boundary $\alpha^2\rho+\alpha u-1=0$ (which is also mapped
to $\hat{\rho}=0$ along with $\rho=0$).  In this situation, the image
in the $(\hat{\rho},\hat{u})$-plane of the part of the initial data
with $u<\alpha^{-1}$ is the same as that of part of the initial data
with $u>\alpha^{-1}$, and this results in a double-peaked image pulse
as shown in Figure~\ref{fig:MultiHump2}.
Finally, if $A_o^2$ exceeds $(c-\alpha^{-1})/\alpha$, then the part of
the coordinate parabola in the $(\hat{\rho},\hat{u})$-plane that is
retraced is the entire left-hand branch for $\hat{u}<c$, and part of
the right-hand branch (along which $\hat{\rho}$ can become arbitrarily
large) is also traced out in the center of the pulse.  Therefore, in
this situation the single-pulse initial data for the MNLS modulation
equations \eqref{modeq} is mapped by $F_+$ to triple-pulse initial data
for the defocusing NLS modulation equations \eqref{eq:moddefocNLS}, as
shown in Figure~\ref{fig:MultiHump3}.
Since the coordinate curves correspond to constant values of one or
the other of the Riemann invariants, the dynamics of all of these
initial conditions under the defocusing NLS modulation equations
\eqref{eq:moddefocNLS} will be given by simple waves. It follows that
the evolution is governed by the scalar inviscid Burgers equation, and
it may thus be shown that the number of shocks that can form
corresponds to the number of critical points of the simple-wave
initial data.  Thus, simply by varying the amplitude of a
single-peaked pulse-like initial condition for the MNLS modulation
equations \eqref{modeq}, one can produce dynamics that lead to either
one, two, or three (dispersively regularized) shocks.

A relation similar to the equivalence between the MNLS modulation
equations \eqref{modeq} and those for the focusing or defocusing NLS
equations has been observed by Kuvshinov and Lakhin
\cite{KuvshinovL94}.  Working with the equivalent derivative NLS
equation, Kuvshinov and Lakhin derive the modulation equations
(Whitham equations) corresponding not to complex exponential plane
waves (genus zero) but rather to periodic waves given by elliptic
functions (genus one).  By comparing the Riemann invariants for these
Whitham equations with the corresponding well-known formulae for the
NLS genus one Whitham equations they deduce complete equivalence
between these two systems of modulation equations.  No doubt this
phenomenon extends to genera greater than one, and a deeper
explanation of these coincidences surely exists at the level of the
algebro-geometric description of arbitrary genus multiphase wave
solutions of the NLS and derivative NLS equations.  We will treat
this problem in this framework in a future publication.

\section{Bounds on the Discrete Spectrum}
\label{sec:hyperbolae}
In this section we establish some basic conditions on the discrete
spectrum associated to the spectral problem \eqref{eq:Laxx} for the
MNLS equation.  As seen in the appendix, for each $k$ with
$\text{Im}\{k^2\}\neq 0$ there are two one-dimensional subspaces of
solutions that decay to zero as $x\rightarrow\pm\infty$
respectively. We refer to {\em eigenvalues} as values of $k$ for which
these two subspaces coincide and {\em eigenfunctions} as the
corresponding solutions $\mathbf{v}$ of \eqref{eq:Laxx} associated to
each eigenvalue that decay to zero in both directions\footnote{This is
  inexact usage, as the spectral parameter $k$ appears nonlinearly in
  the system \eqref{eq:Laxx}.  The concept is, however, completely
  analogous to the usual situation.}. Eigenvalues are exactly the
zeros of of the function $S_{11}(k)$ in the region $\text{Im}\{k^2\}<0$
and thet zeros of the function $S_{22}(k)$ in the region
$\text{Im}\{k^2\}>0$, where $S_{11}(k)$ and $S_{22}(k)$ are defined by
\eqref{eq:Sdets}. The integrabililty of the MNLS equation leads to the
fact that eigenvalues do not vary in time (see
\eqref{eq:Sevolve}). Consequently, we consider the spectral problem at
$t=0$ and set $\phi$ as the initial condition for the Cauchy problem
for the MNLS equation \eqref{MNLScauchy}.  Additionally, we note that
due to the symmetries \eqref{eq:Ssymanti} and \eqref{eq:Ssymholo}, the
eigenvalues have a four-fold symmetry, where if any of $k$, $k^*$,
$-k^*$ or $-k$ is an eigenvalue the others are as well.  This symmetry
allows us to restrict our attention to (say) the first quadrant of the
complex $k$-plane. Any result established for $k$ in the first
quadrant therefore immediately gives analogous results for $k$ in any
other quadrant.

We will be considering the linear differential equation
\eqref{eq:Laxx} with a potential of the form
$\phi(x)=A(x)e^{iS(x)/\e}$ where $A(\cdot)$ and $S(\cdot)$ are
real-valued functions such that $A>0$ and $A$ and $S''$ decay rapidly
to zero for large $|x|$.  At first blush, one difficulty that presents
itself is that there are rapid oscillations in the coefficients of
\eqref{eq:Laxx} when $\e$ is small.  This difficulty is, however,
circumvented by the following device.  If $\mathbf{v}$ is a vector
solution of \eqref{eq:Laxx}, then
$\mathbf{w}:=e^{-(iS(x)/2\e)\sigma_3} \mathbf{v}$ satisfies:
\begin{equation}\label{system6}
2\alpha \e   \frac{d\mathbf{w}}{d x}=i{\bf M}\mathbf{w}\,,
\end{equation}
where
\begin{equation}
{\bf M}:=\begin{bmatrix} -4k^2+1-\alpha S'(x) & 4\alpha k A(x)\\ 4\alpha k A(x) & 4k^2-1+\alpha S'(x)\end{bmatrix}\,.
\label{M}
\end{equation}
Since we have only scaled $ \mathbf{v}$ by exponential factors with
purely imaginary exponents, $ \mathbf{w}$ decays as
$x\rightarrow\pm\infty$ if and only if $ \mathbf{v}$ does, that is,
the discrete spectrum of \eqref{system6} is the same as that of
\eqref{eq:Laxx}.

A formal approach to \eqref{system6} when $\e$ is small is apply the
WKB method; that is, to assume a solution $\mathbf{w}$ of the form
$\mathbf{w}=e^{i\sigma/(2\alpha\e)}( \mathbf{w}_o+\e
\mathbf{w}_1+\cdots)$.  Substituting this expansion for $\mathbf{w}$
into \eqref{system6} and equating terms of the same powers in $\e$
gives a hierarchy of equations relating $\sigma$ and $\{{\bf w}_k\}$.
The leading order (eikonal) equation in $\e$ is:
\begin{equation}\label{eikonal}
{\bf M}  \mathbf{w}_o=\frac{d\sigma}{dx}\mathbf{w}_o\,,  
\end{equation}
and, consequently, $d\sigma/d x$ must be an eigenvalue of
${\bf M}$. This leads us to consider the eigenvalues of ${\bf M}$ as 
smooth functions of $x$.

A simple calculation shows that the eigenvalues of ${\bf M}$ are $\pm\omega$
where
\begin{equation}
\omega(x;k):=\left[16\alpha^2k^2 A(x)^2+(4k^2-1+\alpha S'(x))^2\right]^{1/2}\,.
\label{eq:omega}
\end{equation}
In WKB theory, the so-called turning points are those values of $x$
for which the eigenvalues coincide: $d\sigma/dx=\pm\omega=0$.  Near
such points, WKB theory breaks down as infinite derivatives appear and
approximate solutions become multiply-defined.  Setting
$\omega(x;k)=0$, we see that for a given $k\in\mathbb{C}$ the turning
points $x\in\mathbb{R}$ are the solutions of the equation
\begin{equation}\label{turningpoint1}4i\alpha kA(x)=\pm\left(4k^2-1+\alpha S'(x)\right)\,.
\end{equation}
On the other hand, as $x\in\mathbb{R}$ varies, the equation
\eqref{turningpoint1} defines a parametrized curve $\mathcal{T}$ in
the four quadrants of the complex $k$-plane that we will call the
\emph{turning point curve} associated with the functions $A(\cdot)$
and $S'(\cdot)$. By definition, $k\in\mathcal{T}$ if and only if
$\text{Im}\{k^2\}\neq 0$ and there exists at least one \emph{real}
turning point $x$.  As \eqref{turningpoint1} is quadratic in $k$, we
may solve for $k$ and therefore explicitly present $\mathcal{T}$ as a
parametrized curve:
\begin{equation}
k(x)=\frac{i}{2}\left(s_1\alpha A(x) + s_2\sqrt{\alpha^2A(x)^2+\alpha S'(x)-1}
\right)\,,
\label{eq:quadraticsolve}
\end{equation}
where $s_1$ and $s_2$ are independent signs ($s_j=\pm 1$) yielding
up to four branches of $\mathcal{T}$.  

At this juncture we may make the observation that if $x$ is a point at
which the hyperbolicity (modulational stability) condition
\eqref{stabcond} holds (recall that $\rho:=A^2$ and $u:=S'$), or in
the degenerate case if $\alpha^2\rho+\alpha u-1=0$, then the four
$k$-values produced by the formula \eqref{eq:quadraticsolve} are
purely imaginary and by definition do \emph{not} lie in $\mathcal{T}$.
On the other hand, if $x$ is such that $\alpha^2\rho+\alpha u-1<0$,
then \eqref{eq:quadraticsolve} gives four distinct points
$k\in\mathcal{T}$:
\begin{equation}\begin{split}
\text{Im}\{k\}&=s_1\frac{\alpha}{2}A(x)\, ,\\
\text{Re}\{k\}&=s_2\frac{1}{2}\sqrt{1-\alpha S'(x)-\alpha^2A(x)^2}\, .
\end{split}
\label{eq:returning}\end{equation}
The equations \eqref{eq:returning} when restricted to the values of
$x$ for which the strict inequality $\alpha^2\rho+\alpha u-1<0$ holds
($x$-values of modulational instability) thus give a parametric
representation of the turning point curve $\mathcal{T}$.  In
particular, if the nonstrict modulational stability condition
$\alpha^2A(x)^2+\alpha S'(x)-1\ge 0$ holds for all $x\in\mathbb{R}$,
then by definition the turning point curve is empty:
$\mathcal{T}=\emptyset$.

Throughout this section, we will restrict our attention entirely to
those values of $k\in\mathbb{C}$ with $\text{Im}\{k^2\}\neq 0$ that
\emph{do not} lie on the turning point curve.  For such $k$, the
function $\omega(x;k)$ is well-defined by \eqref{eq:omega} if we
insist on continuity for all $x\in\mathbb{R}$ and impose the boundary
condition that
\begin{equation}
\lim_{x\to +\infty}\omega(x;k)=4k^2-1+\alpha S'_+\,,\quad\quad
\text{where}\quad\quad
S'_+:=\lim_{x\to+\infty}S'(x)\,.
\label{eq:omegabc}
\end{equation}
Suppose that $A(\cdot)>0$ is of class $L^1(\mathbb{R})$ and is
uniformly Lipschitz. In particular, this implies that $A(x)$ decays
to zero\footnote{It is enough to consider the behavior as
  $x\rightarrow +\infty$.  Arguing the contrapositive, suppose that
  $A(x)>0$ is integrable on $\mathbb{R}$ and Lipschitz, and that
  $\limsup_{x\to +\infty} A(x)=M>0$ (we admit the possibility that
  $M=+\infty$).  We will show that the Lipschitz constant of $A$ is
  bounded below by an arbitrarily large number, and hence is infinite.
  Begin by choosing $\delta>0$ to be arbitrarily small.  Let $\{x_n\}$
  be a sequence of real numbers tending to $+\infty$ such that
  $\lim_{n\to\infty}A(x_n)=M>0$.  By passing to a subsequence if
  necessary, we may assume that $x_{n+1}\ge x_n+\delta$ for all $n$.
  We first claim that $I_n:=\inf_{x_n<x<x_n+\delta}A(x)$ tends to zero
  as $n\rightarrow\infty$.  Indeed, were this not the case there would
  exist some $\epsilon>0$ and some $n_o$ such that $I_n\ge\e$ holds
  for all $n\ge n_o$.  On the one hand we would then have
\[
\int_{x_n}^{x_n+\delta}A(x)\,dx\ge\epsilon\delta\,,\quad\quad n\ge n_o\,,
\]
and on the other
\[
\sum_{n=n_0}^\infty\int_{x_n}^{x_n+\delta}A(x)\,dx\le\int_{-\infty}^{+\infty}A(x)\,dx<\infty\,,
\]
which yields a contradiction.  Since $A$ is continuous, there exists
for each $n$ a number $y_n\in [x_n,x_n+\delta]$ such that
$A(y_n)=I_n$.  As $I_n$ tends to zero as $n\rightarrow\infty$, for all
$n$ sufficiently large we will then have both $A(x_n)\ge 2M/3$ (or
$A(x_n)\ge 2$ if $M=+\infty$) and $A(y_n)\le M/3$ (or $A(y_n)\le 1$ if
$M=+\infty$) while $|x_n-y_n|\le\delta$.  Therefore, for all such $n$,
\[
\frac{|A(x_n)-A(y_n)|}{|x_n-y_n|}\ge \frac{C_M}{\delta}\,,
\]
where $C_M$ is a finite positive number depending only on $M$.  As
$\delta$ was arbitrarily small, this shows that the Lipschitz constant
of $A$ is arbitrarily large, and hence infinite.}  as
$x\rightarrow\pm\infty$.  Suppose also that $S'(\cdot)$ is uniformly
Lipschitz and that $S''(\cdot)$ is of class $L^1(\mathbb{R})$
(assuring the existence of limiting values of $S'(x)$ as
$x\rightarrow\pm\infty$).  The function $\omega(x;k)$ defined as
described above then has the following properties whenever
$\text{Im}\{k^2\}\neq 0$ and $k\not\in\mathcal{T}$: $\omega(x;k)$ and
its $x$-derivative $\omega'(x;k)$ are uniformly bounded, and
$\omega(x;k)$ is uniformly bounded away from zero.  Indeed, given the
assumed properties of $A(\cdot)$ and $S'(\cdot)$ it suffices to show
that $\omega(x;k)$ is bounded away from zero.  But this is so since
for $k\not\in\mathcal{T}$ we have $\omega(x;k)\neq 0$ for all
$x\in\mathbb{R}$, while the boundary condition \eqref{eq:omegabc} and
the limit
\begin{equation}
\lim_{x\to -\infty}\omega(x;k)=\pm(4k^2-1+\alpha S'_-)\,,\quad\quad
\text{where}\quad\quad
S'_-:=\lim_{x\to -\infty}S'(x)\,,
\label{eq:omegaminusinfty}
\end{equation}
when combined with the condition that $\text{Im}\{k^2\}\neq 0$ shows that
$\omega(x;k)$ has nonzero limiting values as $x\rightarrow\pm\infty$. 

Given $\omega(x;k)$ defined as above, we set
\begin{equation}
q(x;k):= \frac{2\alpha k A(x)}{\omega(x;k)}\cdot
\frac{d}{dx}\log\left(\frac{A(x)}{\omega(x;k)+4k^2-1+\alpha S'(x)}\right)\,.
\label{eq:q}
\end{equation}
Supposing the same conditions on $A(\cdot)$ and $S'(\cdot)$ as above,
the function $q(x;k)$ is uniformly bounded for $x\in\mathbb{R}$
whenever $\text{Im}\{k^2\}\neq 0$ and $k\not\in\mathcal{T}$ and
$\alpha>0$.  Indeed, writing $q(x;k)$ in the equivalent form
\begin{equation}
q(x;k)=\frac{2\alpha k}{\omega(x;k)}\left(A'(x) -A(x)
\frac{\omega'(x;k)+\alpha S''(x)}{\omega(x;k)+4k^2-1+\alpha S'(x)}\right)\,,
\end{equation}
our attention falls on the denominator in the second term:
$D:=\omega(x;k)+4k^2-1+\alpha S'(x)$.  This denominator cannot vanish
for any $x\in\mathbb{R}$ because this would imply that
$\omega(x;k)^2=(4k^2-1+\alpha S'(x))^2$ which by \eqref{eq:omega} is
in contradiction with the assumptions that $\text{Im}\{k^2\}\neq 0$,
$A(x)>0$, and $\alpha>0$.  According to the boundary condition
\eqref{eq:omegabc}, $D$ also has a nonzero limit as $x\rightarrow
+\infty$ for $\text{Im}\{k^2\}\neq 0$.  A similar situation holds as
$x\rightarrow -\infty$ as long as the limiting value of $\omega(x;k)$
is achieved with the ``$+$'' sign in \eqref{eq:omegaminusinfty}.  It
follows that $q(x;k)$ is bounded for large $|x|$, at
least if the limit \eqref{eq:omegaminusinfty} holds with the
``$+$'' sign.  If \eqref{eq:omegaminusinfty} holds with the ``$-$''
sign, then we must expand $D$ as $x\rightarrow -\infty$:
\begin{equation}
D=\left(4k^2-1+\alpha S'(x)\right)\cdot\left(1-\sqrt{1+\frac{16\alpha^2k^2
A(x)^2}{(4k^2-1+\alpha S'(x))^2}}\right)\,,
\end{equation}
whence it follows for $\text{Im}\{k^2\}\neq 0$ and $\alpha>0$ that
\begin{equation}
\frac{D'}{D}=O\left(\frac{1}{A(x)}\right)\,,\quad
\text{as}\quad x\to -\infty\,,
\end{equation}
so that in this limit $q(x;k)$ is again bounded for large negative
$x$.

Finally, in terms of $\omega(x;k)$ and $q(x;k)$ we define, for
$\text{Im}\{k^2\}\neq 0$ and $k\not\in\mathcal{T}$, the quantity
\begin{equation}
  L_{k}=\sup_{x\in\mathbb{R}}\left|\frac{d}{d x}
\left(\frac{1}{\mathrm{Im}\{\omega(x;k)\}}\right)\right|+
2\sup_{x\in\mathbb{R}}\left|\frac{\mathrm{Re}\{q(x;k)\}}
{\mathrm{Im}\{\omega(x;k)\}}\right|\, .\label{eq:Lko}
\end{equation} 
It is possible to have $L_k=+\infty$.

\subsection{The fundamental condition on eigenvalues.}
Our main result is the following.
\begin{thm}\label{thm:spectralbound} Let $A:\mathbb{R}\to\mathbb{R}_+$ be
  a uniformly Lipschitz function of class $L^1(\mathbb{R})$ and let
  $S':\mathbb{R}\to\mathbb{R}$ be uniformly Lipschitz with
  $S''(\cdot)$ of class $L^1(\mathbb{R})$.  Let $k$ be a fixed complex
  number with $\text{Im}\{k^2\}\neq 0$ and $k\not\in\mathcal{T}$.  If
  $\phi(x):=A(x)e^{iS(x)/\e}$ is the potential in the linear
  differential equation \eqref{eq:Laxx}, then the following statements
  hold:
\begin{itemize}
\item[(a)] If $k$ is an eigenvalue, then
\begin{equation}
\abs{\mathrm{Im}\{k\}}\leq 
\frac{\alpha}{2}\sup_{x\in\mathbb{R}}A(x)\,.
\label{eq:bound1}
\end{equation}
\item[(b)] If $L_{k}<(\alpha\e)^{-1}$ then $k$ is not an eigenvalue.
\end{itemize}
\end{thm}

Our proof of this theorem was inspired by some notes \cite{Percynotes}
we obtained from Percy Deift regarding his proof with Stephanos
Venakides and Xin Zhou of the corresponding ``shadow bound'' estimates
for eigenvalues of the nonselfadjoint Zakharov-Shabat problem.  (We
will explain more about their result, and show how it can be deduced
from ours as a special case, at the end of this section.)  

\begin{proof}


  To prove part (a), suppose that $k$ is an eigenvalue and
  $\mathbf{w}$ is the corresponding eigenfunction. 
  Using \eqref{system6}, we calculate
\begin{equation}\label{Imbound1}
  2\alpha \e  \left(\mathbf{w}^{\dagger}\sigma_3\frac{d\mathbf{w}}{dx}\right)=
  -i\left(4k^2-1+\alpha S'(x)\right)|\mathbf{w}|^2-8\alpha kA(x)
  \, \text{Im} \{  w_1 ^* w_2\}\, ,
\end{equation}
where $|{\bf w}|^2:=|w_1|^2+|w_2|^2$, and we adopt the notation that
will be used throughout the remainder of the paper that $w_1$ and
$w_2$ are the first and second components, respectively, of the vector
$\mathbf{w}$.  
As shown in the appendix (see \eqref{eq:Decay1} and \eqref{eq:Decay2}),
the condition that $A(\cdot)\in L^1(\mathbb{R})$ guarantees that the
eigenfunction $\mathbf{w}\neq 0$ decays exponentially to the zero vector
as $x\rightarrow \pm\infty$.  
Integration by parts then establishes the following identity:
\begin{equation}
  \int_{-\infty}^{+\infty}  \mathbf{w}^{\dagger}\sigma_3 \frac{d\mathbf{w}}{dx}dx=
  \int_{-\infty}^{+\infty}  \left[w_1'  w_1^*-w_2'  w_2^*\right]dx=
  -\left(\int_{-\infty}^{+\infty} \left[ w_1' w_1^*-w_2'w_2\right]^*dx\right)^*\,,
\label{eq:IBPidentity}
\end{equation}
so that all three expressions are purely imaginary.  Therefore,
integration of \eqref{Imbound1} and taking the real part yields
\begin{equation}\label{realeqn}
  0=\text{Im}\{k\}\text{Re}\{k\}\int_{-\infty}^{+\infty}\abs{  \mathbf{w}}^2dx-
\alpha\text{Re}\{k\}\int_{-\infty}^{+\infty}A(x)\,\text{Im}\{  w_1^* w_2\}\,dx\,.
\end{equation}
Since $\text{Re}\{k\}\neq0$, \eqref{realeqn} gives:
\begin{equation}
\text{Im}\{k\}=\frac{\displaystyle{\alpha\int_{-\infty}^{+\infty}A(x)\,
\text{Im}\{  w_1^*w_2\}\,dx}}{\displaystyle{\int_{-\infty}^{+\infty}\abs{  \mathbf{w}}^2\,dx}}\,.\label{eq:Imintegral}
\end{equation}
Applying the inequality $2|\text{Im}\{ w_1^*w_2\}|\leq 2|w_1|| w_2|\leq
\abs{ w_1}^2+\abs{ w_2}^2=\abs{\mathbf{w}}^2$ we obtain:
\begin{equation}
\abs{\text{Im}\{k\}}\leq \frac{\alpha}{2}\sup_{x\in\mathbb{R}}A(x)\,,
\end{equation}
which proves part (a) of Theorem \ref{thm:spectralbound}. 

To prove part (b), suppose that $k$ is an eigenvalue of
\eqref{eq:Laxx}, or equivalently, of \eqref{system6}.
We will now suppose that $L_{k}<(\alpha\e)^{-1}$ and
derive a contradiction. 

Let ${\bf T}(x;k)$ be a $2\times2$ matrix with eigenvectors of $\bf M$
corresponding to (distinct) eigenvalues $\omega(x;k)$ and
$-\omega(x;k)$ as its first and second columns respectively. Note that
${\bf T}(x;k)$ is invertible for all $x\in\mathbb{R}$ because $k$ does not
lie on the turning point curve $\mathcal{T}$.  Make the \emph{gauge
  transformation} $\mathbf{p}=\mathbf{T}(x;k)^{-1} \mathbf{w}$, where by
means of (\ref{system6}), $\mathbf{p}$ satisfies
\begin{equation}\label{pevolution}
  2\alpha\e\frac{d\mathbf{p}}{dx}=-2\alpha\e
  \mathbf{T}(x;k)^{-1}\frac{d\mathbf{T}}{d x}(x;k)  \mathbf{p}+
i\omega(x;k)\sigma_3\mathbf{p}\, . 
\end{equation}
A general eigenvector matrix is only determined up to multiplication on the
right by a diagonal matrix, and we fix the gauge so that 
${\bf T}(x;k)^{-1}\cdot{\bf T}'(x;k)$ is an off-diagonal skew-symmetric
complex matrix depending on $A(x)$, $S'(x)$ and $k$.  Thus, with
the particular eigenvector matrix
\begin{equation}
{\bf T}(x;k)=
\begin{bmatrix}\displaystyle\frac{4\alpha kA(x)}{H(x;k)} &\displaystyle
-\frac{H(x;k)}{2\omega(x;k)}\\\\\displaystyle\frac{H(x;k)}{2\omega(x;k)} &
\displaystyle\frac{4\alpha k A(x)}{H(x;k)}\end{bmatrix}
\end{equation}
where $H(x;k)$ is a continuous function of $x$ satisfying
\begin{equation}
H(x;k)^2=2\omega(x;k)\left(\omega(x;k)+4k^2-1+\alpha S'(x)\right)\,,
\end{equation}
we see that ${\bf T}(x;k)^{-1}{\bf T}'(x;k)$ is off-diagonal as desired:
\begin{equation}
{\bf T}(x;k)^{-1}\frac{d{\bf T}}{dx}(x;k)=iq(x;k)\sigma_2\,,
\end{equation}
where $q(x;k)$ is defined by \eqref{eq:q}.  Moreover, $\det({\bf
  T}(x;k))=1$, and, by the same arguments as were used to deduce the
boundedness of $q(x;k)$, the elements of ${\bf T}(x;k)$ are
uniformly bounded in $x$ (and hence so are those of ${\bf
  T}(x;k)^{-1}$).
Therefore, \eqref{pevolution} becomes
\begin{equation}\label{pevolution2}
2\alpha\e\frac {d\mathbf{p}}{d x}=
i\omega(x;k)\sigma_3\mathbf{p}-2i\alpha\e q(x;k)\sigma_2\mathbf{p}\,.
\end{equation} 
Next we differentiate $\mathbf{p}^{\dagger}\sigma_3 \mathbf{p}$ with
the help of \eqref{pevolution2} and the identities
$\sigma_3^2=\mathbb{I}$, $\sigma_2\sigma_3=i\sigma_1$, and
$\sigma_3\sigma_2=-i\sigma_1$ to find:
\begin{equation}\label{psigma3p}
  2\alpha\e\frac{d}{d x}\left( \mathbf{p}^{\dagger}\sigma_3   \mathbf{p}\right)=
  -2\,\text{Im}\{\omega(x;k)\}
\abs{  \mathbf{p}}^2-4\alpha\e\,\text{Re}\{q(x;k)\}  
\mathbf{p}^{\dagger}\sigma_1  \mathbf{p}\,.
\end{equation}
Since ${\bf T}(x;k)^{-1}$ is uniformly bounded in $x$, ${\bf p}$
decays to zero for large $|x|$.  Dividing through in \eqref{psigma3p}
by $\text{Im}\{\omega(x;k)\}$ and integrating gives
\begin{equation}
\begin{split}
\int_{-\infty}^{+\infty}\abs{\mathbf{p}}^2\,dx&=
-\int_{-\infty}^{+\infty}\frac{\alpha\e}{\text{Im}\{\omega(x;k)\}}\frac{d}{dx}  
\left(\mathbf{p}^{\dagger}\sigma_3  \mathbf{p} \right)\,dx-
\int_{-\infty}^{+\infty}\frac{2\alpha\e\,\text{Re}\{q(x;k)\}}
{\text{Im}\{\omega(x;k)\}}  \mathbf{p}^{\dagger} \sigma_1  \mathbf{p}\,dx\\
&=\int_{-\infty}^{+\infty}\frac{d}{dx}\left(\frac{\alpha\e}
{\text{Im}\{\omega(x;k)\}}\right)  \mathbf{p}^{\dagger}\sigma_3  
\mathbf{p}\,dx-\int_{-\infty}^{+\infty}\frac{2\alpha\e\,\text{Re}\{q(x;k)\}}
{\text{Im}\{\omega(x;k)\}}  \mathbf{p}^{\dagger}\sigma_1  \mathbf{p}\,dx\,,
\end{split}
\end{equation}
where the second equality is due to integration by parts and the decay
of ${\bf p}$.
Since
\begin{equation}
|\mathbf{p}^\dagger\sigma_3\mathbf{p}|=\left|\abs{p_1}^2-\abs{p_2}^2\right|\le
\abs{\mathbf{p}}^2\quad\text{and}\quad
|\mathbf{p}^\dagger\sigma_1\mathbf{p}|=\left|p_2^*p_1+p_1^*p_2\right|
\le 2|p_1||p_2|\le \abs{\mathbf{p}}^2\,,
\end{equation}
we therefore obtain
\begin{equation} \int_{-\infty}^{+\infty}\abs{  \mathbf{p}}^2\,dx
\leq \alpha\e L_{k}\int_{-\infty}^{+\infty}\abs{  \mathbf{p}}^2\,dx\,,
\label{eq:ineq}
\end{equation}
where $L_{k}$ is the constant, independent of $\e$, defined in
\eqref{eq:Lko}. Cancelling the nonzero $L^2(\mathbb{R})$-norm of
$\mathbf{p}$, we arrive at a contradiction.  Therefore $k$ cannot be
an eigenvalue. This establishes part (b) and completes the proof of
Theorem \ref{thm:spectralbound}.
\end{proof}

\subsection{The shadow region.}
For $A(\cdot)$ and $S(\cdot)$ satisfying the conditions of Theorem
\ref{thm:spectralbound}, we have seen that $\omega'(x;k)$ and $q(x;k)$
are uniformly bounded functions of $x\in\mathbb{R}$ whenever $k$ is a
complex number with $\text{Im}\{k^2\}\neq 0$ and $k\not\in\mathcal{T}$.
It follows easily that under these conditions, for each such complex
number $k$ there exists a constant $d_k>0$, independent of $\e$, such
that
\begin{equation}
L_k\le d_k\sup_{x\in\mathbb{R}}\left[\text{Im}\{\omega(x;k)\}\right]^{-2}\,.
\label{eq:Lkbound}
\end{equation} 
It is then of interest to ask for which values of $k$ can one conclude
that $\text{Im}\{\omega(x;k)\}$ is bounded away from zero for all
$x\in\mathbb{R}$, because Theorem \ref{thm:spectralbound} implies that
such a value of $k$ cannot be an eigenvalue for any value of $\e>0$
that is sufficiently small.  Note that $\text{Im}\{k^2\}\neq 0$ implies
that $|\text{Im}\{\omega(x;k)\}|\rightarrow 4|\text{Im}\{k^2\}|\neq 0$ as
$x\rightarrow\pm\infty$.  Consequently, it is enough to show that
$\text{Im}\{\omega(x;k)\}\neq 0$ for all $x\in\mathbb{R}$.  We will now
derive a simple geometric condition in the complex $k$-plane that
implies that $\text{Im}\{\omega(x;k)\}$ is nonzero for all real $x$.

For $C\in\mathbb{R}$, let $\Sigma_C$ denote the hyperbola in the
complex $k$-plane defined by the relation
\begin{equation}
\Sigma_{C}:\left\{k : \text{Re}\{k^2\}+C=
\text{Re}\{k\}^2-\text{Im}\{k\}^2+C=0 \right\}\,.
\label{sigma}
\end{equation}
\begin{lem}
  Given a fixed complex value $k_o$ in the open first quadrant of the
  complex plane, define the value
  $C_o:=-(\mathrm{Re}\{k_o\}^2-\mathrm{Im}\{k_o\}^2)$.  Let
  $I_o$ be the set of $x$ values that correspond to intersection
  points of the turning point curve $\mathcal{T}$ and
  $\Sigma_{C_o}$.  If $2\,\mathrm{Im}\{k_o\}-\alpha A(x)>0$ for all
  $x\in I_o$, then $\mathrm{Im}\{\omega(x;k_o)\}\neq 0$ for all
  $x\in\mathbb{R}$.
 \label{lem:eign}\end{lem}

\begin{proof}
  For any complex number $z$, the condition $\text{Im}\{z\}\neq 0$ can be 
rewritten as a
  condition on $z^2$:
\begin{equation}
\label{eq:conditiononz}
\text{Im}\{z\}\neq 0 \iff \left\{\begin{array}{c} 
\text{Im}\{z^2\}\neq 0\\ \text{ or }\\ \text{Im}\{z^2\}=0 \text{ and } 
\text{Re}\{z^2\}<0\end{array}\right\}.
\end{equation}
From \eqref{eq:omega} we find that $\text{Im}\{\omega(x;k_o)^2\}$ is
given by
\begin{equation}\label{Imomega}
\text{Im}\{\omega(x;k_o)^2\}=
16\,\text{Im}\{k_o\}\text{Re}\{k_o\}\left(2\alpha^2A(x)^2
+\alpha S'(x)-1-4C_o\right)\,.
\end{equation}
Since $k_o$ is in the open first quadrant,
$\text{Re}\{k_o\}\text{Im}\{k_o\}>0$.  Let $J_o$ denote the set of
$x\in\mathbb{R}$ such that $\text{Im}\{\omega(x;k_o)^2\}=0$.

\bigskip

\noindent{\bf Case 1:}  Let $x\in \mathbb{R}\setminus J_o$.  Then
by \eqref{eq:conditiononz} we clearly have $\text{Im}\{\omega(x;k_o)\}\neq 0$.

\bigskip

\noindent{\bf Case 2:} Let $x\in J_o\neq\emptyset$.  Consider the
hyperbola $\Sigma_{C_o}$ defined in \eqref{sigma}.  Each intersection
point $k$ of $\Sigma_{C_o}$ with the turning point curve $\mathcal{T}$
corresponds to a real value of $x$ for which $\alpha^2A(x)^2+\alpha
S'(x)-1<0$ according to the natural parametrization of the latter
curve, given by \eqref{eq:returning}.  By eliminating $\text{Re}\{k\}$
and $\text{Im}\{k\}$ between the two equations of \eqref{eq:returning}
and the definition \eqref{sigma} of the hyperbola $\Sigma_{C_o}$, it
is easy to verify that each real $x$ corresponding to such an
intersection point $k$ is an element of $J_o$.  There may however be
some $x\in J_o$ that do not correspond to intersection points of
$\Sigma_{C_o}$ with the turning point curve $\mathcal{T}$. These are
values of $x$ for which $k=k(x)$ defined by \eqref{eq:quadraticsolve}
lies on the hyperbola $\Sigma_{C_o}$ but is a purely imaginary number
because $\alpha^2A(x)^2+\alpha S'(x)-1\ge 0$, and hence is not in
$\mathcal{T}$ by definition.  Since $\text{Im}\{k_o^2\}\neq 0$ but
$\text{Im}\{\omega(x;k_o)^2\}=0$, \eqref{Imomega} shows that such points
$x$ satisfy the completely equivalent inequality
$\alpha^2A(x)^2-4C_o\le 0$.  Therefore, we define $I_o$ to be the
subset of $J_o$ consisting of $x\in\mathbb{R}$ such that $2\alpha^2
A(x)^2+\alpha S'(x)-1-4C_o=0$ {\em and} $\alpha^2 A(x)^2-4C_o>0$. In
other words $I_o$ consists of all $x\in J_o$ such that $x$ also
corresponds to a point of intersection of the turning point curve
$\mathcal{T}$ and the hyperbola $\Sigma_{C_o}$.  It is also clear,
both directly from the inequality $\alpha^2A(x)^2-4C_o>0$ and also
from the fact that hyperbolae $\Sigma_C$ with $C<0$ contain no purely
imaginary points $k$, that $C_o\le 0$ implies $I_o=J_o$.
 
Let $x$ be a fixed element of $J_o$ and consider
$\text{Re}\{\omega(x;k)^2\}$ restricted to $k$-values lying on the
hyperbola $\Sigma_{C_o}$. This restriction makes
$\text{Re}\{\omega(x;k)^2\}$ a function of $\text{Im}\{k\}^2$:
 \begin{equation}\label{Reomega}
\begin{split}
 \text{Re}\{\omega(x;k)^2\}\Big|_{k\in\Sigma_{C_o}}&=
-64\,\text{Im}\{k\}^4+64C_o\text{Im}\{k\}^2-
4\alpha^2A(x)^2(4C_o-\alpha^2A(x)^2)\\
&=-64\left(\text{Im}\{k\}^2-r_1\right)\left(\text{Im}\{k\}^2-r_2\right)\,,
\end{split}
 \end{equation}
where
\begin{equation}
r_1:=\frac{1}{4}\alpha^2A(x)^2\,,\quad\quad\text{and}\quad\quad
r_2:=\frac{1}{4}(4C_o-\alpha^2A(x)^2)\,.
\end{equation}
As a function of $\text{Im}\{k\}^2$, the right-hand side of
\eqref{Reomega} is a concave-down quadratic.  Case 2 now breaks
into two subcases depending on whether or not $x\in J_o$ satisfies
$x\in I_o$.

\bigskip
 
\noindent {\bf Subcase 2(a):} Let $x\in J_o\backslash I_o$.  Assume
that $\text{Re}\{\omega(x;k_o)^2\}\ge 0$.  Since obviously $k_o$ lies on
the hyperbola $\Sigma_{C_o}$, it follows from \eqref{Reomega} that
$\text{Im}\{k_o\}^2$ must lie in the closed interval whose endpoints are
the two roots $r_1$ and $r_2$, and in particular
\begin{equation}
\text{Im}\{k_o\}^2\le \max(r_1,r_2)\,.
\end{equation}
Eliminating $\text{Im}\{k_o\}^2$ using the definition of $C_o$, 
we therefore find
\begin{equation}
\text{Re}\{k_o\}^2\le\max(r_1-C_o,r_2-C_o)\,.
\label{eq:contra1}
\end{equation}
Now, $r_2-C_o<0$ by definition.  The assumption that $x\not\in
I_o$ however implies further that $r_1-C_o\le 0$.  Therefore,
$\text{Re}\{k_o\}=0$ which contradicts the assumption that $k_o$ lies in
the open first quadrant.  Consequently, for all $x\in
J_o\backslash I_o$, $\text{Re}\{\omega(x;k_o)^2\}<0$ and
therefore $\text{Im}\{\omega(x;k_o)\}\neq 0$ by \eqref{eq:conditiononz}.
 
\bigskip
 
\noindent{\bf Subcase 2(b):} 
Let $x\in I_o$.  It is always the case that $r_1>0$ and for $x\in
I_o$ we also have $r_2<0$, by definition. Since \eqref{Reomega} is
quadratic in $\text{Im}\{k\}^2$ and concave down, and since $k_o\in
\Sigma_{C_o}$, the inequality $\text{Im}\{k_o\}^2>r_1$ implies that
$\text{Re}\{\omega(x;k_o)^2\}<0$ for this value of $x$.  Since $x\in
I_o\subset J_o$, it follows from \eqref{eq:conditiononz} that
(for $k_o$ in the open first quadrant) $2\text{Im}\{k_o\}-\alpha A(x)>0$
implies that $\text{Im}\{\omega(x;k_o)\}\neq 0$.

\bigskip

We have therefore shown that if $x\in\mathbb{R}\setminus I_o$, then
$\text{Im}\{\omega(x;k_o)\}\neq 0$, while (according to subcase 2(b) above)
if $x\in I_o$, we have $\text{Im}\{\omega(x;k_o)\}\neq 0$ as long as
$2\text{Im}\{k_o\}-\alpha A(x)>0$, which completes the proof of the Lemma.
\end{proof}
We might remark that Lemma~\ref{lem:eign} combines with the estimate
\eqref{eq:Lkbound} and part (b) of Theorem~\ref{thm:spectralbound} to
give a weaker ($\e$-dependent) version of part (a) of
Theorem~\ref{thm:spectralbound} because if
$2|\text{Im}\{k_o\}|>\alpha\sup_{x\in\mathbb{R}}A(x)$ then the inequality
$2|\text{Im}\{k_o\}|>\alpha A(x)$ holds in particular for all $x\in I_o$.

Now define
\begin{equation}\label{eigbound1}
C_{\rm min}:=\frac{1}{4}\min_{x\in\mathbb{R}}\left(2\alpha^2 A(x)^2+\alpha S'(x)-1
\right)\quad\quad C_{\rm max}:=\frac{1}{4}\max_{x\in\mathbb{R}}
\left(2\alpha^2 A(x)^2+\alpha S'(x)-1\right)\,.
\end{equation}
\begin{cor} Suppose $A(\cdot)$ and $S'(\cdot)$ satisfy the conditions
  of Theorem~\ref{thm:spectralbound}, and let $k_o$ be a fixed complex
  number such that $\mathrm{Im}\{k_o^2\}\neq0$. If $C_o<C_{\rm min}$
  or $C_o>C_{\rm max}$ then $k_o$ is not an eigenvalue for any $\e>0$
  sufficiently small.  \label{cor:hypbound}
\end{cor}
\begin{proof}
  By means of \eqref{Imomega} it is clear that if $x \in J_o$,
  then $2\alpha^2A(x)^2+\alpha S'(x)-1=4C_o$. Therefore, either of
  the inequalities $C_o<C_{\rm min}$ or $C_o>C_{\rm max}$
  implies that $J_o$ is empty and thus (by Case 1 in the proof of
  Lemma~\ref{lem:eign}) we have $\text{Im}\{\omega(x;k_o)\}\neq 0$ for
  all $x\in\mathbb{R}$.
\end{proof}

Note that $\Sigma_{C_{\rm min}}$ and $\Sigma_{C_{\rm max}}$ are the
osculating hyperbolae of the image of $k(x)$ defined by
\eqref{eq:quadraticsolve}, which in turn is the union of the turning
point curve $\mathcal{T}$ and (possibly) an interval $\mathcal{I}$ of
the imaginary axis.  This gives Corollary~\ref{cor:hypbound}
geometrical meaning since the condition $C_{\rm min}\le C\le C_{\rm
  max}$ is equivalent to the statement that
$\Sigma_C\cap(\mathcal{T}\cup\mathcal{I})\neq\emptyset$.  

A refinement of Corollary~\ref{cor:hypbound} is an immediate consequence
of Lemma~\ref{lem:eign}.
\begin{cor}
  Suppose $A(\cdot)$ and $S'(\cdot)$ satisfy the conditions of
  Theorem~\ref{thm:spectralbound}, and let $k_o$ be a fixed complex
  number such that $\mathrm{Im}\{k_o^2\}\neq0$. If $C_{\rm min}\le C_o\le
  C_{\rm max}$ but $\Sigma_{C_o}\cap \mathcal{T}=\emptyset$, then
  $k_o$ is not an eigenvalue for any $\e>0$ sufficiently
  small.  \label{cor:hypboundII}
\end{cor}
\begin{proof}
  The set $I_o$ is empty, so the proof follows from the estimate
  \eqref{eq:Lkbound} and Lemma~\ref{lem:eign}.
\end{proof}

We may now formulate a more precise geometric condition to exclude the
existence of eigenvalues.  For each $C\in\mathbb{R}$ with
$\Sigma_C\cap\mathcal{T}\neq\emptyset$ we define a number $U_C$ as
follows:
\begin{equation}
U_C:=\sup_{k\in\Sigma_C\cap\mathcal{T}}\text{Im}\{k\}^2\,.
\label{eq:UC}
\end{equation}
Then set
\begin{equation}
\mathcal{S}_C
:=\{k\in\Sigma_C\quad\text{such that}\quad \text{Im}\{k\}^2\le U_C\}\,,
\label{eq:SC}
\end{equation}
and 
\begin{equation}
\mathcal{S}:=\mathop{\bigcup_{C\in[C_{\rm min},C_{\rm max}]}}_{\Sigma_C\cap\mathcal{T}\neq\emptyset}\mathcal{S}_C\,.
\label{eq:shadowS}
\end{equation}
\begin{cor}
\label{cor:shadow}
Assume that $A(\cdot)$ and $S'(\cdot)$ satisfy the conditions of
Theorem~\ref{thm:spectralbound}.  Suppose that $k_o\in\mathbb{C}$ and
$\mathrm{Im}\{k_o^2\}\neq 0$.  If also $k_o\not\in \mathcal{S}$, then
$k_o$ is not an eigenvalue for any $\e>0$ sufficiently small.
\end{cor}
\begin{proof}
  First note that, by Corollaries \ref{cor:hypbound} and
  \ref{cor:hypboundII}, it is sufficient to consider those
  $k_o\not\in\mathcal{S}$ for which $C_{\rm min}\le C_o\le C_{\rm
    max}$ and also $\Sigma_{C_o}\cap\mathcal{T}\neq\emptyset$.  Now,
  clearly $k_o$ lies on the hyperbola $\Sigma_{C_o}$.  However, since
  $\mathcal{S}_{C_o}\subset \mathcal{S}$, we also have that
  $k_o\not\in \mathcal{S}_{C_o}$, so
\begin{equation}
\text{Im}\{k_o\}^2>U_{C_o}\,.
\end{equation}
By the definition of the set $I_o$ and the definition \eqref{eq:returning}
of the turning point curve $\mathcal{T}$, we see from \eqref{eq:UC} that
\begin{equation}
U_{C_o}=\sup_{x\in I_o}\frac{\alpha^2 A(x)^2}{4}\,.
\end{equation}
Lemma \ref{lem:eign} (and its obvious consequences in the remaining
quadrants of the complex $k$-plane) then implies that $k_o$
cannot be an eigenvalue as long as $\e>0$ is sufficiently small.
\end{proof}

At this point we may observe an important consequence of the hyperbolicity
condition \eqref{stabcond}.
\begin{cor}
  Suppose that $A(\cdot)$ and $S(\cdot)$ satisfy the conditions of
  Theorem~\ref{thm:spectralbound} and also that $\alpha^2A(x)^2+\alpha
  S'(x)-1\ge 0$ holds for all $x\in\mathbb{R}$.  The latter condition
  makes the modulation equations \eqref{modeq} (nonstrictly)
  hyperbolic for all $x\in\mathbb{R}$ at $t=0$.  If $k_o$ is any fixed
  complex number with $\mathrm{Im}\{k_o^2\}\neq 0$, then $k_o$ is not an
  eigenvalue for any $\e>0$ sufficiently small.
\end{cor}
\begin{proof}
In this situation, the turning point curve $\mathcal{T}$ is empty, and 
hence so is the set $\mathcal{S}$, so this statement is a direct consequence
of Corollary~\ref{cor:shadow}.
\end{proof}

\begin{figure}[here]
\begin{center}
\includegraphics[scale=0.9]{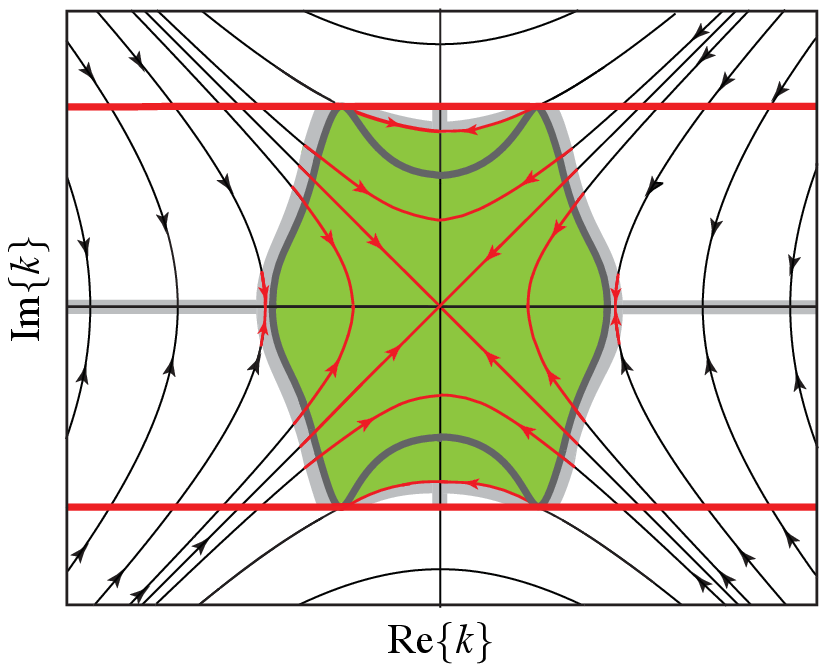}%
\includegraphics[scale=0.9]{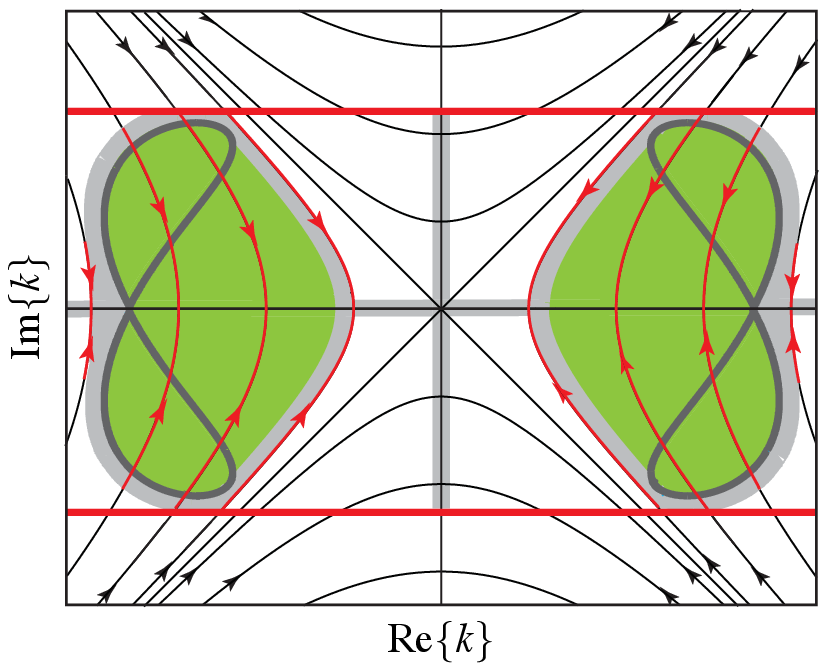}
\end{center}
\caption{\em ``Light'' is projected along the hyperbolae and casts the
shadow on the turning point curve. The shadow region
is shaded green. The collar is shaded grey and the red
horizontal lines correspond to the threshold in part (a) of Theorem
\ref{thm:spectralbound}. }
\label{fig:heuristic}
\end{figure}

The set $\mathcal{S}$ is the ``hyperbolic shadow'' of the turning
point curve $\mathcal{T}$.  One can imagine light being projected
along the hyperbolae $\Sigma_C$ from infinity, and the set
$\mathcal{S}$ is the shadow cast by the turning point curve in this light.
Let $N$ be any open neighborhood containing the axes
$\text{Im}\{k^2\}=0$ and the shadow $\mathcal{S}$.  Then from Corollary
\ref{cor:shadow} and part (a) of Theorem \ref{thm:spectralbound} the
set $F:=N\cap\left\{k\;\;\text{such that}\;\;2|\text{Im}\{k\}|\le
\alpha\sup_{x\in\mathbb{R}}A(x)\right\}$ contains all of the eigenvalues if
$\e>0$ is small enough.  In particular, the set $N$ may be taken to be
the union of $\mathcal{S}$ with a thin ``collar'' surrounding the
boundary of $\mathcal{S}$ and the axes.  If one wants to contain the
discrete spectrum for all $0<\e\le \e_{\rm max}$, then the collar must
be made thicker for larger $\e_{\rm max}$.  The region $F$ is perhaps
best understood visually.  Figure \ref{fig:heuristic} contains two
heuristic plots of $F$ in relation to the turning point curve and the
family of hyperbolae $\Sigma_C$.

The need for a ``collar'' arises from our methods of proof: we found
an upper bound for $L_k$ (see \eqref{eq:Lkbound}) and argued that to
ensure that the eigenvalue exclusion inequality $L_k<(\alpha\e)^{-1}$
holds for $\e$ small enough we simply checked for finiteness of the
upper bound.  The bound \eqref{eq:Lkbound} involved a constant $d_k$
that depended on $k$ in an unspecified way.  With better accounting of
the $k$-dependence of upper bounds for $L_k$ it is possible to obtain
a sharper condition on eigenvalue exclusion that explicitly depends on
$\e$; in other words, one may estimate the width of the ``collar''.
As a simple example of such a calculation, we present the following
result.
\begin{cor}
  Assume the same conditions on $A(\cdot)$ and $S'(\cdot)$ as in
  Theorem~\ref{thm:spectralbound}.  There exist constants $K_\alpha>0$
  and $C_\alpha>0$ independent of $\e$, such that whenever $|k|\ge
  K_\alpha$ and
\begin{equation}
|\mathrm{Im}\{k\}|\ge \frac{C_\alpha\e^{1/2}}{|\mathrm{Re}\{k\}|}\,,
\end{equation}
then $k$ is not an eigenvalue.
\label{cor:collar}
\end{cor}
This Corollary says that the width of the collar about the real $k$-axis
scales like $\e^{1/2}$ when $|k|$ is large.  Moreover, it shows that the
collar becomes thinner when $|\text{Re}\{k\}|$ is larger.  
\begin{proof}
Another upper bound for $L_k$ is clearly
\begin{equation}
L_k\le \sup_{x\in\mathbb{R}}|\omega'(x;k)|\cdot\sup_{x\in\mathbb{R}}
\frac{1}{|\text{Im}\{\omega(x;k)\}|^2}+2\sup_{x\in\mathbb{R}}|q(x;k)|
\cdot\sup_{x\in\mathbb{R}}\frac{1}{|\text{Im}\{\omega(x;k)\}|}\,.
\end{equation}
When $k$ is large in magnitude, we have the following asymptotics holding
uniformly with respect to $x\in\mathbb{R}$:
\begin{equation}
\begin{split}
\omega(x;k)&=4k^2+O(1)\,,\\
\omega'(x;k)&=\frac{1}{2\omega(x;k)}\left(32\alpha^2k^2A(x)A'(x)+
2(4k^2-1+\alpha S'(x))\alpha S''(x)\right)=O(1)\,,\\
q(x;k)&=O(|k|^{-1})\,.
\end{split}
\end{equation}
From the above it obviously follows that
$\text{Im}\{\omega(x;k)\}=4\text{Im}\{k^2\}+O(1)$ uniformly in $x$ for
large $k$.  When $\text{Im}\{k^2\}$ is small, this estimate is not
satisfactory, so to improve it we note that $\sqrt{1+x+iy}=f+ig$ with
\begin{equation}
\begin{split}
f&=\sqrt{\frac{1}{2}\left(\sqrt{(1+x)^2+y^2}+(1+x)\right)}\\
g&=\text{sgn}(y)\sqrt{\frac{1}{2}\left(\sqrt{(1+x)^2+y^2}-(1+x)\right)}
\end{split}
\end{equation}
where $x$ and $y$ are real and sufficiently small, and all square roots
are positive.  It follows that as $z=x+iy$ tends to zero, 
\begin{equation}
\sqrt{1+z}=1+O(|z|)+iO(y)
\end{equation}
where all order symbols denote real-valued functions.  Applying this result
to 
\begin{equation}
\omega(x;k)=4k^2\sqrt{1+(2\alpha^2A(x)^2+\alpha S'(x)-1)\frac{1}{2k^2}+
(1-\alpha S'(x))^2\frac{1}{16k^4}}\,,
\end{equation}
and using the facts that
\begin{equation}
\text{Im}\{k^{-2}\}=-\frac{\text{Im}\{k^2\}}{|k|^4}
\quad\quad\text{and}\quad\quad
\text{Im}\{k^{-4}\}=-\frac{2\text{Re}\{k^2\}\text{Im}\{k^2\}}{|k|^8}\,,
\end{equation}
we see that
\begin{equation}
\text{Im}\{\omega(x;k)\}=4\,\text{Im}\{k^2\}\cdot\left(1+O(|k|^{-2})\right)
\end{equation}
holds uniformly for $x\in\mathbb{R}$.  We therefore have shown that
\begin{equation}
L_k=O\left(\frac{1}{\text{Im}\{k^2\}^2}\right)+O\left(\frac{1}{|k|}
\cdot\frac{1}{|\text{Im}\{k^2\}|}\right)=
O\left(\frac{1}{\text{Im}\{k^2\}^2}\right)+O\left(\frac{|\text{Im}\{k^2\}|}{|k|}
\cdot\frac{1}{\text{Im}\{k^2\}^2}\right)\,.
\end{equation}
Since part (a) of Theorem~\ref{thm:spectralbound} allows us to
restrict attention to a strip in which $|\text{Im}\{k\}|$ is uniformly bounded,
it then follows that 
\begin{equation}
L_k=O\left(\frac{1}{\text{Im}\{k^2\}^2}\right)
\end{equation}
as $|k|$ tends to infinity in this strip.  The proof is then complete upon
recalling part (b) of Theorem~\ref{thm:spectralbound}.
\end{proof}

\begin{figure}
\begin{minipage}[h]{8cm}
\begin{center}
\scalebox{.7}{\includegraphics{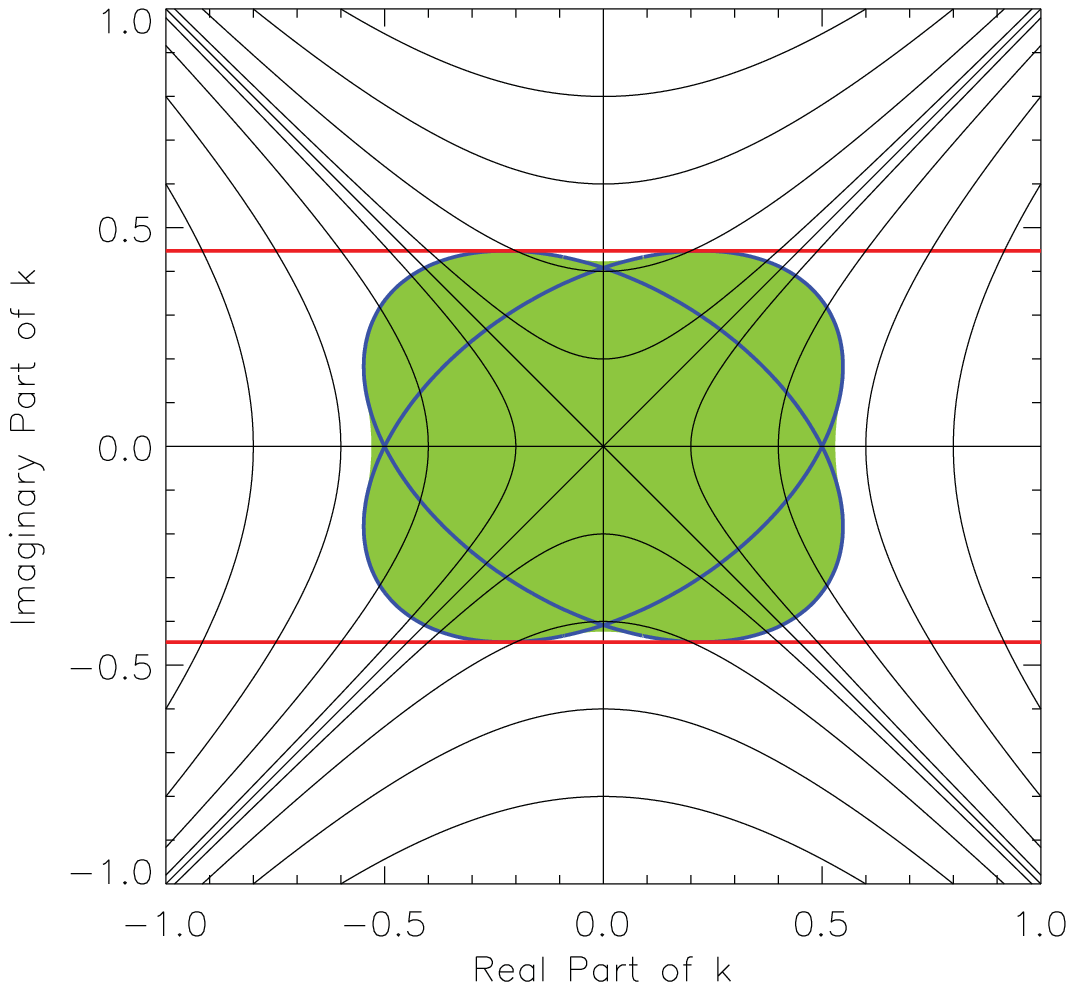}}
\caption{\em The turning point curve $\mathcal{T}$ and its hyperbolic
  shadow $\mathcal{S}$ for $\alpha=0.8944$.}\label{fig:1}
\end{center}
\end{minipage}
\hfill
\begin{minipage}[h]{8cm}
\begin{center}
\scalebox{.7}{\includegraphics{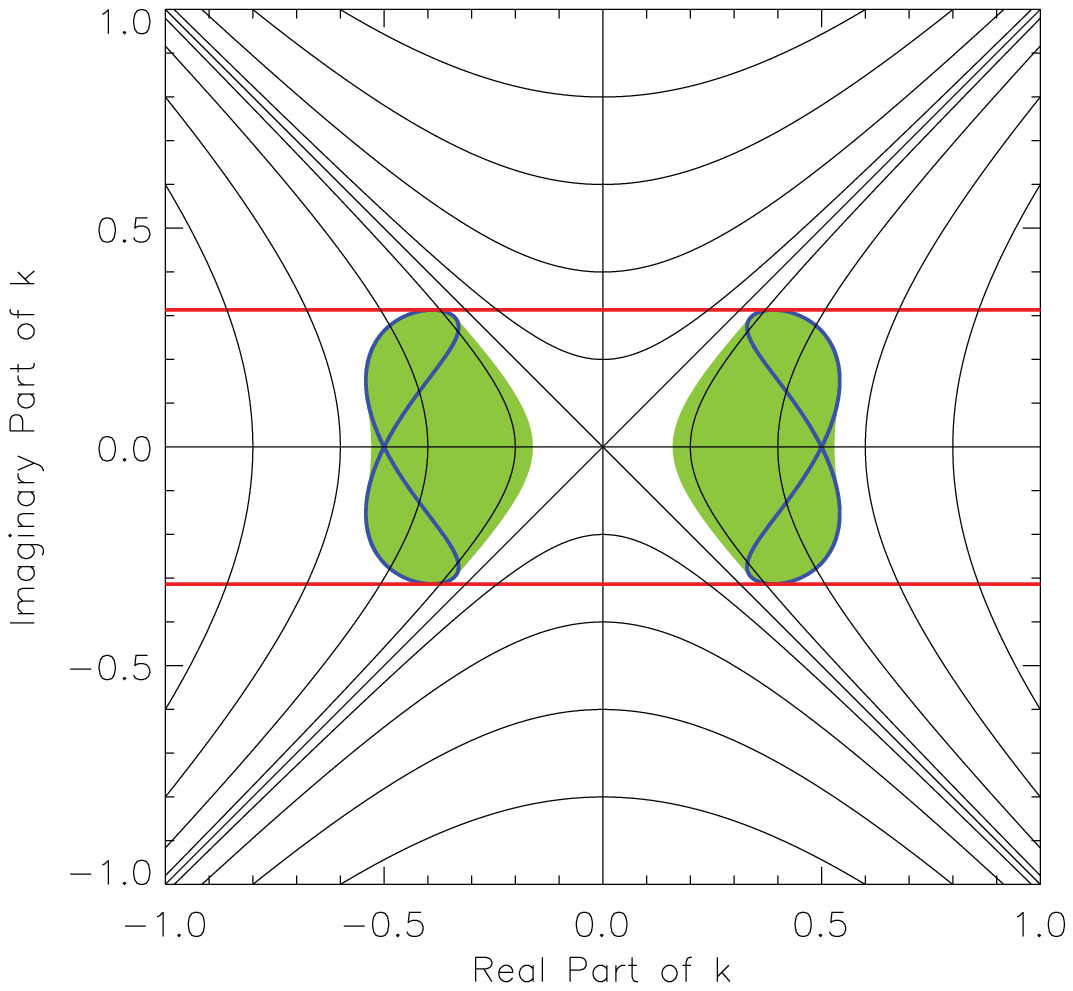}}
\caption{\em The turning point curve $\mathcal{T}$ and its hyperbolic
shadow $\mathcal{S}$ for $\alpha=0.6261$.}\label{fig:2}
\end{center}
\end{minipage}
\end{figure}

\begin{figure}[here]
\begin{center}
\scalebox{.7}{\hspace{-.4in}\includegraphics{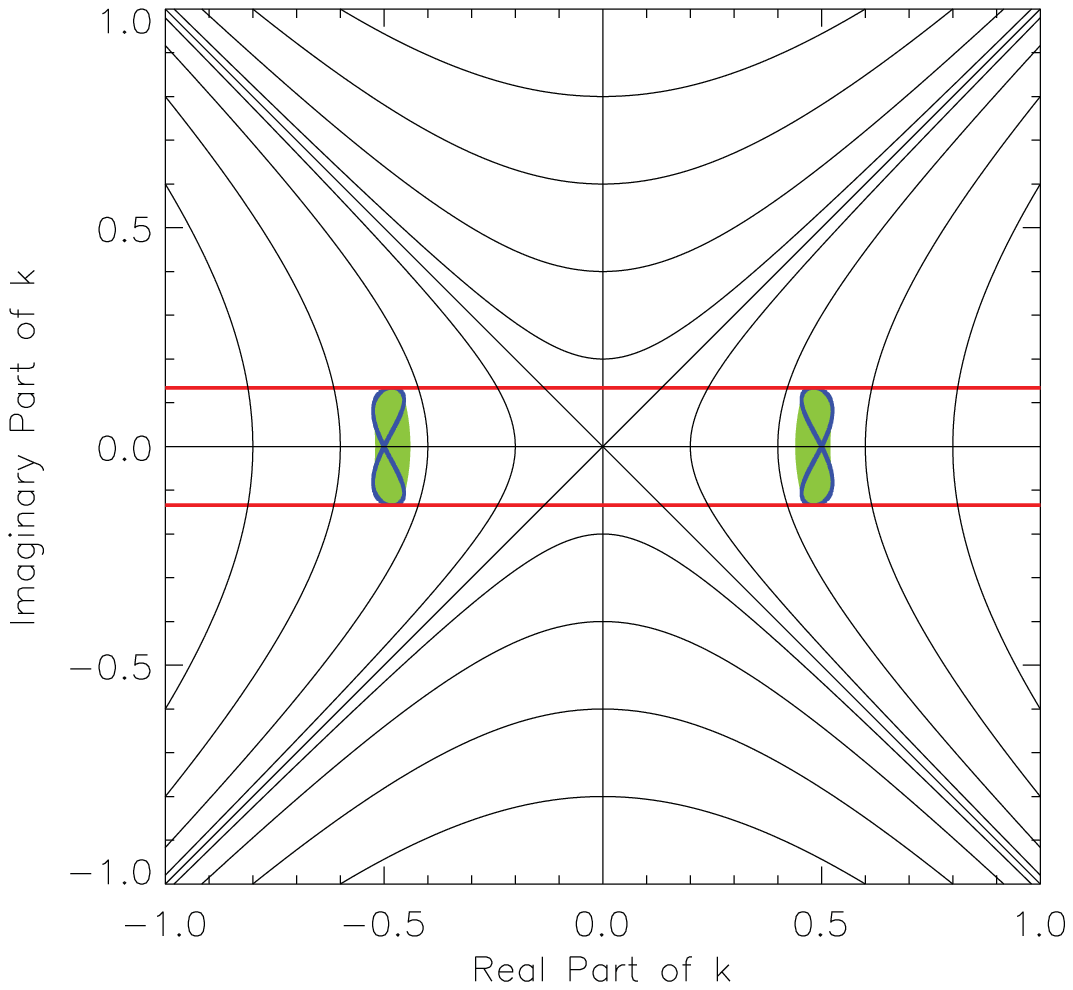}%
\includegraphics{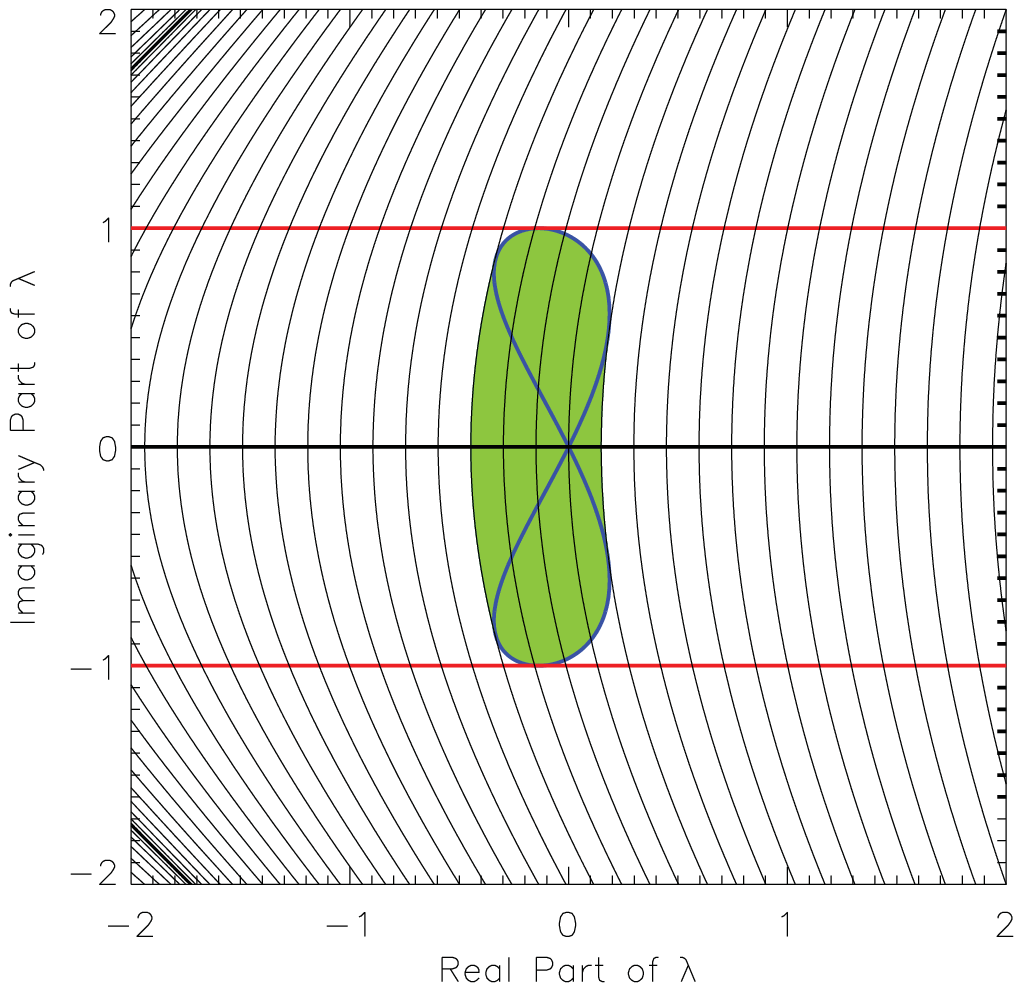}}
\caption{\em The turning point curve $\mathcal{T}$ and its hyperbolic
  shadow $\mathcal{S}$ for $\alpha=0.2683$.  Left: the $k$-plane.
  Right: the $\lambda$-plane.}\label{fig:3}
\end{center}
\end{figure}

\begin{figure}[here]
\scalebox{.7}{\hspace{-.4in}\includegraphics{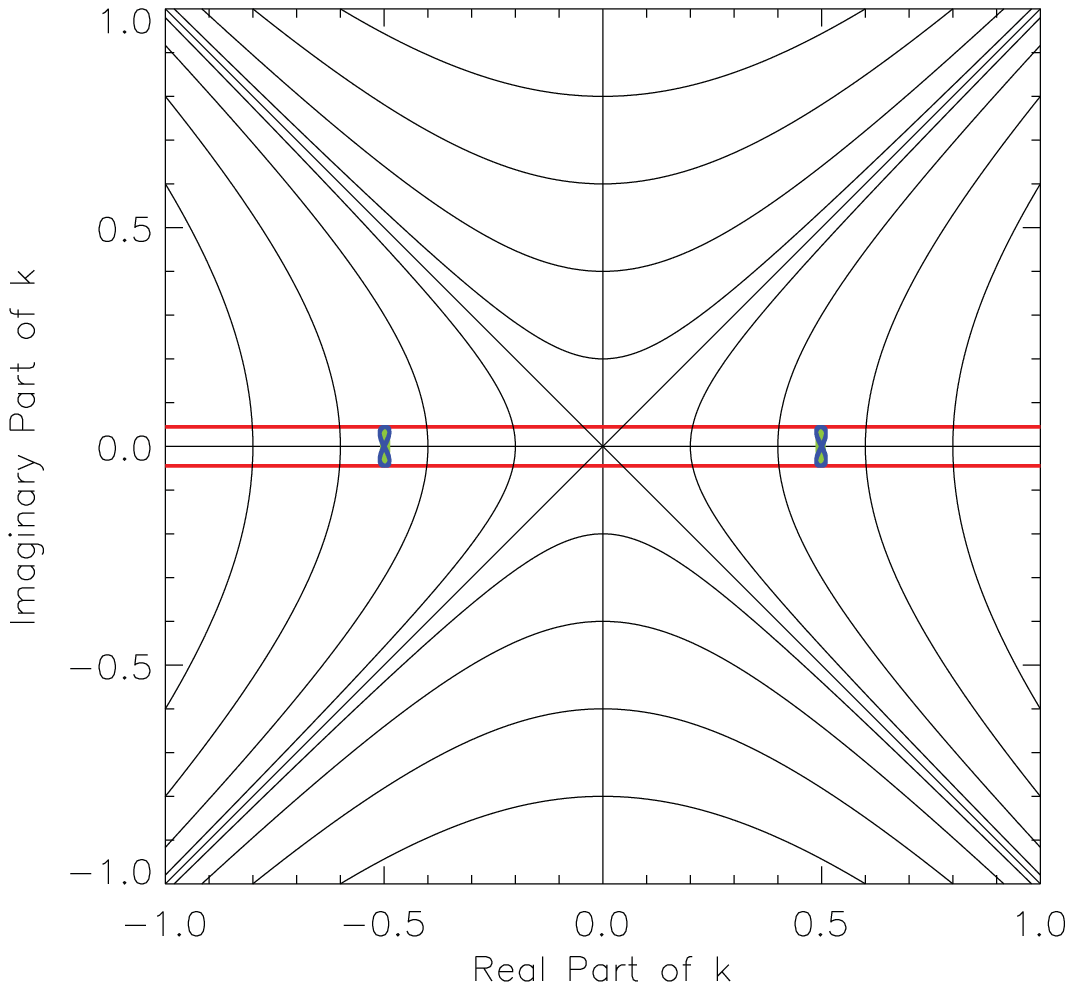}%
\includegraphics{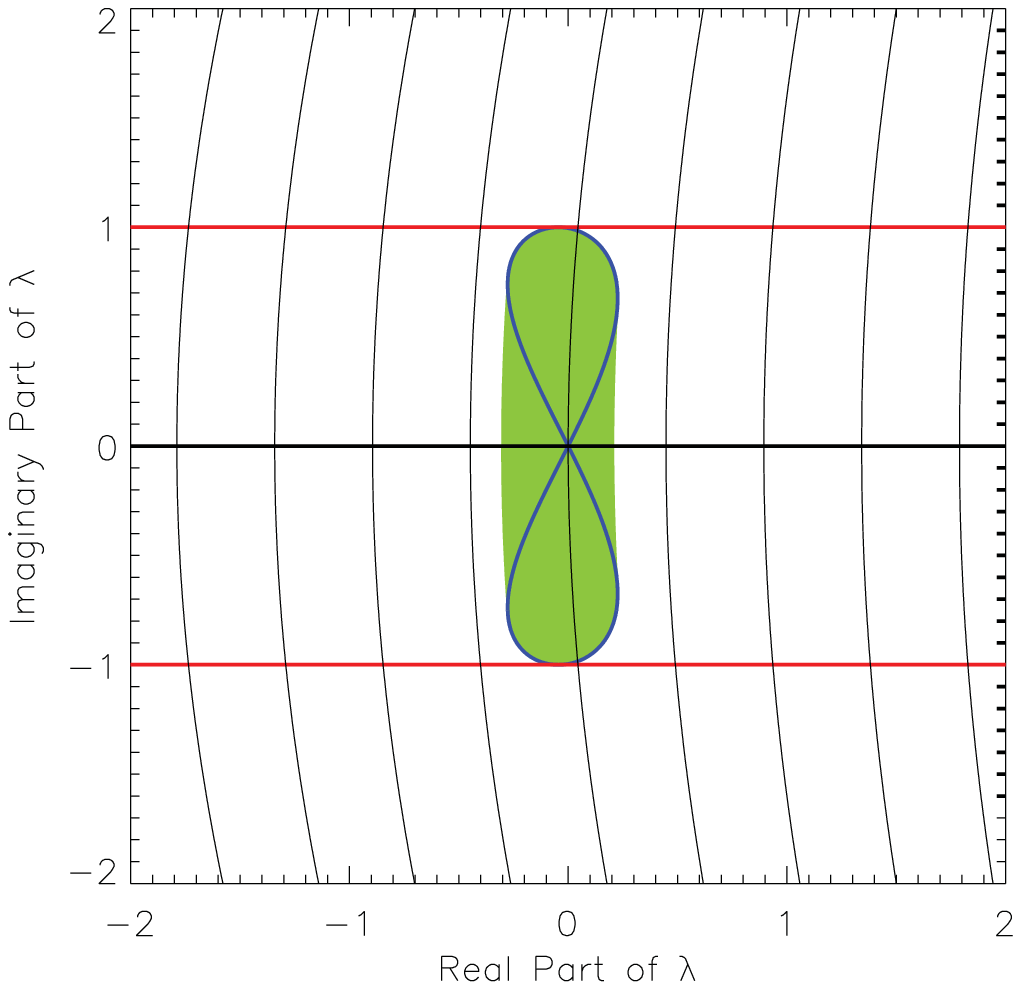}}
\caption{\em The turning point curve $\mathcal{T}$ and its hyperbolic shadow
$\mathcal{S}$ for $\alpha=0.0894$.  Left:  the $k$-plane.  
Right:  the $\lambda$-plane.}\label{fig:4}
\end{figure}

What do shadows $\mathcal{S}$ look like for functions $A(\cdot)$ and
$S(\cdot)$ that might arise in applications?  Figures
\ref{fig:1}--\ref{fig:4} each contain plots of turning point curves
$\mathcal{T}$ (blue), several of the hyperbola $\Sigma_{C}$ for
various values of $C$ (black), the bounds on $\text{Im}\{k\}$ from from
part (a) of Theorem~\ref{thm:spectralbound} (red), and the shadow
region $\mathcal{S}$ (green). The figures are all for the potentials
$\phi(x)=A(x)e^{iS(x)/\e}$ with $A(x)=\mathrm{sech}(x)$ and
$S'(x)=\mathrm{sech}(x)\tanh(x)$.  (This example was also considered
in the context of the Zakharov-Shabat spectral problem in \cite{B96}
and \cite{M01}.)  The four figures illustrate the dependence of the
shadow region $\mathcal{S}$ on the parameter $\alpha$. (For clarity,
these figures do not illustrate the $\e$-dependent ``collar''.)

Figures \ref{fig:3} and \ref{fig:4} each consist of two plots of the
same thing; one displayed in the $k$-plane and one displayed in the
$\lambda$-plane, where $\lambda$ and $k$ are related by
\eqref{eq:klambda} as explained in the appendix.  When $\alpha$ is
small, as in these figures, one expects features of the spectrum when
viewed in the $\lambda$-plane to resemble known features of the
spectrum for the nonselfadjoint Zakharov-Shabat spectral problem,
which arises by a formal limit of $\alpha\to 0$.  The hyperbolae
$\Sigma_C$ tend to vertical lines in this limit, and part of the
turning point curve tends to a fixed curve in the $\lambda$-plane.
Indeed, substituting for $k$ in terms of $\lambda$ from
\eqref{eq:klambda} in the relations \eqref{eq:returning} defining the
turning point curve $\mathcal{T}$, one may select the branch of
$\mathcal{T}$ for which $\text{Re}\{k\}\approx 1/2$ for $\alpha$ small
and pass to the limit of $\alpha\to 0$ with $\lambda$ and $x$ held
fixed, to obtain the limiting curve
\begin{equation}
\begin{split}
\text{Im}\{\lambda\}&=\pm A(x)\,,\\
\text{Re}\{\lambda\}&=-\frac{1}{2}S'(x)\,.
\end{split}
\label{eq:T0}
\end{equation}
We refer to this limiting curve in the complex $\lambda$-plane as
$\mathcal{T}_o$.  Deift, Venakides, and Zhou\footnote{This is an
  unpublished result, see \cite{B96}, \cite{M01}} have established
results analogous to ours in the context of the nonselfadjoint
Zakharov-Shabat eigenvalue problem; in these results the curve
$\mathcal{T}_o$ plays the role of the $\alpha$-dependent curve
$\mathcal{T}$ and the family of vertical lines $\text{Re}\{\lambda\}=C$
plays the role of the family of hyperbolae $\Sigma_C$.  Thus we see
that the known ``shadow bound'' result for the Zakharov-Shabat problem
can be deduced from our results as a limiting special case.

The shadow bound for the Zakharov-Shabat eigenvalue problem implies
that if $S''(\cdot)\equiv 0$ (so that $S'(\cdot)\equiv C_o$ for some
constant $C_o$), then the eigenvalues lie very close to the real axis or
the vertical segment with endpoints $-C_o/2\pm iA_{\rm max}$, $A_{\rm
  max}:=\sup_{x\in\mathbb{R}}A(x)$, in the complex $\lambda$-plane.
Klaus and Shaw \cite{KS} have proved that if in addition $A(\cdot)$ is
a real function with a single critical point (necessarily a local
maximum) then the eigenvalues of the nonselfadjoint Zakharov-Shabat
spectral problem are purely imaginary numbers regardless of the value
of $\e>0$.  Such a result represents a much more precise confinement
of the eigenvalues than is afforded by the shadow bound with its
attendant $\e$-dependent ``collar''.  It is an open question whether
there exists such a result in the context of the MNLS spectral problem
\eqref{eq:Laxx}.  A reasonable guess might be to draw an analogy
between the vertical ``rays'' making the shadow in the Zakharov-Shabat
problem with the hyperbolic ``rays'' making the shadow in the MNLS
spectral problem and thus suppose that if the functions $A(\cdot)$ and
$S'(\cdot)$ are such that the turning point curve $\mathcal{T}$ is
contained within a single hyperbola $\Sigma_C$ for some $C<0$:
\begin{equation}
2\alpha^2A(x)^2+\alpha S'(x)-1=4C<0\,,\quad\quad\forall x\in\mathbb{R}
\label{eq:confine}
\end{equation}
(this is the MNLS analogue of the condition $S'(\cdot)\equiv C_o$ in
the Zakharov-Shabat case\footnote{In fact, upon setting $C=(\alpha
  C_o-1)/4$ in \eqref{eq:confine} and taking the limit $\alpha\to 0$
  one obtains exactly the relation $S'(\cdot)\equiv C_o$.}, which makes the
corresponding turning point curve $\mathcal{T}_o$ collapse onto the
imaginary axis of the $\lambda$-plane), and if an appropriate analogue
of the Klaus-Shaw critical point condition is satisfied, then the
eigenvalues all lie exactly for all $\e>0$ on the hyperbola
$\Sigma_C$.  Noting that the Klaus-Shaw critical point condition
implies that the limiting turning point curve $\mathcal{T}_o\subset
i\mathbb{R}$ is traced out according to its natural parametrization
exactly twice as $x$ increases from $-\infty$ to $+\infty$, we might
further conjecture that the analogue of this condition for the MNLS
problem is that $\mathcal{T}\subset\Sigma_C$ is covered exactly twice
as $x$ varies.  But, from \eqref{eq:returning} we see that this occurs
for $A(\cdot)$ and $S'(\cdot)$ satisfying \eqref{eq:confine} exactly
when the amplitude function $A(\cdot)$ is positive and has a single
critical point, necessarily a local maximum; that is, we expect that
the same exact monotonicity condition will be required of $A(\cdot)$
as in the Zakharov-Shabat case to obtain exact spectral confinement to
a curve, but that the relation $S'(\cdot)\equiv C_o$ must be replaced
by \eqref{eq:confine}.

\section{A Family of Special Initial Data}
\label{sec:hypergeometric}
In this section, we investigate in detail the spectral problem corresponding
to the MNLS equation (see \eqref{eq:Laxx} in the appendix) for initial data
of the form $\phi(x,0)=A(x)e^{iS(x)/\e}$ where we assume that the amplitude
and phase functions are given by
\begin{equation}
A(x)=\nu\,\text{sech}(x)\quad\quad\text{and}\quad\quad
S(x)=S_0+\int_0^x\left[\delta+\mu\tanh(x)\right]\,dx\,,
\label{eq:specialform}
\end{equation}
where $\nu>0$, $\delta$, $\mu$, and $S_0$ are arbitrary real
parameters.  Following the algorithm described in the appendix, our
aim is to calculate all of the scattering data for this family.  This
requires solving the linear system \eqref{eq:Laxx}, which we will do
by showing that for this special initial data it is essentially a
hypergeometric equation.  Our analysis will be valid for all values of
the data parameters $\nu$, $\delta$, $\mu$, and $S_0$ and for all
values of $\alpha\ge 0$.  More significantly, our analysis will also
be valid for all $\e>0$ which will allow us to carry out a complete
semiclassical analysis of the inverse (Riemann-Hilbert) problem in a
subsequent publication.  In connection with this latter application,
the family of special initial data under consideration here is
particularly interesting because it will turn out that for certain
choices of parameters in the family and certain values of $\alpha$ we
can be in any of the cases where the condition \eqref{stabcond} is
satisfied for all $x\in\mathbb{R}$, is not satisfied for any
$x\in\mathbb{R}$, or satisfied for some $x$ and not others.  Thus we
have the possibility of providing a completely rigorous asymptotic
explanation of the effect of the stability threshold on the
semiclassical dynamics of the MNLS equation.

The idea of seeking potentials $\phi$ for which the spectral problem
\eqref{eq:Laxx} might be hypergeometric is an old idea for
Schr\"odinger operators.  This technique was applied to the
nonselfadjoint Zakharov-Shabat operator (upon which the integrable
theory of the focusing NLS equation is based) first by Satsuma and
Yajima \cite{SY} who considered the case
$\phi(x,0)=\nu\,\text{sech}(x)$, and subsequently by Tovbis and
Venakides \cite{TV} who generalized the technique to potentials of the
general form\footnote{Tovbis and Venakides actually assumed that
  $\delta=0$ but elementary symmetries of the Zakharov-Shabat problem
  corresponding to the Galilean invariance of the focusing NLS
  equation imply that their results are easily transformed to
  completely handle the case of $\delta\neq 0$ as well.} we consider
here.  Our work in this section shows that potentials of
Tovbis-Venakides form are hypergeometric also for the more complicated
spectral problem \eqref{eq:Laxx} corresponding to the MNLS equation.
Moreover, our results generalize those of Tovbis and Venakides in the
sense that we recover their formulae by taking a suitable limit
corresponding to $\alpha\rightarrow 0$.

\subsection{Reduction to hypergeometric form.}
Making the substitution ${\bf w}=e^{-iS(x)\sigma_3/(2\e)}{\bf v}$ in the linear
system \eqref{eq:Laxx} where $\phi=A(x)e^{iS(x)/\e}$ has the effect of removing
the oscillatory terms from the coefficient matrix.  Thus, with the amplitude
and phase functions $A$ and $S$ given by \eqref{eq:specialform} we arrive
at the following equation satisfied by ${\bf w}$:
\begin{equation}
2\e\frac{d\mathbf{w}}{dx}=
\begin{bmatrix} 2\Omega-i\mu\tanh(x)
& 4ik\nu\,\text{sech}(x)\\  4ik\nu\,\text{sech}(x) 
& -2\Omega+i\mu\tanh(x)
\end{bmatrix} \mathbf{w}\,, 
\label{eq:specialw}
\end{equation}
where
\begin{equation}
\Omega:=\Lambda-\frac{i\delta}{2}=\frac{1}{2i\alpha}(4k^2+\alpha\delta-1)\,.
\end{equation}
The key observation we make, in direct analogy with \cite{SY} and \cite{TV},
is that the change of independent variable $y=\tanh(x)$ together with the
substitution
\begin{equation}
{\bf a}:=\begin{bmatrix}1 & 0\\0 & \sqrt{1-y^2}\end{bmatrix}{\bf w}=
\begin{bmatrix}1 & 0\\0 & \text{sech}(x)\end{bmatrix}{\bf w}
\end{equation}
reduces \eqref{eq:specialw} to a linear first-order system with
rational (in the new independent variable $y$) coefficients.  Indeed,
${\bf a}$ satisfies
\begin{equation}
2\e(1-y^2)\frac{d{\bf a}}{dy}=\begin{bmatrix}
2\Omega-i\mu y & 4ik\nu\\4ik\nu(1-y^2) & -2\Omega+(i\mu-2\e) y
\end{bmatrix}{\bf a}\,.
\label{eq:speciala}
\end{equation}
This $2\times 2$ system has exactly three singular points, $y=\pm 1$
and $y=\infty$.  Moreover, all three are regular singular points,
which essentially makes \eqref{eq:speciala} a hypergeometric differential
equation \cite{CarrierKP83}.  To make this clearer, we follow some steps
to convert \eqref{eq:speciala} to a more standard form.

The method of Frobenius applies near each of the (regular singular) points
$y=\pm 1$.  That is, for $y$ in a neighborhood of $y=\pm 1$ there is a solution
${\bf a}(y)$ that has a convergent expansion of the form
\begin{equation}
{\bf a}=(1\mp y)^{\rho_\pm}\sum_{n=0}^\infty (1\mp y)^n{\bf a}_{n\pm}\,,
\end{equation}
where ${\bf a}_{n\pm}$ are vector-valued coefficients.  Substituting
this series into \eqref{eq:speciala} and gathering the coefficients of
like powers of $(1\pm y)$ leads to a hierarchy of algebraic equations
relating the coefficients ${\bf a}_{n\pm}$ and the exponent
$\rho_\pm$.  The equation arising at leading order is
\begin{equation}
\mp 4\e\rho_\pm{\bf a}_{0\pm}=\begin{bmatrix}2\Omega\mp i\mu & 4ik\nu\\0 & 
-2\Omega\pm (i\mu-2\e)\end{bmatrix}{\bf a}_{0\pm}\,.
\end{equation}
In order for there to exist nontrivial series solutions (subsequent
coefficients vanish for generic exponents if ${\bf a}_{0\pm}={\bf 0}$)
it is therefore necessary that $\mp 4\e\rho_\pm$ agree with one of the
eigenvalues of the matrix on the right-hand side.  This is the
\emph{indicial equation} of Frobenius theory.  The exponents therefore
are
\begin{equation}
\rho_\pm=\frac{i\mu}{4\e}\mp\frac{\Omega}{2\e}\quad\quad\text{or}\quad\quad
\rho_\pm=\frac{1}{2}-\left(\frac{i\mu}{4\e}\mp\frac{\Omega}{2\e}\right)\,.
\end{equation}
The case where $k^2\in\mathbb{R}$ will be of particular interest
shortly, and for such $k$ the two exponents never differ by an
integer, which means that a full basis of solutions near each of the
two singular points $y=\pm 1$ may be obtained in the form of
convergent Frobenius series using both exponents in each case.  

The purpose of calculating the Frobenius exponents is to show that by making
the substitution
\begin{equation}
{\bf b}=(1+y)^{-(2\Omega+i\mu)/(4\e)}(1-y)^{(2\Omega-i\mu)/(4\e)}{\bf a}
\label{eq:shift}
\end{equation}
the differential equation \eqref{eq:speciala} is converted into
another having the same three singular points, all regular singular,
but with the property that near each of the points $y=\pm 1$ one of
the Frobenius exponents has been shifted to zero.  Thus
the equation for ${\bf b}$ equivalent to that for ${\bf a}$ via
\eqref{eq:shift}, 
\begin{equation}
\e (1-y^2)\frac{d{\bf b}}{dy}=\begin{bmatrix}0 & 2ik\nu\\
2ik\nu(1-y^2) & -2\Omega+(i\mu-\e)y\end{bmatrix}{\bf b}\,,
\label{eq:specialb}
\end{equation}
is guaranteed to have one solution that is analytic at $y=-1$ and
another that is analytic near $y=+1$.  The zero that has appeared in
the coefficient matrix makes it easy to eliminate one of the
components and arrive at a second-order differential equation; but the
process is made even easier by first making the substitution
\begin{equation}
{\bf c}=\begin{bmatrix}1 & 0\\0 & (1-y^2)^{-1}\end{bmatrix}{\bf b}
\end{equation}
leading to the the coupled system 
\begin{equation}
\e\frac{dc_1}{dy}=2ik\nu c_2\quad\quad\text{and}\quad\quad
\e(1-y^2)\frac{dc_2}{dy}=2ik\nu c_1 +[(i\mu+\e)y-2\Omega]c_2\,.
\label{eq:specialc}
\end{equation}
At this point it is trivial to eliminate $c_2$ and obtain a
second-order differential equation for $c_1$ alone.  The resulting
equation for $c_1$ is a standard hypergeometric equation
\cite{CarrierKP83} in the independent variable $z=(1+y)/2$ with
parameters
\begin{equation}
\begin{split}
A&=\frac{1}{\e}\left(\frac{i\mu}{2}+\sqrt{16k^2\nu^2-\mu^2}\right)\,,\\
B&=\frac{1}{\e}\left(\frac{i\mu}{2}-\sqrt{16k^2\nu^2-\mu^2}\right)\,,\\
C&=\frac{1}{\e}\left(\Omega+\frac{i\mu}{2}\right)+\frac{1}{2}\,.
\end{split}
\end{equation}
The standard hypergeometric function $F(A,B;C;z)$ is therefore one solution
for $c_1$.  

\subsection{Euler transforms and integral representations of Jost solutions.}
From the point of view of the scattering theory outlined in the
appendix, it is necessary to obtain appropriate bases of solutions of
the equation \eqref{eq:Laxx}, equivalent here to the hypergeometric
problem \eqref{eq:specialc} and use them to compute certain Wronskian
determinants as functions of $k$.  In principle, all of this
information is available in the vast literature on special functions
(of which \cite{AndrewsAR} and \cite{CarrierKP83} are but two
sources).  Rather than simply quoting the results we need, we proceed
directly by deriving integral representations for the required
solutions and using these to calculate the scattering data.  This
approach has the advantage that we will have the integral
representations available for subsequent perturbative analysis of the
spectral problem \eqref{eq:Laxx} in the semiclassical limit based on
Langer transformations.  See \cite{BronskiMM} for an example of this
kind of analysis.

Integral representations for the Jost solutions may be obtained with the
help of \emph{Euler transforms} \cite{Hille76}.
Suppose $c_{1}$ and $c_{2}$ have the integral representations:
\begin{equation}
c_{1}(y)=\int_{\Sigma}C_1(t)(t-y)^{\beta_1}\,dt \quad\quad 
\text{and}\quad\quad c_{2}(y)=\int_{\Sigma}C_2(t)(t-y)^{\beta_2}\,dt\,,
\end{equation}
where the contour $\Sigma$, the exponents $\beta_j\in\mathbb{C}$ and
the functions $C_j(\cdot)$ (the Euler transforms of $c_j(\cdot)$) are
to be chosen so that these expressions satisfy \eqref{eq:specialc}.
Assuming that the contour $\Sigma$ is independent of $y$ and noting that
therefore
\begin{equation}
\frac{dc_{1}}{dy}=-\beta_1\int_{\Sigma}C_1(t)(t-y)^{\beta_1-1}\,dt\,, 
\end{equation}
the first equation of \eqref{eq:specialc} is satisfied by choosing
\begin{equation}
\label{betaeqn}
\beta_1:=\beta_2+1 \quad\quad\text{and}\quad\quad
C_1(t):=-\frac{2ik\nu}{(\beta_2+1)\e}C_2(t)\,.
\end{equation} 
For convenience, we write $C(t)=C_2(t)$ and $\beta=\beta_2$.  
With these choices, the second equation of \eqref{eq:specialc} becomes
\begin{multline}
-\e\beta(1-y^2)\int_\Sigma C(t)(t-y)^{\beta-1}\,dt = 
\frac{4k^2\nu^2}{(\beta+1)\e}\int_\Sigma C(t)(t-y)^{\beta+1}\,dt\\
{}+\left[(i\mu+\e)y-2\Omega\right]\int_\Sigma C(t)(t-y)^{\beta}\,dt\,.
\end{multline}
The aim in satisfying an equation such as this is to equate integrands rather
than integrals.  As a first step, we may of course move the polynomial-in-$y$
coefficients inside of the integrals; in doing this we write $y$ in the form 
$y=t-(t-y)$ so that we may continue to view the integrands as functions of
the variables $t$ and $t-y$.  Thus we obtain
\begin{multline}
\int_\Sigma\e\beta (1-t^2)C(t)(t-y)^{\beta-1}\,dt+\int_\Sigma
\left[(2\e\beta+\e+i\mu)t-2\Omega\right]C(t)(t-y)^{\beta}\,dt \\
{}=\int_\Sigma\left[\e\beta+\e+i\mu-\frac{4k^2\nu^2}{(\beta+1)\e}\right]
C(t)(t-y)^{\beta+1}\,dt\,.
\label{eq:threeexponents}
\end{multline}
The next step is to use integration by parts to get all of the
exponents of $t-y$ to be the same.  Then comparison of the integrands
will lead to a linear differential equation for $C(t)$, the Euler
transform of that for $c_2(y)$.  Since there are three consecutive
exponents, two integrations by parts will be required in general to
achieve this, and we will arrive at a second-order equation for
$C(t)$.  This would not be advantageous, as we began (essentially)
with such an equation.  We want to reduce the order.

The point is that by choosing $\beta$ appropriately we may eliminate one
of the three consecutive exponents in \eqref{eq:threeexponents}, a step
that will lead to a first-order equation for $C(t)$.  Indeed, the quantity
in square brackets on the right-hand side of \eqref{eq:threeexponents}
is a constant depending on $\beta$, and it vanishes if we choose one of
the two values
\begin{equation}\label{beta2}
\beta=-1+\frac{1}{2\e}\left(-i\mu\pm i\sqrt{\mu^2-16k^2\nu^2}\right)\,.
\end{equation}
With either of these two choices for $\beta$ (as long as $\beta\neq
-1$, and we will discuss this exceptional case once we make a proper
definition \eqref{eq:betadefine} of $\beta$ below), the right-hand
side of \eqref{eq:threeexponents} vanishes.  Integrating by parts in
the first integral on the left-hand side of \eqref{eq:threeexponents},
we therefore find that
\begin{equation}
\int_\Sigma\left[-\e\frac{d}{dt}\left[(1-t^2)C(t)\right] + \left[(2\e\beta+\e+i\mu)t-2\Omega
\right]C(t)\right](t-y)^\beta\,dt=0\,,
\end{equation}
assuming the contour $\Sigma$ is chosen so that the boundary terms
vanish (we will check this later).  
We solve this equation by equating the integrand to zero, yielding the
first-order equation for $C(t)$:
\begin{equation}
\e\frac{d}{dt}\left[(1-t^2)C(t)\right] = \left[(2\e\beta+\e+i\mu)t-2\Omega
\right]C(t)\,.
\end{equation}
The Euler transform $C(t)$ is therefore (up to a multiplicative constant)
a branch of the multi-valued function
\begin{equation}
C(t)=(1-t)^{\gamma_+}(1+t)^{\gamma_-}\,,\quad\quad\text{where}\quad\quad
\gamma_\pm=-\frac{1}{2\e}(2\e\beta+3\e+i\mu\mp 2\Omega)\,.
\end{equation}

\begin{figure}[h]
\begin{center}
\includegraphics{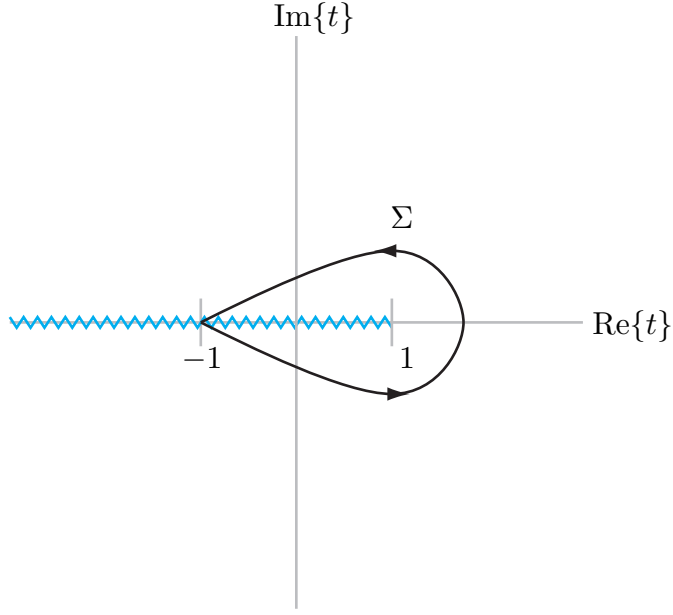}
\end{center}
\caption{\em The contour $\Sigma$ in the complex $t$-plane.  The branch
cuts of the integrand all lie on the wavy half-line.}
\label{fig:SigmaContour}
\end{figure}

We now suppose that $k^2\in\mathbb{R}$ and $\mu^2-16k^2\nu^2>0$, which
puts $k$ either on the imaginary axis or in the real interval
$(-|\mu|/(4\nu),|\mu|/(4\nu))$.  We arbitrarily make the concrete choice
of the $+$ sign in \eqref{beta2} (and interpret the square root there
as being positive).  The task at hand is to choose the contour
$\Sigma$ so that we obtain integral representations for the Jost
solution ${\bf j}^{(1)}_+(x;k)$ defined by the chain of
transformations ${\bf j}^{(1)}_+={\bf v}\mapsto{\bf w}\mapsto{\bf
  a}\mapsto{\bf b}\mapsto{\bf c}$ and the boundary condition
\begin{equation}
\lim_{x\rightarrow+\infty}{\bf j}_+^{(1)}(x;k)e^{-\Lambda x/\e}=\begin{bmatrix}
1\\0\end{bmatrix}\,.
\label{eq:j1plusBC}
\end{equation}
We therefore attempt to represent the components of ${\bf
  j}_+^{(1)}(x;k)=(J_{11+}(x;k),J_{21+}(x;k))^T$ in the form
\begin{equation}
\begin{split}
J_{11+}(x;k)&=-\frac{2ik\nu e^{iS_0/(2\e)}}{(\beta+1)\e}e^{\Lambda x/\e}
\int_\Sigma C(t)(t-y)^{\beta+1}\,dt\,,
\\
J_{21+}(x;k)&=e^{-iS(x)/\e}\,\text{sech}(x)e^{iS_0/(2\e)}e^{\Lambda x/\e}
\int_\Sigma C(t)(t-y)^{\beta}\,dt\,.
\end{split}
\label{eq:Jints}
\end{equation}
We make these formulae concrete in two steps.  First we select the branch of
$C(t)$ as follows:
\begin{equation}
C(t)=C_0(t-1)^{\gamma_+}(t+1)^{\gamma_-}
\end{equation}
where $C_0$ is a constant to be determined, and where the principal
branch of the power functions is intended; that is, $-\pi<\arg(t\pm
1)<\pi$.  Similarly we choose $-\pi<\arg(t-y)<\pi$.  Next, we select
the contour $\Sigma$ illustrated in Figure~\ref{fig:SigmaContour}.
Our assumptions on $k$ imply that $\text{Re}\{\beta\}=-1$ and
$\text{Re}\{\gamma_\pm\}=-1/2$.  The integrals in \eqref{eq:Jints} are
therefore both convergent at $t=\pm 1$ for $-1<y<1$.  Moreover, we have
\begin{equation}
(1-t^2)C(t)(t-y)^\beta=O\left((t\mp 1)^{1/2}\right)\,,\quad\quad
\text{as $t\rightarrow \pm 1$}\,,
\end{equation}
so the neglect of boundary terms in the integration by parts argument
is indeed justified after the fact in this situation.

To choose the constant $C_0$ and verify the boundary condition
\eqref{eq:j1plusBC} we need to calculate the behavior of the integrals
appearing in \eqref{eq:Jints} in the limit $y\rightarrow 1$, corresponding
to $x\rightarrow +\infty$.  It is easy to see that both of these integrals
are in fact analytic in a neighborhood of $y=1$, and a simple dominated
convergence argument shows that
\begin{equation}
\lim_{y\rightarrow 1}\int_\Sigma (t-1)^{\gamma_+}(t+1)^{\gamma_-}(t-y)^{\beta+j}\,dt=
\int_\Sigma (t-1)^{\gamma_++\beta+j}(t+1)^{\gamma_-}\,dt\,,\quad\quad j=0,1\,.
\end{equation}
(The factor $(t-y)^{\beta+j}$ is uniformly bounded on $\Sigma$
independently of $y$ near $y=1$, and the remaining factor
$(t-1)^{\gamma_+}(t+1)^{\gamma_-}$ is in $L^1(\Sigma)$, so indeed
dominated convergence applies.)  Again recalling
$\text{Re}\{\gamma_-\}=-1/2$ we see that these are both finite under our
assumptions on $k$.  This confirms the boundary condition
\eqref{eq:j1plusBC} up to the determination of the constant $C_0$.  To
get this constant, note that these limiting integrals are essentially
beta integrals, which is to say that they are ratios of gamma
functions.  In the case $j=1$ the limiting integrand is integrable at
$t= 1$ as well as $t=-1$.  Thus the contour $\Sigma$ may be collapsed
to opposite sides of the branch cut giving
\begin{equation}
\begin{split}
\int_\Sigma (t-1)^{\gamma_++\beta+1}(t+1)^{\gamma_-}\,dt&=
-2i\sin(\pi(\gamma_++\beta+1))
\int_{-1}^{+1}(1-t)^{\gamma_++\beta+1}(t+1)^{\gamma_-}\,dt\\
&=-2^{3+\gamma_++\gamma_-+\beta}i\sin(\pi(\gamma_++\beta+1))\int_0^1u^{\gamma_-}
(1-u)^{\gamma_++\beta+1}\,du\\
&=-2^{3+\gamma_++\gamma_-+\beta}i\sin(\pi(\gamma_++\beta+1))
\frac{\Gamma(\gamma_-+1)\Gamma(\gamma_++\beta+2)}{\Gamma(\gamma_++\gamma_-+\beta+3)}\\
&=2^{3+\gamma_++\gamma_-+\beta}\pi i\frac{\Gamma(\gamma_-+1)}{\Gamma(-1-\gamma_+-\beta)\Gamma(\gamma_++\gamma_-+\beta+3)}\,,
\end{split}
\end{equation}
where we have used the ``beta-gamma'' identity 
\begin{equation}
\int_0^1u^{A-1}(1-u)^{B-1}\,du=\frac{\Gamma(A)\Gamma(B)}{\Gamma(A+B)}\,,\quad\quad \text{Re}\{A\}>0\,,\quad\text{Re}\{B\}>0
\end{equation}
and Euler's reflection formula 
\begin{equation}
\sin(\pi z)\Gamma(z)\Gamma(1-z)=\pi
\label{eq:Eulerreflection}
\end{equation}
(see \cite{AndrewsAR}).  
According to this calculation, the boundary
condition \eqref{eq:j1plusBC} requires that we choose
\begin{equation}
C_0=\frac{(\beta+1)\e e^{-iS_0/(2\e)}}{2^{4+\gamma_++\gamma_-+\beta}\pi k\nu}
\frac{\Gamma(\gamma_++\gamma_-+\beta+3)\Gamma(-1-\gamma_+-\beta)}
{\Gamma(\gamma_-+1)}\,,
\label{eq:C0}
\end{equation}
and with this choice ${\bf j}_+^{(1)}(x;k)$ is determined uniquely.  

This completes our construction of the Jost solution ${\bf
  j}_+^{(1)}(x;k)$.  The three remaining Jost solutions may also be
found explicitly by following similar arguments, but we will be able
to calculate all of the scattering data from our knowledge of ${\bf
  j}_+^{(1)}(x;k)$ alone.  In the case of the specific class of
potentials $\phi=\phi(x;\nu,S_0,\mu,\e)$ under consideration here it
is easy to confirm that for each fixed $k\in\mathbb{C}$, whenever
${\bf v}(x;k)$ satisfies the differential equation \eqref{eq:Laxx} with
$\phi=\phi(x;\nu,S_0,\mu,\e)$, then the vector
\begin{equation}
{\bf v}^\sharp(x;k):=\begin{bmatrix}0 & -1\\1 & 0\end{bmatrix}{\bf v}(-x;k)
\end{equation}
satisfies the same differential equation for the same
$k\in\mathbb{C}$, but for a different potential in the same family,
namely $\phi^\sharp=\phi(x;\nu,-S_0,-\mu,\e)$.  In this way, we see
that, for example,
\begin{equation}
{\bf j}_-^{(2)}(x;k)=\begin{bmatrix}0 & -1\\1 & 0\end{bmatrix}
{\bf j}_+^{(1)}(-x;k)\Big|_{S_0\to -S_0,\;\mu\to -\mu}\,.
\end{equation}
Since our formulae for  ${\bf j}_+^{(1)}(x;k)$ are valid for all values of
the parameters, we also have immediate access to integral representations
for the components of ${\bf j}_-^{(2)}(x;k)$.

\subsection{Continuous scattering data.}
We now use the integral representation for ${\bf j}_+^{(1)}(x;k)$ to
obtain the entries of the scattering matrix ${\bf S}(k)$ and establish
their basic properties.  We also will derive formulae for the reflection
coefficient $r(k)$ and the reduced reflection coefficient $\rho(z)$.

As described in the appendix (see \eqref{eq:Sdets} in particular), the
scattering matrix entries $S_{11}(k)$ and $S_{21}(k)$ may be defined
in terms of ${\bf j}_+^{(1)}(x;k)$ by the limits
\begin{equation}
S_{11}(k):=\lim_{x\rightarrow -\infty} e^{-\Lambda x/\e}J_{11+}(x;k)
\end{equation}
and
\begin{equation}
S_{21}(k):=\lim_{x\rightarrow -\infty}e^{\Lambda x/\e}J_{21+}(x;k)\,.
\end{equation}
To calculate $S_{11}(k)$ for $k^2\in\mathbb{R}$ with $\mu^2-4k^2\nu^2>0$ we
must therefore evaluate the limit
\begin{equation}
S_{11}(k)=-\frac{2ik\nu C_0e^{iS_0/(2\e)}}{(\beta+1)\e}\lim_{y\rightarrow -1}
\int_\Sigma (t-1)^{\gamma_+}(t+1)^{\gamma_-}(t-y)^{\beta+1}\,dt\,.
\label{eq:S11first}
\end{equation}
The integral here is not analytic in a neighborhood of $y=-1$, but
since $(t-y)^{\beta+1}$ is uniformly bounded for $t$ and $y$ both near
$-1$ for the values of $k$ under consideration, dominated convergence
again applies and allows us to take the limit under the integral.
Thus
\begin{equation}
S_{11}(k)=-\frac{2ik\nu C_0e^{iS_0/(2\e)}}{(\beta+1)\e}
\int_\Sigma (t-1)^{\gamma_+}(t+1)^{\gamma_-+\beta+1}\,dt\,.
\end{equation}
Collapsing the contour to the real line and expressing the resulting
beta integral in terms of gamma functions yields 
\begin{equation}
\begin{split}
S_{11}(k)&=\frac{2^{\gamma_++\gamma_-+\beta+4}\pi k\nu C_0e^{iS_0/(2\e)}}{(\beta+1)\e}
\frac{\Gamma(\gamma_-+\beta+2)}{\Gamma(-\gamma_+)
\Gamma(\gamma_++\gamma_-+\beta+3)}\\
&=\frac{\Gamma(\gamma_-+\beta+2)\Gamma(-1-\gamma_+-\beta)}
{\Gamma(-\gamma_+)\Gamma(\gamma_-+1)}\,.
\end{split}
\label{eq:S11last}
\end{equation}

Similarly, to calculate $S_{21}(k)$ for $k^2\in\mathbb{R}$ with
$\mu^2-4k^2\nu^2>0$ we must evaluate the limit
\begin{equation}
S_{21}(k)=C_0e^{iS_0/(2\e)}\lim_{x\rightarrow -\infty}\left[e^{-iS(x)/\e}\,\text{sech}(x)e^{2\Lambda x/\e}\int_\Sigma (t-1)^{\gamma_+}(t+1)^{\gamma_-}(t-\tanh(x))^\beta\,dt
\right]\,.
\label{eq:S21first}
\end{equation}
Here we cannot immediately pass to the limit in the integral. However,
noting that the integrand is integrable at $t=\infty$ we may 
first deform the contour $\Sigma$ to the top and bottom of the branch
cut for real $t<-1$, giving
\begin{equation}
\int_\Sigma(t-1)^{\gamma_+}(t+1)^{\gamma_-}(t-y)^\beta\,dt = 
2i\sin(\pi(\gamma_++\gamma_-+\beta))\int_{-\infty}^{-1}(1-t)^{\gamma_+}(-1-t)^{\gamma_-}(y-t)^\beta\,dt\,.
\end{equation}
Next, composing an appropriate $y$-dependent scaling with a M\"obius
transformation we make the substitution
\begin{equation}
t=-1-(1+y)\frac{u}{1-u}
\end{equation}
giving
\begin{multline}
\int_{-\infty}^{-1}(1-t)^{\gamma_+}(-1-t)^{\gamma_-}(y-t)^\beta\,dt\\
{}=
-2^{\gamma_+}(1+y)^{1+\beta+\gamma_-}
\int_0^1\left[1-u+\frac{1+y}{2}u\right]^{\gamma_+}u^{\gamma_-}(1-u)^{-\gamma_+-\gamma_--\beta-2}\,du\,.
\end{multline}
Now we may apply a dominated convergence argument to the resulting integral,
since $\text{Re}\{\gamma_\pm\}=-1/2$ and $\text{Re}\{\beta\}=-1$ implies that
\begin{equation}
\left|\left[1-u+\frac{1+y}{2}u\right]^{\gamma_+}\right|\le 
(1-u)^{\text{Re}\{\gamma_+\}}=(1-u)^{-1/2}
\end{equation}
and 
\begin{equation}
\left|u^{\gamma_-}(1-u)^{-\gamma_+-\gamma_--\beta-2}\right|=u^{\text{Re}\{\gamma_-\}}
(1-u)^{-\text{Re}\{\gamma_++\gamma_-+\beta+2\}}=u^{-1/2}\,.
\end{equation}
Therefore
\begin{equation}
\begin{split}
\lim_{y\rightarrow -1}
\int_0^1\left[1-u+\frac{1+y}{2}u\right]^{\gamma_+}u^{\gamma_-}(1-u)^{-\gamma_+-\gamma_--\beta-2}\,du&=\int_0^1 u^{\gamma_-}(1-u)^{-\gamma_--\beta-2}\,du\\
&=\frac{\Gamma(\gamma_-+1)\Gamma(-\gamma_--\beta-1)}
{\Gamma(-\beta)}\,.
\end{split}
\label{eq:S21last}
\end{equation}
Since by direct calculation with $y=\tanh(x)$ we have
\begin{equation}
\lim_{x\rightarrow -\infty}\left[e^{-iS(x)/\e}\,\text{sech}(x)e^{2\Lambda x/\e}
(1+y)^{1+\beta+\gamma_-}\right]=e^{-iS_0/\e}2^{i\mu/(2\e)-\Omega/\e+1/2}\,,
\end{equation}
we find
\begin{equation}
\begin{split}
S_{21}(k)&=-C_0e^{-iS_0/(2\e)}2^{-\beta}i\sin(\pi(\gamma_++\gamma_-+\beta))\frac{
\Gamma(\gamma_-+1)\Gamma(-\gamma_--\beta-1)}{\Gamma(-\beta)}\\
&=C_0e^{-iS_0/(2\e)}2^{-\beta}i\sin(\pi(\gamma_++\gamma_-+\beta+3))\frac{
\Gamma(\gamma_-+1)\Gamma(-\gamma_--\beta-1)}{\Gamma(-\beta)}\\
&=\frac{i\e e^{-iS_0/\e}2^{i\mu/\e}}{2k\nu}(\beta+1)\frac{\Gamma(-\gamma_--\beta-1)\Gamma(-\gamma_+-\beta-1)}{\Gamma(-\gamma_+-\gamma_--\beta-2)\Gamma(-\beta)}\\
&=-
\frac{i\e e^{-iS_0/\e}2^{i\mu/\e}}{2k\nu}\frac{\Gamma(-\gamma_--\beta-1)\Gamma(-\gamma_+-\beta-1)}{\Gamma(-\gamma_+-\gamma_--\beta-2)\Gamma(-\beta-1)}
\,,
\end{split}
\end{equation}
where we used \eqref{eq:Eulerreflection} upon substituting for $C_0$
from \eqref{eq:C0} and the factorial identity
$\Gamma(z+1)=z\Gamma(z)$.

All of our analysis so far has assumed that $k^2\in\mathbb{R}$ with
$\mu^2-16k^2\nu^2>0$.  If instead we have $k^2\in\mathbb{R}$ with
$16k^2\nu^2-\mu^2>0$, then we claim that all of our results remain valid
if we define
\begin{equation}
\beta=-1+\frac{1}{2\e}\left(-i\mu-\sqrt{16k^2\nu^2-\mu^2}\right)\,,\quad\quad
16k^2\nu^2-\mu^2>0\,.
\end{equation}
In this case we have the inequalities $\text{Re}\{\beta\}<-1$ and
$\text{Re}\{\gamma_\pm\}>-1/2$.  In fact, all of the preceding arguments
remain intact exactly as written with the exception of that
leading from \eqref{eq:S11first} through \eqref{eq:S11last} and that
leading from \eqref{eq:S21first} through \eqref{eq:S21last}.  We emphasize
that the \emph{results} of these calculations remain the same; only their
proofs must be altered. 

We may combine the formulae for the two sign cases for $16k^2\nu^2-\mu^2$
by introducing the function
\begin{equation}
R(k):=\begin{cases}-i\sqrt{\mu^2-16k^2\nu^2}\,,\quad\quad&16k^2\nu^2<\mu^2\\
\sqrt{16k^2\nu^2-\mu^2}\,,\quad\quad&16k^2\nu^2>\mu^2\,,
\end{cases}
\end{equation}
where in each situation the positive square root is meant, and by writing
\begin{equation}
\beta:=
-1-\frac{1}{2\e}\left(i\mu+R(k)\right)\,,\quad\quad\text{Im}\{k^2\}=0\,.  
\label{eq:betadefine}
\end{equation}
At this point, we should resolve the question of what happens if
$\beta=-1$.  From the above definitions it follows we may only have
$\beta=-1$ for $\text{Im}\{k^2\}=0$ if $\mu\ge 0$, and in this case we
must have $k=0$.  When $k=0$ the differential equations for $c_1(y)$
and $c_2(y)$ decouple and become explicitly solvable in closed-form
(Euler transforms are not required).  One can check that the
construction of Jost solutions for $k=0$ agrees with what we have done
for $k\neq 0$ in the sense of taking limits as $k\rightarrow 0$ along
the axes.  Moreover, the behavior at $k=0$ is consistent with what is
shown in the appendix (see \eqref{eq:Yjjzero}, \eqref{eq:Ytildezero},
\eqref{eq:Sjjzero}, and \eqref{eq:Soffdiagzero}) to hold for more
general potentials.

\begin{figure}[h]
\begin{center}
\includegraphics{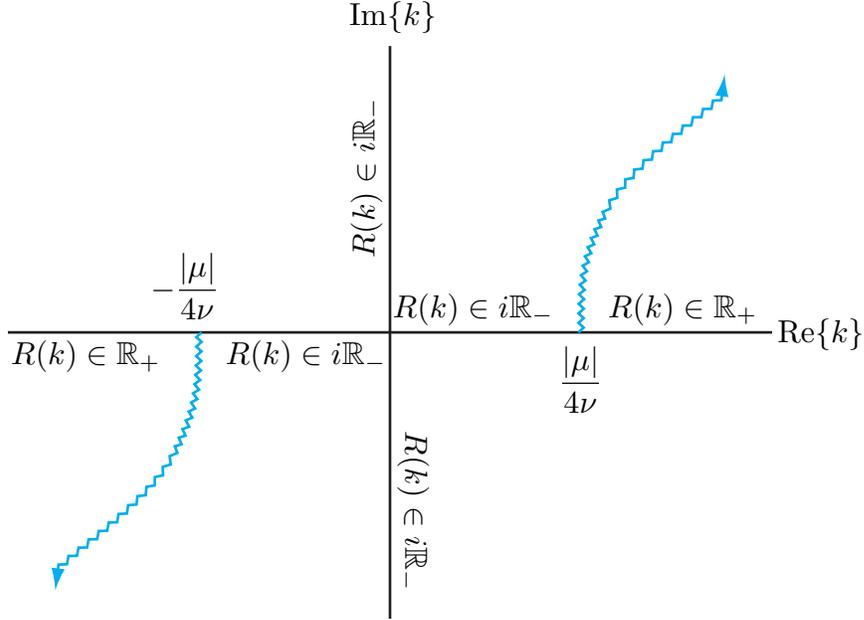}
\end{center}
\caption{\em The branch cuts of the analytic continuation of $R(k)$ from
the axes are shown with wavy curves.}
\label{fig:SqrtBranchCuts}
\end{figure}

The function $R(k)$ has an analytic continuation from the axes
$\text{Im}\{k^2\}=0$ to the second and fourth quadrants, and we define
$R(k)$ in the complex $k$-plane by introducing appropriate branch cuts
as illustrated in Figure~\ref{fig:SqrtBranchCuts}.
Since the quantity $R(k)^2=16k^2\nu^2-\mu^2$ lies in the lower half-plane when
$\text{Im}\{k^2\}<0$, it can be easily shown by taking the appropriate branch
of the square root consistent with the definition of $R(k)$ for $k$ in 
the second quadrant, that for such $k$ we have $\text{Re}\{R(k)\}>0$.

In terms of this function, we then have
\begin{equation}
S_{11}(k)=\frac{\Gamma(\frac{1}{2}-\frac{\Omega}{\e}-\frac{i\mu}{2\e})
\Gamma(\frac{1}{2}-\frac{\Omega}{\e}+\frac{i\mu}{2\e})}
{\Gamma(\frac{1}{2}-\frac{\Omega}{\e}-\frac{R(k)}{2\e})
\Gamma(\frac{1}{2}-\frac{\Omega}{\e}+\frac{R(k)}{2\e})}\,,
\label{eq:S11special}
\end{equation}
and
\begin{equation}
S_{21}(k)=-\frac{i\e e^{-iS_0/\e}2^{i\mu/\e}}{2k\nu}
\frac{\Gamma(\frac{1}{2}+\frac{i\mu}{2\e}+\frac{\Omega}{\e})
\Gamma(\frac{1}{2}+\frac{i\mu}{2\e}-\frac{\Omega}{\e})}
{\Gamma(\frac{i\mu}{2\e}-\frac{R(k)}{2\e})\Gamma(\frac{i\mu}{2\e}
+\frac{R(k)}{2\e})}\,.
\label{eq:S21special}
\end{equation}
The definition of $R(k)$ makes the arguments of all gamma functions in
\eqref{eq:S11special} analytic in $k$ for $\text{Im}\{k^2\}<0$.  This
same condition on $k$ then also implies that
$\text{Re}\{1/2-\Omega/\e\pm
i\mu/(2\e)\}=1/2-2\text{Im}\{k^2\}/(\e\alpha)>1/2$, which verifies the
expected fact that $S_{11}(k)$ extends from the axes to an analytic
function of $k$ for $\text{Im}\{k^2\}<0$.  While not guaranteed by the
general theory, one can nonetheless see directly from
\eqref{eq:S11special} that $S_{11}(k)$ is even in $R(k)$ and hence
single-valued and meromorphic in the whole complex $k$-plane (there
are poles if $\text{Im}\{k^2\}>0$).  As is consistent with the general
theory, $S_{21}(k)$ does not have an analytic extension to any of the
four quadrants; however again we can see from \eqref{eq:S21special}
that in this case $S_{21}(k)$ is even in $R(k)$ and hence
single-valued in the complex $k$-plane.  Its only singularities are
simple poles lying in all four quadrants.

Applying the symmetry relations \eqref{eq:Ssymanti} established in the
appendix, we then obtain from \eqref{eq:S11special} a formula for
$S_{22}(k)$:
\begin{equation}
S_{22}(k)=\frac{\Gamma(\frac{1}{2}+\frac{\Omega}{\e}+\frac{i\mu}{2\e})
\Gamma(\frac{1}{2}+\frac{\Omega}{\e}-\frac{i\mu}{2\e})}
{\Gamma(\frac{1}{2}+\frac{\Omega}{\e}-\frac{R(k^*)^*}{2\e})
\Gamma(\frac{1}{2}+\frac{\Omega}{\e}+\frac{R(k^*)^*}{2\e})}\,,
\label{eq:S22special}
\end{equation}
which extends to an analytic function for $\text{Im}\{k^2\}>0$, and from
\eqref{eq:S21special} a formula for $S_{12}(k)$:
\begin{equation}
\begin{split}
S_{12}(k)&=-\frac{i\e e^{iS_0/\e}2^{-i\mu/\e}}{2k\nu}
\frac{\Gamma(\frac{1}{2}-\frac{i\mu}{2\e}-\frac{\Omega}{\e})
\Gamma(\frac{1}{2}-\frac{i\mu}{2\e}+\frac{\Omega}{\e})}
{\Gamma(-\frac{i\mu}{2\e}-\frac{R(k^*)^*}{2\e})\Gamma(-\frac{i\mu}{2\e}
+\frac{R(k^*)^*}{2\e})}\\
&=-\frac{i\e e^{iS_0/\e}2^{-i\mu/\e}}{2k\nu}
\frac{\Gamma(\frac{1}{2}-\frac{i\mu}{2\e}-\frac{\Omega}{\e})
\Gamma(\frac{1}{2}-\frac{i\mu}{2\e}+\frac{\Omega}{\e})}
{\Gamma(-\frac{i\mu}{2\e}-\frac{R(k)}{2\e})\Gamma(-\frac{i\mu}{2\e}
+\frac{R(k)}{2\e})}\,,\quad\quad \text{Im}\{k^2\}=0
\,.
\end{split}
\end{equation}
(The second line follows because for $\text{Im}(k^2)=0$, either
$R(k^*)^*=R(k)$ or $R(k^*)^*=-R(k)$.)  The (reflection coefficient)
function $r(k)=-S_{12}(k)/S_{22}(k)$ is therefore 
\begin{equation}
r(k)=\frac{i\e e^{iS_0/\e}2^{-i\mu/\e}}{2k\nu}\cdot
\frac{\Gamma(\frac{1}{2}-\frac{i\mu}{2\e}-\frac{\Omega}{\e})}
{\Gamma(\frac{1}{2}+\frac{i\mu}{2\e}+\frac{\Omega}{\e})}\cdot
\frac{\Gamma(\frac{1}{2}+\frac{\Omega}{\e}-\frac{R(k)}{2\e})
\Gamma(\frac{1}{2}+\frac{\Omega}{\e}+\frac{R(k)}{2\e})}
{\Gamma(-\frac{i\mu}{2\e}-\frac{R(k)}{2\e})
\Gamma(-\frac{i\mu}{2\e}+\frac{R(k)}{2\e})}\,,\quad\text{Im}\{k^2\}=0\,.
\end{equation}
Since $R(0)=-i|\mu|$, it is easy to see that $r(k)$ is an odd
function of $k$ that is regular at $k=0$ (in fact in this case
$r(k)$ is analytic at $k=0$).  Therefore the ``reduced'' reflection
coefficient $\rho(z)$ defined by \eqref{eq:rhoreflectiondefine} is a
continuous function of $z\in\mathbb{R}$ given explicitly by 
\begin{equation}
\rho(z)=-\frac{i\e e^{iS_0/\e}2^{-i\mu/\e}}{2z\nu}\cdot
\frac{\Gamma(\frac{1}{2}-\frac{i\mu}{2\e}-\frac{\Omega}{\e})}
{\Gamma(\frac{1}{2}+\frac{i\mu}{2\e}+\frac{\Omega}{\e})}\cdot
\frac{\Gamma(\frac{1}{2}+\frac{\Omega}{\e}-\frac{R}{2\e})
\Gamma(\frac{1}{2}+\frac{\Omega}{\e}+\frac{R}{2\e})}
{\Gamma(-\frac{i\mu}{2\e}-\frac{R}{2\e})
\Gamma(-\frac{i\mu}{2\e}+\frac{R}{2\e})}\,,\quad\text{Im}\{z\}=0\,,
\label{eq:rhospecial}
\end{equation}
where in terms of $z=-k^2$ we have
\begin{equation}
\Omega=-\frac{1}{2i\alpha}(4z+1-\alpha\delta)\,,
\end{equation}
and
\begin{equation}
R=\begin{cases}-i\sqrt{\mu^2+16z\nu^2}\,,\quad\quad&-16z\nu^2<\mu^2\\
\sqrt{-16z\nu^2-\mu^2}\,,\quad\quad&-16z\nu^2>\mu^2\,.
\end{cases}
\end{equation}

Note that for $\alpha> 0$ the reduced reflection coefficient is never
(that is, for no values of the parameters $\nu$, $\mu$, $\delta$, and
$S_0$ of the potential, nor for any $\e>0$) identically zero as a
function of $z\in\mathbb{R}$.  This should be strongly contrasted with
the case $\alpha=0$, in which one considers the same initial data in
the nonselfadjoint Zakharov-Shabat problem.  For that spectral problem
one of the main points made by Satsuma and Yajima \cite{SY} is that
when $\mu=0$ there exists a sequence of values of $\e>0$ tending to
zero for which the reflection coefficient vanishes identically.  This
leads to the powerful concept of the ``higher-order solitons'' ---
exact solutions with simple initial data that can be found by a finite
number of algebraic steps as the Riemann-Hilbert problem degenerates
into a finite-dimensional linear algebra problem.  At least in this
family of special initial data there are no higher-order solitons or
reflectionless potentials when $\alpha>0$ and therefore the
Zakharov-Shabat problem must be exchanged for the more complicated
problem \eqref{eq:Laxx}.

\subsection{Discrete scattering data.}
As discussed in the appendix, the discrete spectrum for the problem
\eqref{eq:Laxx} corresponds to the zeros of $S_{11}(k)$ with
$\text{Im}\{k^2\}<0$ and those of $S_{22}(k)$ with $\text{Im}\{k^2\}>0$.
By the symmetries \eqref{eq:Ssymanti} and \eqref{eq:Ssymholo} it
suffices to consider the zeros of $S_{11}(k)$ with $\text{Im}\{k\}>0$
and $\text{Re}\{k\}<0$.  From \eqref{eq:S11special} one can see that
zeros of $S_{11}(k)$ occur where the arguments of the gamma functions
in the denominator are non-positive integers.  Since for $k$ in the
second quadrant we have both $\text{Re}\{R(k)\}>0$ and
$\text{Re}\{\Omega\}<0$, only one of the gamma functions can contribute
zeros of $S_{11}(k)$ in this quadrant.  Therefore, the zeros of
$S_{11}(k)$ with $\text{Im}\{k\}>0$ and $\text{Re}\{k\}<0$ are precisely
those values of $k=k_n$ such that
\begin{equation}\label{eigenvalueformula}
\Omega +\frac{1}{2}R(k)=\left(n+\frac{1}{2}\right)\e\,,
\end{equation}
where $n$ is a non-negative integer.  It is easy to see that $n$
coincides with $\gamma_+$.  
Isolating $R(k)$ and squaring, we thus find that the desired values
$k$ in the second quadrant are\footnote{No superfluous
  solutions are introduced because we have already shown that no
  solutions in the second quadrant can be obtained by changing the
  sign of $R(k)$.} the solutions with $\text{Im}\{k\}>0$ and
$\text{Re}\{k\}<0$ of the equation
\begin{equation}
w^2+(iL\alpha+\nu^2\alpha^2)w+\frac{1}{16}\left(4\nu^2\alpha^2(1-\alpha\delta)
-4\alpha^2 L^2
-\alpha^2\mu^2\right)=0\,,\quad w:=k^2-\frac{1}{4}(1-\alpha\delta)\,,
\label{eq:wquad}
\end{equation}
where $L:=-\e(n+1/2)<0$.  The discrete spectrum may thus be obtained
by solving this quadratic equation for $w$ and keeping only those
solutions for which $\text{Im}\{w\}<0$ (corresponding to the second
quadrant for $k$).  This results in a finite number (possibly zero) of
roots of $S_{11}(k)$ in the second quadrant of the complex $k$-plane.

If we view $L$ as a general negative real parameter rather than a
discrete one, a viewpoint that is increasingly accurate in the
semiclassical limit $\e\downarrow 0$, we can eliminate the parameter
$L$ between the real and imaginary parts of \eqref{eq:wquad} and hence
find a curve $C$ in the $(x,y)=(\text{Re}\{w\},\text{Im}\{w\})$-plane that
is independent of $\e$ but that necessarily contains the eigenvalues
for all $\e$.  The eigenvalues for this problem lie \emph{exactly} on
$\e$-independent curves\footnote{Note however, that these curves do
  not turn out to be hyperbolae of the family $\Sigma_C$.  This should
  perhaps not come as any surprise, since for the special initial data
  under consideration the turning point curve $\mathcal{T}$ does not
  coincide with a hyperbola $\Sigma_C$ either.  See the discussion at
  the very end of Section~\ref{sec:hyperbolae}.}.

Indeed, in terms of $x$ and $y$, the real part of \eqref{eq:wquad} is
\begin{equation}
\label{eq:wreal}
x^2-y^2-\alpha L y+\nu^2\alpha^2x
-\frac{\alpha^2 L^2}{4}+\frac{1}{4}\nu^2\alpha^2(1-\alpha\delta)-
\frac{\alpha^2\mu^2}{16}=0\,,
\end{equation}
and the imaginary part of \eqref{eq:wquad} is linear in $L$ and can be
solved to yield
\begin{equation}
\label{eq:wimag}
L=-\frac{y}{\alpha}\cdot\frac{2x+\alpha^2\nu^2}{x}\,.
\end{equation}
Substitution of \eqref{eq:wimag} into \eqref{eq:wreal} eliminates $L$ and
yields the relation
\begin{equation}
\label{eq:curve}
x^2Q(x)=
\frac{\nu^4\alpha^4y^2}{4}\,,\quad\quad Q(x):=
x^2+\nu^2\alpha^2x+\frac{1}{4}\nu^2\alpha^2(1-\alpha\delta)
-\frac{\alpha^2\mu^2}{16}\,.
\end{equation}
The curve $C$ defined by \eqref{eq:curve} therefore necessarily contains
all eigenvalues $k$ in the second quadrant when written in terms of $w$.

However, not all of the curve $C$ is relevant, because we need to take
into account the inequalities $y<0$ (corresponding to $k$ being in the
second quadrant) and $L<0$.  The latter condition can be translated
easily into a condition on $x$ with the help of \eqref{eq:wimag}:
since $y<0$ and $L<0$ we must have $x$ and $2x+\alpha^2\nu^2$ being of
opposite sign, which forces the conditions $-\alpha^2\nu^2/2<x<0$.
Therefore in addition to lying on the curve $C$, to correspond to an
eigenvalue for sufficiently small $\e$, $w$ must also lie in the
semi-infinite strip $S$ defined by
\begin{equation}
S:=\{(x,y)\in\mathbb{R}^2\quad\text{such that}\quad
\text{$y<0$ and $-\alpha^2\nu^2/2<x<0$}\}\,.
\end{equation}

\begin{figure}[h]
\begin{center}
\includegraphics{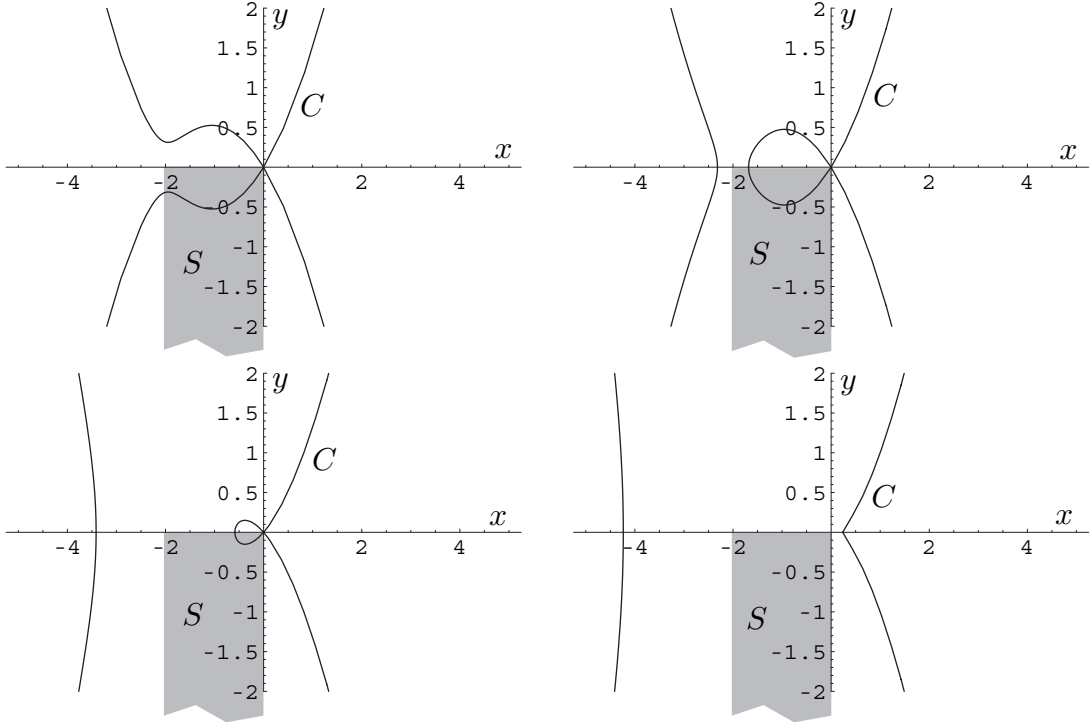}
\end{center}
\caption{\em Loss of eigenvalues with the variation of $\delta$.  In
  all plots, $\alpha=1$, $\mu=4$, and $\nu=2$, which makes the
  threshold for intersection of $C$ and $S$ correspond to $\delta=0$.
  Upper left: $\delta=-4.1$.  Upper right: $\delta=-3.9$.  Lower left:
  $\delta=-2$.  Lower right: $\delta=1$.}
\label{fig:deltas}
\end{figure}

We now claim that the curve $C$ intersects the strip $S$ if and only
if $\mu^2<4\nu^2(1-\alpha\delta)$.  Indeed, if this condition holds,
then $Q(0)>0$, in which case the local structure of the curve $C$ near
$x=y=0$ is a simple crossing of two real branches of opposite slopes.
Therefore in this situation there is a branch of $C$ near the origin
with $x<0$ and $y<0$, a region that is contained in the strip $S$.  On
the other hand, if $Q(0)<0$, then there are two distinct real branches
of $C$, one with $x>0$ and one with $x<-\alpha^2\nu^2$, both of which
are disjoint from $S$.  See Figure~\ref{fig:deltas}.
Therefore, if $Q(0)<0$ there are no
eigenvalues for any $\e>0$, while if $Q(0)>0$ then there exist
eigenvalues for all sufficiently small $\e>0$.

In the context of the particular family of potentials $\phi$ under
consideration here, it is interesting and useful to compare the
threshold condition for creation of eigenvalues with the condition on
the parameters $\nu$, $\mu$, and $\delta$ such that the stability
condition in \eqref{stabcond} is satisfied for all $x\in\mathbb{R}$.
In this specific situation, \eqref{stabcond} becomes
\begin{equation}\label{stabcond2}
\alpha\mu\tanh(x)+\alpha\delta+\alpha^2\nu^2\text{sech}^2(x)-1>0\,.
\end{equation}
Depending on the sign of $\mu$, the minimum of the expression on the
left of the inequality in \eqref{stabcond2} occurs either as $x\to
+\infty$ or as $x\to -\infty$.  The minimum value is
$-\alpha\abs{\mu}-(1-\alpha\delta)$. Consequently, if
$\abs{\mu}<-(1-\alpha\delta)/\alpha$ then the condition
\eqref{stabcond2} is satisfied for all $x\in\mathbb{R}$. Furthermore,
if $\abs{\mu}\ge -(1-\alpha\delta)/\alpha$ (in particular if
$1-\alpha\delta>0$), then the stability condition \eqref{stabcond2}
fails for some $x\in\mathbb{R}$.

We may summarize the possibilities for global-in-$x$ stability and existence
of discrete spectrum for sufficiently small $\e$ in the following way:
\begin{itemize}
\item $1-\alpha\delta<0$:  No eigenvalues exist for any $\e>0$.
\begin{itemize}
\item  If  $\abs{\mu}<-(1-\alpha\delta)/\alpha$ then we have modulational
stability for all $x\in\mathbb{R}$.
\item If $\abs{\mu}>-(1-\alpha\delta)/\alpha$ modulational instability
  occurs for some $x\in\mathbb{R}$.
\end{itemize}
\item $1-\alpha\delta>0$ : Regardless of the value of $\mu$,
  modulational instability occurs for some $x\in\mathbb{R}$.
\begin{itemize}
\item If $\abs{\mu}<2\abs{\nu}\sqrt{1-\alpha\delta}$ then eigenvalues
  exist for sufficiently small $\e>0$.
\item If $\abs{\mu}>2\abs{\nu}\sqrt{1-\alpha\delta}$ then eigenvalues
  do not exist for any $\e>0$.
\end{itemize}
\end{itemize}

A complete calculation of the discrete scattering data must include
the determination of the proportionality constants $\gamma_j$
associated with each of the eigenvalues $k_j$ in the second quadrant
of the complex $k$-plane.  As described in the appendix (see
\eqref{eq:proportionality}), this constant is defined by the relation
${\bf j}_+^{(1)}(x;k_j)=\gamma_j{\bf j}_-^{(2)}(x;k_j)$ holding as an
identity for all $x\in\mathbb{R}$.  Even though it is possible to
write down an integral representation for ${\bf j}_-^{(2)}(x;k_j)$, it
is not convenient to calculate the constants $\{\gamma_j\}$ in this
way (with the exception of certain problems where it is known that the
reflection coefficient vanishes identically; see \cite{KMM} or
\cite{SG}).  However, another way to proceed in special cases such as
this is to examine the consequences of the scattering relation
\begin{equation}
{\bf j}_+^{(1)}(x;k)=S_{11}(k){\bf j}_-^{(1)}(x;k) + S_{21}(k){\bf j}_-^{(2)}(x;k)\,,\quad\quad k^2\in\mathbb{R}\,.
\label{eq:scatteringrelation}
\end{equation}
It is not generally the case that this equation admits analytic
continuation away from the axes, but in the special case
of the class of potentials $\phi$ under consideration, 
analytic continuation is indeed possible.   By continuation along a path
beginning on the real $k$-axis at a value of $k$ with $16k^2\mu^2>\nu^2$, 
one can show that the
(nonphysical) continuation of ${\bf j}_-^{(1)}(x;k)$ into the second
quadrant is generically regular at $k=k_j$.  The only poles of this continuation
occur if
\begin{equation}
\frac{i\mu}{2}+\Omega=-\left(m+\frac{1}{2}\right)\e\,, 
\label{eq:phantom}
\end{equation}
where $m$ is a nonnegative integer.  This relation also gives poles $k_m$ of
$S_{21}(k)$ in the second quadrant lying on the hyperbola
\begin{equation}
\text{Re}\{k_m\}^2-\text{Im}\{k_m\}^2 =\frac{1}{4}(1-\alpha\delta+\alpha\mu)\,.
\label{eq:phantomhyperbola}
\end{equation}
We call these poles \emph{phantom poles}.  
As long as the eigenvalue $k_j$ in the second quadrant does not (accidentally)
coincide with a phantom pole, 
we learn from \eqref{eq:scatteringrelation} and the eigenvalue condition
$S_{11}(k_j)=0$ that $\gamma_j=S_{21}(k_j)$.  This formula is thus
generically valid for the special class of potentials under
consideration.  

This expression for the proportionality constants has added
significance when interpreted in terms of the modified proportionality
constants $c_j^0:=\gamma_j/S_{11}'(k_j)$ as one then has
\begin{equation}
c_j^0=\mathop{\text{Res}}_{k=k_j}r(k^*)^*\,,
\end{equation}
where $r(k)$ is the reflection coefficient
$r(k)=-S_{12}(k)/S_{22}(k)$, or equivalently in terms of the
``reduced'' reflection coefficient,
\begin{equation}
-2c_j^0=\mathop{\text{Res}}_{z=-k_j^2}\rho(z^*)^*\,.
\end{equation}
Such relations relating the pole data to residues of jump matrix data
are indispensible in the analysis of Riemann-Hilbert problems of
inverse scattering; see, for example \cite{TovbisVZ04}.  The main idea
is to note that the jump matrix in the jump condition
\eqref{eq:reducedjump} for the ``reduced'' Riemann-Hilbert problem has
a factorization:
\begin{equation}
\begin{bmatrix}1-z|\rho(z)|^2 & -z\rho(z)\\
\rho(z)^* & 1\end{bmatrix}=
\begin{bmatrix} 1 & -z\rho(z)\\ 0 & 1\end{bmatrix}
\begin{bmatrix} 1 & 0 \\ \rho(z)^* & 1\end{bmatrix}\,,\quad\quad
z\in\mathbb{R}\,.
\label{eq:factorize}
\end{equation}
The function $\rho(z^*)^*$, with $\rho(z)$ defined by
\eqref{eq:rhospecial} and meromorphically continued to the whole
complex $z$-plane, has two types of poles in the upper half $z$-plane:
the eigenvalues $z_n=-k_n^2$ with $k_n$ satisfying
\eqref{eigenvalueformula}, and the phantom poles $z_m=-k_m^2$ with
$k_m$ satisfying \eqref{eq:phantom}.  If the curves supporting
the eigenvalue poles and the phantom poles are disjoint, 
then the first and second matrix factors on the right-hand side of 
\eqref{eq:factorize} may be deformed into the lower and upper half
$z$-plane respectively in such a way as to remove the eigenvalue poles
from the Riemann-Hilbert problem without introducing any new
phantom poles.  Unlike in the Zakharov-Shabat problem with
hypergeometric potential studied by Tovbis and Venakides \cite{TV},
there are parameter values in the MNLS case for which the curves
supporting the eigenvalue poles and phantom poles intersect.  Note
that if $\mu>0$ or $\mu<-\alpha^2\nu^2/2$, the interaction of
eigenvalue poles and phantom poles is prevented because the hyperbola
\eqref{eq:phantomhyperbola} supporting the phantom poles corresponds
to a vertical line in the $w$-plane that lies to the right or left,
respectively, of the strip $S$.  Phantom poles may only interfere with
the eigenvalue poles if $-\alpha^2\nu^2/2\le\mu\le 0$.

\begin{figure}[h]
\begin{center}
\includegraphics{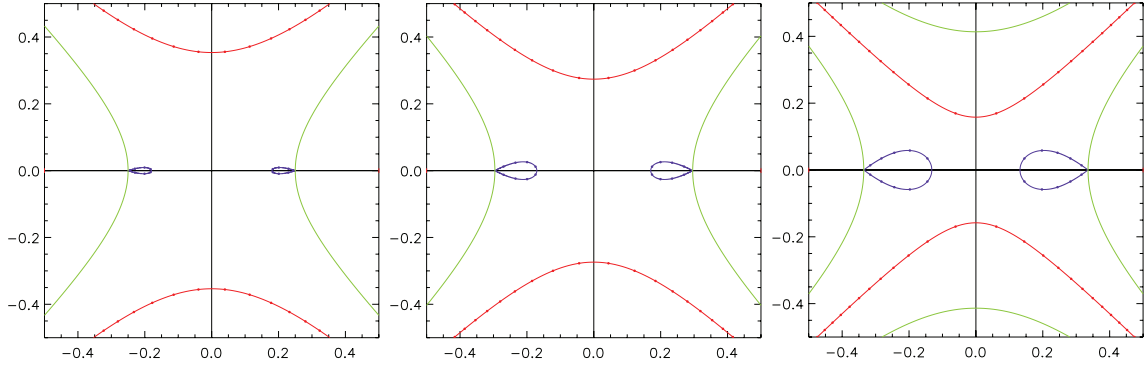}
\end{center}
\caption{\em The eigenvalues in the complex $k$-plane are shown with
  purple dots on the curve of the same color that supports them for
  all $\e>0$.  The green hyperbolae are the images of the left and
  right sides of the strip $S$ in the $w$-plane.  The red hyperbola is
  the curve that supports phantom poles shown as red dots for
  $\mu=-0.5$ (that supporting phantom poles for $\mu=0.5$ is not shown
  but is necessarily outside the region between the bounding green
  hyperbolae).  Left: $\alpha=1.5$.  Center: $\alpha=1.3$.  Right:
  $\alpha=1.1$.  In this figure and all that follow in this section,
  $\nu=0.6846$, $\delta=0.5$, $|\mu|=0.5$, and $\e=0.075$.}
\label{fig:alphaseq1}
\end{figure}

To illustrate these phenomena, we fix the values $\e=0.075$,
$\nu=0.6846$, $\delta=0.5$, and $|\mu|=0.5$ (the discrete spectrum
only depends on $|\mu|$ while the phantom poles are sensitive to the
sign of $\mu$ as well), and allow $\alpha$ to vary.  According to the
criterion established above, eigenvalues will exist for sufficiently
small $\e>0$ only for $0\le\alpha<(1-\mu^2/(4\nu^2))/\delta\approx
1.733$.  In Figure~\ref{fig:alphaseq1} we show the emergence of
eigenvalues as $\alpha$ is decreased below the threshold value of
$\alpha\approx 1.733$.

\begin{figure}[h]
\begin{center}
\includegraphics{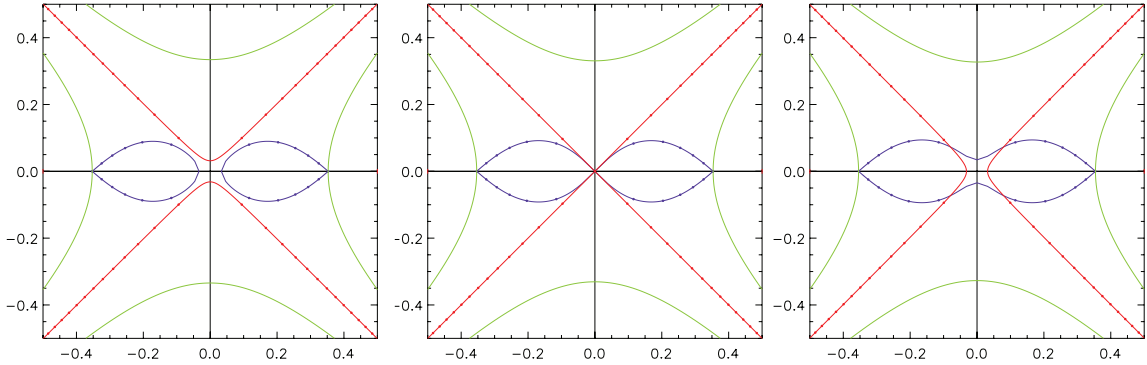}
\end{center}
\caption{\em The eigenvalues and phantom poles in the complex
  $k$-plane showing onset of interaction between eigenvalues and
  phantom poles.  Left: $\alpha=1.004$.  Center: $\alpha=1$.
  Right: $\alpha=0.996$.}
\label{fig:alphaseq2}
\end{figure}

Note that when the hyperbola \eqref{eq:phantomhyperbola} that supports
the phantom poles passes through the origin in the $k$-plane, that is,
when $\alpha=1/(\delta-\mu)$, we also have $Q(-(1-\alpha\delta)/4)=0$,
which implies that the curve supporting the eigenvalues also passes
through the origin in the $k$-plane.  This situation leads to the
interaction of the eigenvalues with the phantom poles when $\mu=-0.5$,
as shown in Figure~\ref{fig:alphaseq2}.

\begin{figure}[h]
\begin{center}
\includegraphics{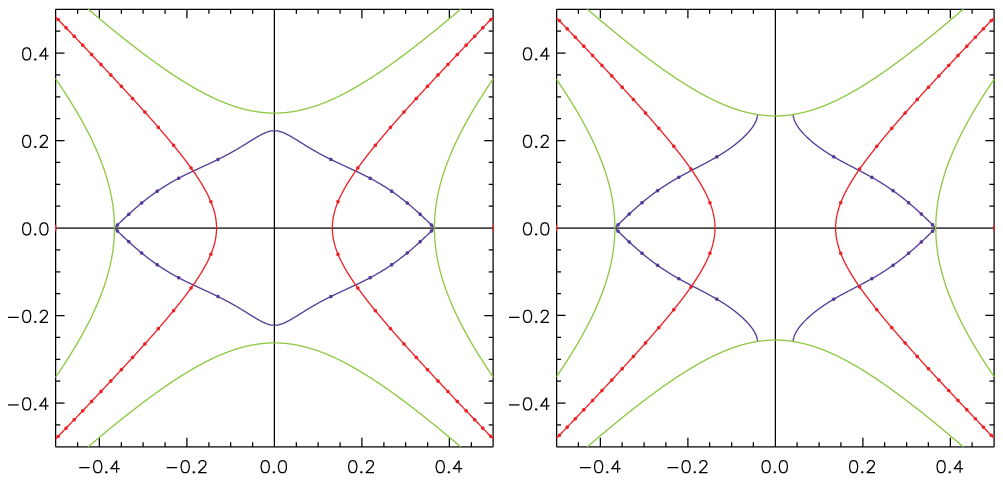}
\end{center}
\caption{\em The eigenvalues and phantom poles in the complex
  $k$-plane showing the appearance of a spectral gap.  Left:
  $\alpha=0.930$.  Right: $\alpha=0.924$.}
\label{fig:alphaseq3}
\end{figure}

\begin{figure}[h]
\begin{center}
\includegraphics{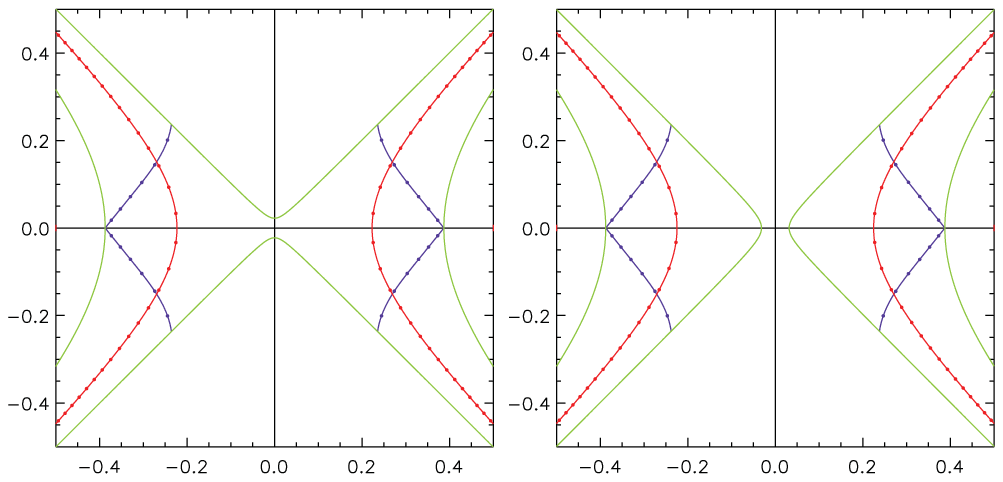}
\end{center}
\caption{\em The eigenvalues and phantom poles in the complex
  $k$-plane showing a transition in one of the bounding hyperbolae.  Left:
  $\alpha=0.801$.  Right: $\alpha=0.798$.}
\label{fig:alphaseq4}
\end{figure}

Further decrease of $\alpha$ causes a new phenomenon:  the separation
of the curve supporting the eigenvalues from the imaginary axis.  
Figure~\ref{fig:alphaseq3} shows the onset of a ``spectral gap'' between
the discrete spectrum and the continuous spectrum on the imaginary axis
occurring at approximately $\alpha=0.927$.
The parameter $\alpha$ is decreased further in
Figure~\ref{fig:alphaseq4}, 
where we see that the passage through the origin of the hyperbola
corresponding to the left-hand side of the strip $S$ in the $w$-plane
(this occurs exactly when
$\alpha=-(\delta\pm\sqrt{\delta^2+8\nu^2})/(4\nu^2)$) leads to only a
small quantitative difference in the eigenvalues and phantom poles
and the curves that support them.

\begin{figure}[h]
\begin{center}
\includegraphics{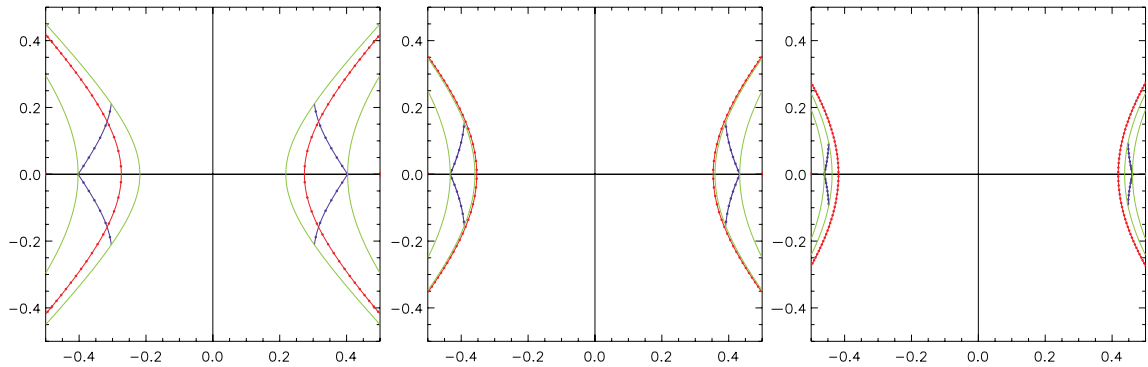}
\end{center}
\caption{\em The eigenvalues and phantom poles in the complex
  $k$-plane showing the phantom poles moving away from the
  eigenvalues as $\alpha$ is decreased.  Left: $\alpha=0.7$.  Center:
  $\alpha=0.5$.  Right: $\alpha=0.3$.}
\label{fig:alphaseq5}
\end{figure}

Figure~\ref{fig:alphaseq5} 
shows the final essential transition that occurs as $\alpha$ is
decreased to zero: the departure of the hyperbola supporting the
phantom poles for $\mu=-0.5$ from the region containing the eigenvalue
curve.  This transition occurs exactly when $\alpha=-\mu/(2\nu^2)$,
which is approximately $\alpha=0.533$ for $\mu=-0.5$ and $\nu=0.6846$.

\begin{figure}[h]
\begin{center}
\includegraphics{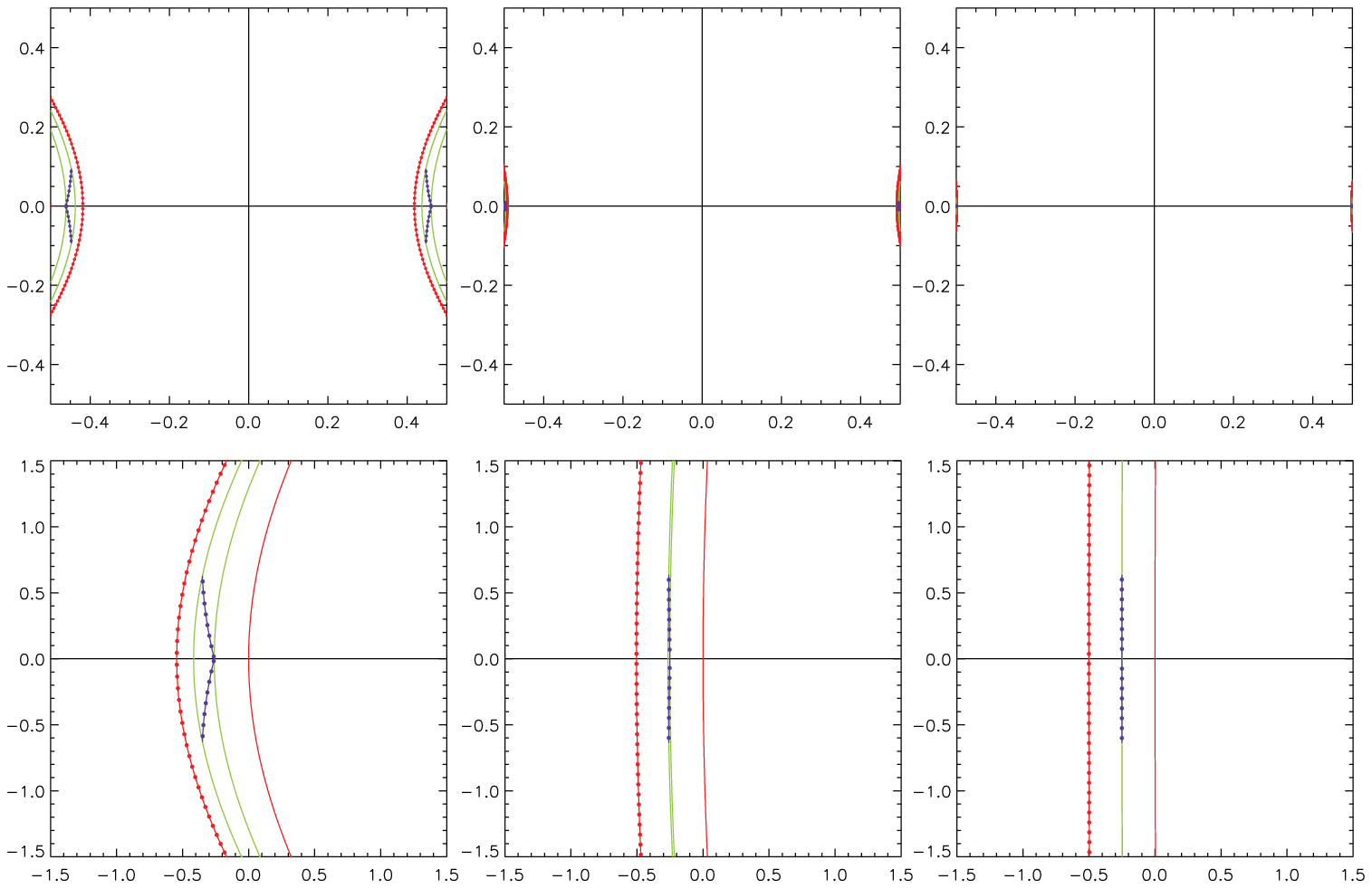}
\end{center}
\caption{\em The eigenvalues and phantom poles in the complex
  $k$-plane (top) and $\lambda$-plane (bottom).  The red hyperbolae
  contain the phantom poles for $\mu=-0.5$ (shown with phantom pole
  locations superimposed) and for $\mu=0.5$ (shown without
  superimposed phantom poles).  Left: $\alpha=0.3$.  Center:
  $\alpha=0.03$.  Right: $\alpha=0.003$.}
\label{fig:alphaseq6}
\end{figure}

When $\alpha$ becomes very small, one expects some correspondence
between our results and those of Tovbis and Venakides \cite{TV} for
the nonselfadjoint Zakharov-Shabat eigenvalue problem which is the
formal limit of the MNLS spectral problem when viewed in the correct
variable $\lambda$, which is defined in terms of $k$ and $\alpha$ by
\eqref{eq:klambda}.  Figure~\ref{fig:alphaseq6} 
shows the discrete spectrum and phantom poles as $\alpha$ tends to
zero, both in the complex $k$-plane and the complex $\lambda$-plane.
In this limit, the discrete spectrum all accumulates near $k=\pm 1/2$,
and the transformation \eqref{eq:klambda} blows up the region near
$k=1/2$ to reveal a nontrivial limiting structure.  This limiting
structure is in agreement with the results of Tovbis and Venakides
\cite{TV} in that the limiting discrete spectrum consists of
equally-spaced points on the vertical line
$\text{Re}\{\lambda\}=-\delta/2$ with the phantom poles lying on a
distinct vertical line: $\text{Re}\{\lambda\}=-\delta/2+\mu/2$.

\section{Acknowledgements}
This paper is dedicated to the memory of Alwyn C. Scott, a great
teacher and mentor, and above all a true pioneer of nonlinear science.
Both authors would like to thank Jared Bronski, J. Nathan Kutz,
Gregory Lyng, Ken McLaughlin, and Arthur Vartanian for stimulating
conversations on these topics, and we would like to especially
acknowledge E. V. Doctorov, who introduced one of us to this problem,
and Percy Deift, who kindly showed us his notes \cite{Percynotes} of
his calculation with Stephanos Venakides and Xin Zhou of the ``shadow
bound'' for the Zakharov-Shabat problem.  J. DiFranco and P. D. Miller
were both partially supported by a Focused Research Group grant from
the National Science Foundation, DMS-0354373 during the preparation of
this manuscript.

\newpage

\section*{Appendix: Scattering and Inverse Scattering for the MNLS 
Equation}
\subsection*{Lax pair.  Zero-curvature representation.}
Let $k\in\mathbb{C}$ be a parameter and define
\begin{equation}
\Lambda:=-\frac{2i}{\alpha}\left(k^2-\frac{1}{4}\right)\,.
\end{equation}
Also given a complex-valued function $\phi$ of
$(x,t)\in\mathbb{R}^2$, let
\begin{equation}
{\bf L}=\begin{bmatrix}\Lambda & 2ik\phi\\
2ik\phi^* & -\Lambda\end{bmatrix}
\end{equation}
and
\begin{equation}
{\bf B}=\begin{bmatrix} i\Lambda^2+2ik^2|\phi|^2 & -2k\Lambda\phi-k\e\phi_x
-2i\alpha k|\phi|^2\phi\\
-2k\Lambda\phi^*+k\e\phi_x^*-2i\alpha k|\phi|^2\phi^* &
-i\Lambda^2-2ik^2|\phi|^2\end{bmatrix}\,.
\end{equation}
The simultaneous linear equations 
\begin{equation}
\e\frac{\partial{\bf v}}{\partial x}={\bf L}{\bf v}
\label{eq:Laxx}
\end{equation}
and
\begin{equation}
\e\frac{\partial{\bf v}}{\partial t}={\bf B}{\bf v}
\label{eq:Laxt}
\end{equation}
for a two-component vector unknown ${\bf v}(x,t;k)$ are said to
comprise the \emph{Lax pair} for the MNLS equation \eqref{MNLScauchy}.
For a general function $\phi(x,t)$ the vector ${\bf v}(x,t;k)$ is
overdetermined by the two equations \eqref{eq:Laxx} and
\eqref{eq:Laxt}.  The MNLS equation for $\phi(x,t)$ arises as the
compatibility condition for the two equations of the Lax pair.
Indeed, by cross-differentiation one sees that the \emph{zero curvature
condition}
\begin{equation}
\e\frac{\partial{\bf L}}{\partial t}-\e\frac{\partial{\bf B}}{\partial x} +
[{\bf L},{\bf B}]={\bf 0}
\label{eq:ZCC}
\end{equation}
on the matrices ${\bf L}$ and ${\bf B}$ is required for there to exist
a full basis of simultaneous solutions of \eqref{eq:Laxx} and
\eqref{eq:Laxt}.  Furthermore, the matrix equation \eqref{eq:ZCC} is
equivalent to the MNLS equation governing $\phi(x,t)$ as long as
$k\neq 0$ (the condition \eqref{eq:ZCC} holds trivially if $k=0$).

While one might think of the MNLS equation \eqref{MNLScauchy} as a
perturbation of the focusing NLS equation when $\alpha$ is small, the
singular dependence on $\alpha$ in the spectral problem
\eqref{eq:Laxx} suggests that at the level of inverse scattering we
are considering a very singular perturbation.  The spectral problem
\eqref{eq:Laxx} can be made to look more like the Zakharov-Shabat
eigenvalue problem if one introduces the alternate spectral variable $\lambda$
defined by the relation
\begin{equation}
\lambda=\frac{1}{\alpha}(2k-1)\quad\quad\text{or, equivalently,}\quad\quad
k=\frac{1}{2}(\alpha\lambda+1)\,.
\label{eq:klambda}
\end{equation}
These relations are clearly singular as $\alpha\rightarrow 0$.
Substituting for $k$ in \eqref{eq:Laxx} we may nonetheless hold
$\lambda$ fixed and formally pass to the limit of $\alpha\to 0$.  The
coefficient matrix thus becomes that of the nonselfadjoint
Zakharov-Shabat problem \cite{ZS72}:
\begin{equation}
\mathop{\lim_{\alpha\to 0}}_{\text{$\lambda$ fixed}}{\bf L}=
\begin{bmatrix}-i\lambda & i\phi\\i\phi^* & i\lambda\end{bmatrix}\,.
\end{equation}
This relation does not help us to deal with the inverse scattering
transform for the MNLS equation, but it does help us to compare some
results with known results for the Zakharov-Shabat scattering problem,
by looking at what happens to the spectrum of \eqref{eq:Laxx} for
small $\alpha$ in the $\lambda$-plane rather than the $k$-plane.

\subsection*{Jost solutions and the scattering matrix.}
Until further notice, we consider $t$ to be fixed and suppress all
explicit dependence on $t$.  Assume that $\phi$ is a complex-valued
function of $x$ that satisfies rapidly decreasing boundary conditions
as $x\rightarrow\pm\infty$.  Then as $x\rightarrow\pm\infty$ the
differential equation \eqref{eq:Laxx} has purely oscillatory (rather
than exponentially growing and decaying) solutions if and only if
$k^2\in\mathbb{R}$, and thus the real and imaginary axes of the
complex $k$-plane form the analogue\footnote{It is just an analogue
  because the system is not of the form $\mathcal{L}u=ku$ for any
  linear operator $\mathcal{L}$; that is, the spectral parameter $k$
  enters in an essentially nonlinear fashion.} of continuous spectrum
for the differential equation \eqref{eq:Laxx}.

The Jost solutions of \eqref{eq:Laxx} are, for $k^2\in\mathbb{R}$, the
columns of two fundamental solution matrices ${\bf J}_\pm(x;k)$ for
this equation having simple asymptotics as $x\rightarrow\pm\infty$.
By definition for $k^2\in\mathbb{R}$ the matrices ${\bf J}_\pm(x;k)$
satisfy
\begin{equation}
\e\frac{\partial{\bf J}_\pm}{\partial x}={\bf L}{\bf J}_\pm\,,\quad\quad
\lim_{x\rightarrow\pm\infty}
{\bf J}_\pm(x;k)e^{-\Lambda x\sigma_3/\e}=\mathbb{I}\,.
\end{equation}
Since the trace of ${\bf L}$ is zero, the Wronskian of any two vector
solutions of \eqref{eq:Laxx} is independent of $x$, so by taking determinants
in the boundary conditions we see that 
\begin{equation}
\det({\bf J}_\pm(x;k))=1
\label{eq:Jdet}
\end{equation}
which shows that if the Jost solution matrices exist for a given $k$
with $k^2\in\mathbb{R}$ then they are both fundamental solution
matrices for \eqref{eq:Laxx}.  Hence they are necessarily related by a 
constant (independent of $x$) matrix of coefficients:
\begin{equation}
{\bf J}_+(x;k)={\bf J}_-(x;k){\bf S}(k)\,.
\label{eq:Smatrix}
\end{equation}
The matrix ${\bf S}(k)$ for $k^2\in\mathbb{R}$ is called the
\emph{scattering matrix} for \eqref{eq:Laxx} corresponding to the
coefficient function $\phi$.  From \eqref{eq:Jdet} and
\eqref{eq:Smatrix} it follows that
\begin{equation}
\begin{aligned}
S_{11}(k)&=\det\left({\bf j}_+^{(1)}(x;k),{\bf j}_-^{(2)}(x;k)\right)\,,\\
S_{21}(k)&=\det\left({\bf j}_-^{(1)}(x;k),{\bf j}_+^{(1)}(x;k)\right)\,,
\end{aligned}
\quad\quad
\begin{aligned}
S_{12}(k)&=\det\left({\bf j}_+^{(2)}(x;k),{\bf j}_-^{(2)}(x;k)\right)\,,\\
S_{22}(k)&=\det\left({\bf j}_-^{(1)}(x;k),{\bf j}_+^{(2)}(x;k)\right)\,,
\end{aligned}
\label{eq:Sdets}
\end{equation}
where ${\bf j}_\pm^{(1)}(x;k)$ and ${\bf j}_\pm^{(2)}(x;k)$ denote the
first and second columns respectively of the matrix ${\bf J}_\pm(x;k)$.

\subsection*{Neumann series representations.  Analyticity properties.}
The Jost solutions may be constructed from integral equations
equivalent to the differential equation and boundary conditions they
satisfy.  
We begin by introducing the matrices ${\bf Y}_\pm(x;k)$ related to
the Jost matrices ${\bf J}_\pm(x;k)$ by
\begin{equation}
{\bf Y}_\pm(x;k)=e^{-\Lambda x\sigma_3/\e}{\bf J}_\pm(x;k)\,.
\end{equation}
By direct calculation using the differential equation \eqref{eq:Laxx}
satisfied by ${\bf J}_\pm(x;k)$ we find that
\begin{equation}
\e\frac{\partial{\bf Y}_\pm}{\partial x} = \begin{bmatrix}
0 & 2ik\phi(x)e^{-2\Lambda x/\e}\\ 2ik\phi(x)^*e^{2\Lambda x/\e} & 0
\end{bmatrix}{\bf Y}_\pm\,.
\end{equation}
Also, with the use of the assumption that $k^2\in\mathbb{R}$, the boundary
conditions to be satisfied by ${\bf J}_\pm(x;k)$ are equivalent to the
requirement that
\begin{equation}
\lim_{x\rightarrow\pm\infty}{\bf Y}_\pm(x;k)=\mathbb{I}\,.
\end{equation}
We therefore seek ${\bf Y}_\pm(x;k)$ as solutions of the following
integral equations:
\begin{equation}
{\bf Y}_\pm(x;k)=\mathbb{I}+\frac{2ik}{\e}\int_{\pm\infty}^x
\begin{bmatrix}
0 & \phi(y)e^{-2\Lambda y/\e}\\ \phi(y)^*e^{2\Lambda y/\e} & 0
\end{bmatrix}{\bf Y}_\pm(y;k)\,dy\,.
\label{eq:matrixinteqn}
\end{equation}
Iterating \eqref{eq:matrixinteqn} once leads to uncoupled scalar
integral equations for the four matrix elements of ${\bf Y}_\pm(x;k)$.

In particular, the diagonal elements satisfy
\begin{equation}
\begin{split}
Y_{11\pm}(x;k)&=1+\int_{\pm\infty}^x K(x,z;k)Y_{11\pm}(z;k)\,dz\\
Y_{22\pm}(x;k)&=1+\int_{\pm\infty}^x K(x,z;k^*)^*Y_{22\pm}(z;k)\,dz\,,
\end{split}
\label{eq:Ydiaginteqns}
\end{equation}
where the kernel is
\begin{equation}
K(x,z;k):=-\frac{4k^2}{\e^2}\phi(z)^*\int_z^x\phi(y)e^{-2\Lambda(y-z)/\e}\,dy\,.
\label{eq:kernel}
\end{equation}
Note that the kernel is an even function of the complex variable $k$.
Assuming that $\phi\in L^1(\mathbb{R})$:
\begin{equation}
\|\phi\|_1:=\int_{-\infty}^{+\infty}|\phi(x)|\,dx<\infty\,,
\end{equation}
we easily get the estimate
\begin{equation}
|K(x,z;k)|\le\frac{4|k|^2}{\e^2}\|\phi\|_1\cdot |\phi(z)|\,,\quad\quad
\text{if $\text{sgn}(x-z)\cdot\text{Im}\{k^2\}\ge 0$\,.}
\label{eq:Kbound}
\end{equation}
Seeking to approximate the diagonal matrix elements by iteration
starting from the initial guess $Y_{jj\pm}(x;k)\equiv 1$ leads to solutions of
\eqref{eq:Ydiaginteqns} in the form of Neumann series with $m$th term
given by an $m$-fold integral:
\begin{equation}
N_{m\pm}(x;k):=\int_{\pm\infty}^xK(x,z_m;k)\int_{\pm\infty}^{z_m}K(z_m,z_{m-1};k)
\cdots\int_{\pm\infty}^{z_2}K(z_2,z_1;k)\,dz_1\cdots dz_m\,,
\end{equation}
(or, in the case of $Y_{22\pm}(x;k)$, $N_{m\pm}(x;k^*)^*$).  We also
define $N_{0\pm}(x;k)\equiv 1$.  It follows from \eqref{eq:Kbound} that
\begin{equation}
\begin{split}
|N_{m-}(x;k)|&\le \left[\frac{4|k|^2}{\e^2}\|\phi\|_1\right]^m\int_{-\infty}^x
|\phi(z_m)|\int_{-\infty}^{z_m}|\phi(z_{m-1})|\cdots\int_{-\infty}^{z_2}|\phi(z_1)|\,
dz_1\cdots dz_m \\&= \frac{1}{m!}\left[\frac{4|k|^2}{\e^2}\|\phi\|_1
\int_{-\infty}^x|\phi(z)|\,dz\right]^m\,,\quad\quad\text{if $\text{Im}\{k^2\}\ge 0$\,,}
\end{split}
\end{equation}
and likewise that
\begin{equation}
|N_{m+}(x;k)|\le 
\frac{1}{m!}\left[\frac{4|k|^2}{\e^2}\|\phi\|_1
\int_x^{+\infty}|\phi(z)|\,dz\right]^m\,,\quad\quad
\text{if $\text{Im}\{k^2\}\le 0$\,.}
\end{equation}
By comparison with the exponential series it follows that the Neumann
series
\begin{equation}
\begin{split}
Y_{11\pm}(x;k)&=\sum_{m=0}^\infty N_{m\pm}(x;k)\,,\quad\quad 
\pm\text{Im}\{k^2\}\le 0\,,\\
Y_{22\pm}(x;k)&=\sum_{m=0}^\infty N_{m\pm}(x;k^*)^*\,,\quad\quad
\pm\text{Im}\{k^2\}\ge 0\,,
\end{split}
\label{eq:NeumannSeriesDiagonal}
\end{equation}
all converge uniformly in compact subsets of the indicated regions of
the complex $k$-plane and furnish there the unique solutions of the
Volterra integral equations \eqref{eq:Ydiaginteqns}.  Since the kernel
is analytic and even in $k$ it then follows by uniform convergence
that for each fixed $x\in\mathbb{R}$ all four of these functions are
even in $k$ and continuous in the indicated regions, also being
analytic in the interior.  Moreover we have the uniform ($L^\infty$)
estimate
\begin{equation}
|Y_{jj\pm}(x;k)|\le \exp\left(\frac{4|k|^2}{\e^2}\|\phi\|_1^2\right)
\label{eq:Linfty}
\end{equation}
that is valid for $j=1,2$ whenever $k^2$ is in the indicated half-plane
of convergence.  Together with \eqref{eq:Kbound} and the assumption
that $\phi\in L^1(\mathbb{R})$ this estimate shows that 
\begin{equation}
\lim_{x\rightarrow\pm\infty}Y_{jj\pm}(x;k)=1
\label{eq:Decay1}
\end{equation}
holds for each fixed $k$ in the region of convergence.  Now let
$x\in\mathbb{R}$ be fixed and consider how $Y_{jj\pm}(x;k)$ behaves as
$k\rightarrow 0$ from each sector of convergence in the $k$-plane.
Clearly the $L^\infty(\mathbb{R})$ estimate \eqref{eq:Linfty} is
uniform for small $k$ as is the bound \eqref{eq:Kbound} of the kernel
$K$.  Moreover, since \eqref{eq:Kbound} also shows that $K\rightarrow
0$ as $k\rightarrow 0$ pointwise in $x$ and $z$, a dominated
convergence argument applied to the right-hand side of
\eqref{eq:Ydiaginteqns} shows that
\begin{equation}
\lim_{k\rightarrow 0}Y_{jj\pm}(x;k)=1
\label{eq:Yjjzero}
\end{equation}
where the limit is taken from any direction within one of the sectors
of convergence.  Finally, we consider how $Y_{jj\pm}(x;k)$ behaves as
$k\rightarrow\infty$ within each sector of convergence in the
$k$-plane.  To get the desired estimate, we have to note that the
apparent $k^2$ growth of the kernel $K$ should be compensated for by the
exponential behavior of the integral, so we could first integrate by 
parts in the definition \eqref{eq:kernel} to get
\begin{equation}
K(x,z;k)=\frac{i\alpha k^2}{\e (k^2-1/4)}\phi(z)^*\left[\phi(x)e^{-2\Lambda (x-z)/\e}-\phi(z)-\int_z^x\frac{\partial\phi}{\partial y}(y)e^{-2\Lambda(y-z)/\e}\,dy
\right]\,.
\end{equation}
The fraction outside the brackets is now uniformly bounded as
$k\rightarrow\infty$, and assuming that $\partial\phi(x)/\partial x$ is in
$L^1(\mathbb{R})$ (which implies that
\begin{equation}
\|\phi\|_\infty:=\sup_{x\in\mathbb{R}}|\phi(x)|
\end{equation}
is finite, so $\phi\in L^\infty(\mathbb{R})$ as well), 
then we may replace the estimate
\eqref{eq:Kbound} by
\begin{equation}
|K(x,z;k)|\le\frac{\alpha}{\e}\frac{|k|^2}{|k^2-1/4|}\left(2\|\phi\|_\infty+
\|\phi'\|_1\right)
\cdot|\phi(z)|\,,\quad\quad\text{if $\text{sgn}(x-z)\cdot\text{Im}\{k^2\}\ge 0$}
\,.
\label{eq:Kbound2}
\end{equation}
This bound is uniform in $k$ large.  Furthermore, it now follows from
dominated convergence that
\begin{equation}
\lim_{k\rightarrow\infty}K(x,z;k)=-i\frac{\alpha}{\epsilon}|\phi(z)|^2
\end{equation}
with the limit being taken in a strict subsector of a sector for which
\eqref{eq:Kbound2} holds.  Assuming therefore that $\phi\in
L^1(\mathbb{R})$ and that $\phi'\in L^1(\mathbb{R})$, we have by
dominated convergence that
\begin{equation}
\lim_{k\rightarrow\infty}N_{m\pm}(x;k)=\frac{1}{m!}\left[-i\frac{\alpha}{\e}\int_{\pm\infty}^x|\phi(z)|^2\,dz\right]^m\,,
\end{equation}
with the limit being taken in the appropriate subsector (note that the
presumed conditions on $\phi$ also imply that $\phi\in
L^2(\mathbb{R})$ so the right-hand side is finite).  It further
follows by dominated convergence applied to the infinite Neumann
series \eqref{eq:NeumannSeriesDiagonal} that
\begin{equation}
\begin{split}
\lim_{k\rightarrow\infty}Y_{11+}(x;k)&=\exp\left(i\frac{\alpha}{\e}\int_x^{+\infty}
|\phi(z)|^2\,dz\right)\,,\\
\lim_{k\rightarrow\infty}Y_{11-}(x;k)&=\exp\left(-i\frac{\alpha}{\e}\int_{-\infty}^x
|\phi(z)|^2\,dz\right)\,,\\
\lim_{k\rightarrow\infty}Y_{22+}(x;k)&=\exp\left(-i\frac{\alpha}{\e}\int_x^{+\infty}
|\phi(z)|^2\,dz\right)\,,\\
\lim_{k\rightarrow\infty}Y_{22-}(x;k)&=\exp\left(i\frac{\alpha}{\e}\int_{-\infty}^x
|\phi(z)|^2\,dz\right)\,,
\end{split}
\label{eq:Ydiagasymp}
\end{equation}
with the limit being taken in an arbitrary strict subsector of the
sector of existence in each case.

Similarly, by iterating \eqref{eq:matrixinteqn} once we find that the functions
\begin{equation}
\tilde{Y}_{12\pm}(x;k):=e^{2\Lambda x/\e}Y_{12\pm}(x;k)\quad\quad\text{and}
\quad\quad
\tilde{Y}_{21\pm}(x;k):=-e^{-2\Lambda x/\e}Y_{21\pm}(x;k)
\end{equation}
satisfy the integral equations
\begin{equation}
\begin{split}
\tilde{Y}_{12\pm}(x;k)&=f_\pm(x;k)+\int_{\pm\infty}^x \tilde{K}(x,z;k)\tilde{Y}_{12\pm}(z;k)\,dz\\
\tilde{Y}_{21\pm}(x;k)&=f_\pm(x;k^*)^*+\int_{\pm\infty}^x \tilde{K}(x,z;k^*)^*\tilde{Y}_{21\pm}(z;k)\,dz\,,
\end{split}
\label{eq:tildeYinteqns}
\end{equation}
where the modified kernel is 
\begin{equation}
\tilde{K}(x,z;k):=K(x,z;k)e^{-2\Lambda(z-x)/\e}=
-\frac{4k^2}{\e^2}\phi(z)^*\int_z^x\phi(y)e^{-2\Lambda(y-x)/\e}\,dy\,,
\end{equation}
and where
\begin{equation}
f_\pm(x;k):=\frac{2ik}{\e}\int_{\pm\infty}^x\phi(y) e^{-2\Lambda (y-x)/\e}\,dy \,.
\end{equation}
As before, the assumption that $\phi\in L^1(\mathbb{R})$ gives 
\begin{equation}
|\tilde{K}(x,z;k)|\le\frac{4|k|^2}{\e^2}\|\phi\|_1\cdot |\phi(z)|\,,\quad\quad
\text{if $\text{sgn}(x-z)\cdot\text{Im}\{k^2\}\le 0$\,,}
\label{eq:Ktildebound}
\end{equation}
and we also have
\begin{equation}
|f_\pm(x;k)|\le\frac{2|k|}{\e}\|\phi\|_1\,,\quad\quad
\text{if $\pm\text{Im}\{k^2\}\ge 0$\,,}
\label{eq:fbound}
\end{equation}
estimates that are uniformly valid for small $k$.
The further assumption that $\phi'\in L^1(\mathbb{R})$ gives 
\begin{equation}
|\tilde{K}(x,z;k)|\le\frac{\alpha}{\e}\frac{|k|^2}{|k^2-1/4|}
\left(2\|\phi\|_\infty+
\|\phi'\|_1\right)
\cdot|\phi(z)|\,,\quad\quad\text{if $\text{sgn}(x-z)\cdot\text{Im}\{k^2\}\le 0$}
\,,
\label{eq:Ktildebound2}
\end{equation}
and 
\begin{equation}
|f_\pm(x;k)|\le\frac{\alpha}{2}\frac{|k|}{|k^2-1/4|}(\|\phi\|_\infty+\|\phi'\|_1)\,,\quad\quad
\text{if $\pm\text{Im}\{k^2\}\ge 0$\,,}
\label{eq:fbound2}
\end{equation}
estimates that are uniformly valid for large $k$.  Using \eqref{eq:Ktildebound}
and \eqref{eq:fbound} we see that the following Neumann series representations
converge uniformly for $k$ in compact subsets of the indicated sectors:
\begin{equation}
\begin{split}
\tilde{Y}_{12\pm}(x;k)&=\sum_{m=0}^\infty \tilde{N}_{m\pm}(x;k)\,,\quad\quad
\pm\text{Im}\{k^2\} \ge 0\,,\\
\tilde{Y}_{21\pm}(x;k)&=\sum_{m=0}^\infty \tilde{N}_{m\pm}(x;k^*)^*\,,\quad\quad
\pm\text{Im}\{k^2\} \le 0\,,
\end{split}
\end{equation}
where 
\begin{equation}
\tilde{N}_{m\pm}(x;k):=
\int_{\pm\infty}^x\tilde{K}(x,z_m;k)\int_{\pm\infty}^{z_m}\tilde{K}(z_m,z_{m-1};k)
\cdots\int_{\pm\infty}^{z_2}\tilde{K}(z_2,z_1;k)f_\pm(z_1;k)\,dz_1\cdots dz_m\,.
\end{equation}
The functions $\tilde{N}_{m\pm}(x;k)$ are all odd functions of $k$.
Thus, $\tilde{Y}_{12+}(x;k)$ and $\tilde{Y}_{21-}(x;k)$ are odd and
analytic for $\text{Im}\{k^2\}>0$ and continuous for $\text{Im}\{k^2\}\ge
0$.  Similarly, $\tilde{Y}_{12-}(x;k)$ and $\tilde{Y}_{21+}(x;k)$ are
odd and analytic for $\text{Im}\{k^2\}<0$ and continuous for $\text{Im}\{k^2\}\le
0$.  If $g(x;k)$ denotes any of these four functions in their sectors
of continuity, we have the uniform estimate
\begin{equation}
|g(x;k)|
\le \frac{2|k|}{\e}\|\phi\|_1\cdot\exp\left(\frac{4|k|^2}{\e}\|\phi\|^2_1\right)\,.
\end{equation}
A dominated convergence argument applied to \eqref{eq:tildeYinteqns} with
the help of this estimate then shows that for all $k$ in the sectors of
existence,
\begin{equation}
\lim_{x\rightarrow\pm\infty}\tilde{Y}_{12\pm}(x;k)=0\quad\quad\text{and}\quad\quad
\lim_{x\rightarrow\pm\infty}\tilde{Y}_{21\pm}(x;k)=0\,,
\label{eq:Decay2}
\end{equation}
and similarly that
\begin{equation}
\lim_{k\rightarrow 0}\tilde{Y}_{12\pm}(x;k)=0\quad\quad\text{and}\quad\quad
\lim_{k\rightarrow 0}\tilde{Y}_{21\pm}(x;k)=0
\label{eq:Ytildezero}
\end{equation}
with the limits being taken from within the closed sectors of
continuity.  Also, using \eqref{eq:Ktildebound2} and
\eqref{eq:fbound2} we see that $kg(x;k)$ is bounded as
$k\rightarrow\infty$ within the closed sector of continuity, and by dominated
convergence,
\begin{equation}
\lim_{k\rightarrow\infty}k\tilde{N}_{m\pm}(x;k)=\frac{\alpha}{2}
\phi(x)\frac{1}{m!}\left[i
\frac{\alpha}{\e}\int_{\pm\infty}^x|\phi(z)|^2\,dz\right]^m\,,
\end{equation}
with the limit being taken in any strictly smaller subsector of
$\pm\text{Im}\{k^2\}>0$.  It follows that
\begin{equation}
\begin{split}
\lim_{k\rightarrow\infty}k\tilde{Y}_{12+}(x;k)&
=\frac{\alpha}{2}\phi(x)\exp\left(-i\frac{\alpha}{\e}
\int_x^{+\infty}|\phi(z)|^2\,dz\right)\,,\\
\lim_{k\rightarrow\infty}k\tilde{Y}_{12-}(x;k)&
=\frac{\alpha}{2}\phi(x)\exp\left(i\frac{\alpha}{\e}
\int_{-\infty}^x|\phi(z)|^2\,dz\right)\,,\\
\lim_{k\rightarrow\infty}k\tilde{Y}_{21+}(x;k)&
=\frac{\alpha}{2}\phi(x)^*\exp\left(i\frac{\alpha}{\e}
\int_x^{+\infty}|\phi(z)|^2\,dz\right)\,,\\
\lim_{k\rightarrow\infty}k\tilde{Y}_{21-}(x;k)&
=\frac{\alpha}{2}\phi(x)^*\exp\left(-i\frac{\alpha}{\e}
\int_{-\infty}^x|\phi(z)|^2\,dz\right)\,,
\end{split}
\label{eq:Yoffdiagasymp}
\end{equation}
again with the limit being taken from within a strict subsector of the
sector of convergence.

From this analysis we see that when $\phi$ and $\phi'$ are in $L^1(\mathbb{R})$,
the Jost solutions 
\begin{equation}
{\bf j}_+^{(1)}(x;k)=\begin{bmatrix}Y_{11+}(x;k)\\-\tilde{Y}_{21+}(x;k)\end{bmatrix}e^{\Lambda x/\e}\quad\quad\text{and}\quad\quad
{\bf j}_-^{(2)}(x;k)=\begin{bmatrix}\tilde{Y}_{12-}(x;k)\\Y_{22-}(x;k)\end{bmatrix}e^{-\Lambda x/\e}
\end{equation}
are analytic for $\text{Im}\{k^2\}< 0$ and continuous for
$\text{Im}\{k^2\}\le 0$.  For fixed $k$ with $\text{Im}\{k^2\}<0$ they
decay exponentially to zero as $x\rightarrow +\infty$ and
$x\rightarrow -\infty$ respectively.  Similarly, the Jost solutions
\begin{equation}
{\bf j}_-^{(1)}(x;k)=\begin{bmatrix}Y_{11-}(x;k)\\-\tilde{Y}_{21-}(x;k)\end{bmatrix}e^{\Lambda x/\e}\quad\quad\text{and}\quad\quad
{\bf j}_+^{(2)}(x;k)=\begin{bmatrix}\tilde{Y}_{12+}(x;k)\\Y_{22+}(x;k)\end{bmatrix}e^{-\Lambda x/\e}
\end{equation}
are analytic for $\text{Im}\{k^2\}> 0$ and continuous for
$\text{Im}\{k^2\}\ge 0$.  For fixed $k$ with $\text{Im}\{k^2\}>0$ they
decay exponentially to zero as $x\rightarrow -\infty$ and
$x\rightarrow +\infty$ respectively.  From \eqref{eq:Sdets} it then
follows that $S_{11}(k)$ is analytic for $\text{Im}\{k^2\}<0$ and
continuous for $\text{Im}\{k^2\}\le 0$ while $S_{22}(k)$ is analytic for
$\text{Im}\{k^2\}>0$ and continuous for $\text{Im}\{k^2\}\ge 0$.  
From \eqref{eq:Sdets} we learn in addition that
\begin{equation}
\lim_{k\rightarrow 0}S_{11}(k)=1\quad\quad\text{and}\quad\quad
\lim_{k\rightarrow 0}S_{22}(k)=1
\label{eq:Sjjzero}
\end{equation}
with the limit being taken from within a closed sector of analyticity
in each case.  Also
\begin{equation}
\mathop{\lim_{k\rightarrow 0}}_{\text{Im}\{k^2\}=0}S_{12}(k)=0\quad\quad
\text{and}\quad\quad
\mathop{\lim_{k\rightarrow 0}}_{\text{Im}\{k^2\}=0}S_{21}(k)=0\,.
\label{eq:Soffdiagzero}
\end{equation}
Furthermore, all elements of ${\bf S}(k)$ are bounded for $k$ large,
and we also have
\begin{equation}
\lim_{k\rightarrow\infty}S_{11}(k)=\exp\left(i\frac{\alpha}{\e}\|\phi\|_2^2
\right)\quad\quad\text{and}\quad\quad
\lim_{k\rightarrow\infty}S_{22}(k)=\exp\left(-i\frac{\alpha}{\e}\|\phi\|_2^2
\right)
\label{eq:Sasymp}
\end{equation}
where 
\begin{equation}
\|\phi\|_2^2:=\int_{-\infty}^{+\infty}|\phi(x)|^2\,dx<\infty
\end{equation}
and the limit is taken from within a strict subsector of the sector
of analyticity in each case.

Let $k_j$ be a zero of $S_{11}(k)$ with $\text{Im}\{k_j^2\}\le 0$.
Since $S_{11}(k)$ is a Wronskian, it follows that the Jost solutions
${\bf j}_+^{(1)}(x;k_j)$ and ${\bf j}_-^{(2)}(x;k_j)$ are
proportional, with a constant of proportionality that we denote by
$\gamma_j$:
\begin{equation}
{\bf j}_+^{(1)}(x;k_j)=\gamma_j{\bf j}_-^{(2)}(x;k_j)\,,\quad\quad\forall x\in
\mathbb{R}\,.
\label{eq:proportionality}
\end{equation}
Moreover, if we have the strict inequality $\text{Im}\{k_j^2\}<0$ then
the left and right hand sides of \eqref{eq:proportionality} both
represent solutions that decay exponentially to zero as
$x\rightarrow\pm\infty$.  In other words, these values of $k$ are
precisely the analogues\footnote{See the previous footnote.} of
discrete spectrum for this problem.  We further assume that $\phi$ is such
that:
\begin{itemize}
\item $S_{11}(k)$ has a finite number of zeros.
\item $S_{11}(k)$ does not vanish for $\text{Im}\{k^2\}=0$.
\item All zeros of $S_{11}(k)$ are simple; that is, $S_{11}(k)=0$ implies that
$S_{11}'(k)\neq 0$.
\end{itemize}

Directly from the Neumann series representations, we observe the following
symmetries of the Jost solutions:
\begin{equation}
{\bf j}_\pm^{(1)}(x;k^*)^*=\begin{bmatrix}0 & 1\\-1 &0\end{bmatrix}
{\bf j}_\pm^{(2)}(x;k)\,,
\label{eq:Janti}
\end{equation}
an antiholomorphic symmetry, and
\begin{equation}
{\bf j}_\pm^{(1)}(x;-k)=\begin{bmatrix}1 & 0 \\ 0 & -1\end{bmatrix}
{\bf j}_\pm^{(1)}(x;k)\quad\quad\text{and}\quad\quad
{\bf j}_\pm^{(2)}(x;-k)=\begin{bmatrix}-1  &0\\0 & 1\end{bmatrix}
{\bf j}_\pm^{(2)}(x;k)\,,
\label{eq:Jholo}
\end{equation}
a holomorphic symmetry.  From the definitions \eqref{eq:Sdets} we then have
\begin{equation}
S_{22}(k)=S_{11}(k^*)^*\quad\quad\text{and}\quad\quad S_{21}(k)=-S_{12}(k^*)^*
\label{eq:Ssymanti}
\end{equation}
and
\begin{equation}
S_{11}(-k)=S_{11}(k)\,,\quad S_{22}(-k)=S_{22}(k)\,,\quad
S_{12}(-k)=-S_{12}(k)\,,\quad\text{and}\quad S_{21}(-k)=-S_{21}(k)\,.
\label{eq:Ssymholo}
\end{equation}
Consequently the discrete spectrum for this problem consists of
quartets of points of the form $(k,-k,k^*,-k^*)$, distinct because (by
assumption) $\text{Im}\{k^2\}\neq 0$.  The proportionality relation
\eqref{eq:proportionality} in conjunction with the symmetries of the
Jost solutions then implies also that
\begin{equation}
\begin{split}
{\bf j}_+^{(1)}(x;-k_j)&=-\gamma_j{\bf j}_-^{(2)}(x;-k_j)\,,\\
{\bf j}_+^{(2)}(x;k_j^*)&=-\gamma_j^*{\bf j}_-^{(1)}(x;k_j^*)\,,\\
{\bf j}_+^{(2)}(x;-k_j^*)&=\gamma_j^*{\bf j}_-^{(1)}(x;-k_j^*)\,.
\end{split}
\label{eq:proportionality2}
\end{equation}

\subsection*{Riemann-Hilbert problem of inverse scattering.}
Consider the matrix ${\bf M}(k;x)$ defined for $x\in\mathbb{R}$ in
terms of the Jost solutions as follows:
\begin{equation}
{\bf M}(k;x):=\begin{cases}\displaystyle
\begin{bmatrix}\displaystyle
\frac{e^{-\Lambda x/\e}}{S_{11}(k)}{\bf j}_+^{(1)}(x;k)\,, & 
\displaystyle e^{\Lambda x/\e}{\bf j}_-^{(2)}(x;k)\end{bmatrix}\,,\quad\quad 
& \text{Im}\{k^2\}<0\,,\\\\
\displaystyle
\begin{bmatrix}\displaystyle
e^{-\Lambda x/\e}{\bf j}_-^{(1)}(x;k)\,, & \displaystyle
\frac{e^{\Lambda x/\e}}{S_{22}(k)}
{\bf j}_+^{(2)}(x;k)\end{bmatrix}\,,\quad\quad &\text{Im}\{k^2\}>0\,.
\end{cases}
\label{eq:Mdefine}
\end{equation}
It follows from \eqref{eq:Sdets} that $\det({\bf M}(k;x))=1$ at every point in
its domain of definition (which implicitly excludes the discrete spectrum).
Let the (fourfold) discrete spectrum be denoted $D$.  
The following other properties of ${\bf M}(k;x)$ are fundamental:
\begin{itemize}
\item[]{\bf Analyticity:} ${\bf M}(k;x)$ is analytic for
  $\text{Im}\{k^2\}\neq 0$ and $k\not\in D$ and takes continuous
  boundary values on the axes $\text{Im}\{k^2\}=0$ from each of the four
  sectors of its analyticity.  Moreover, ${\bf M}(k;x)$ is uniformly
  bounded for large $k$.
\item[]{\bf Jump Condition:}  Let ${\bf M}_\pm(k;x)$ denote the boundary value
taken from the region where $\pm\text{Im}\{k^2\}<0$:
\begin{equation}
{\bf M}_\pm(k;x):=\mathop{\lim_{z\rightarrow k}}_{\pm\text{Im}\{z^2\}<0}
{\bf M}(z;x)\,,\quad\quad\text{Im}\{k^2\}=0\,.
\label{eq:Mboundaryvalues}
\end{equation}
Then the boundary values are related by the formula
\begin{equation}
{\bf M}_+(k;x)={\bf M}_-(k;x){\bf V}(k;x)\,,\quad\quad\text{Im}\{k^2\}=0\,,
\end{equation}
where
\begin{equation}
{\bf V}(k;x)=e^{\Lambda x\sigma_3/\e}{\bf V}_0(k)
e^{-\Lambda x\sigma_3/\e}\,,
\end{equation}
and
\begin{equation}
{\bf V}_0(k):=
\begin{bmatrix}
\displaystyle
1-\frac{S_{12}(k)S_{21}(k)}{S_{11}(k)S_{22}(k)} &\displaystyle
 -\frac{S_{12}(k)}{S_{22}(k)}\\
\displaystyle \frac{S_{21}(k)}{S_{11}(k)} & 1\end{bmatrix}\,.
\end{equation}
Note that upon taking into account the symmetries \eqref{eq:Ssymanti}
and \eqref{eq:Ssymholo} we see that in terms of the function
\begin{equation}
r(k):=-\frac{S_{12}(k)}{S_{22}(k)}
\label{eq:rdefine}
\end{equation}
we may write ${\bf V}_0(k)$ in the form
\begin{equation}
{\bf V}_0(k)=\begin{bmatrix}1\pm |r(k)|^2 & r(k)\\\pm r(k)^* & 1
\end{bmatrix}\,,\quad\quad \pm k^2>0\,.
\end{equation}
\item[]{\bf Singularities:}  The matrix ${\bf M}(k;x)$ has simple poles at
the points of the finite set $D$.  If $k_j\in D$ with $\text{Im}\{k_j\}>0$ and
$\text{Re}\{k_j\}<0$, then 
\begin{equation}
\mathop{\text{Res}}_{k=k_j}{\bf M}(k;x)=\lim_{k\rightarrow k_j}
{\bf M}(k;x)\begin{bmatrix}0 & 0\\c_j(x) & 0\end{bmatrix}
\end{equation}
where 
\begin{equation}
c_j(x):=c_j^0 e^{-2\Lambda_jx/\e}\,,\quad\quad
c_j^0:=\frac{\gamma_j}{S_{11}'(k_j)}\,,\quad\quad
\Lambda_j:=\Lambda\Big|_{k=k_j}\,.
\end{equation}
Similarly, 
\begin{equation}
\mathop{\text{Res}}_{k=-k_j}{\bf M}(k;x)=\lim_{k\rightarrow -k_j}
{\bf M}(k;x)\begin{bmatrix}0 & 0\\c_j(x) & 0\end{bmatrix}\,,
\end{equation}
\begin{equation}
\mathop{\text{Res}}_{k=k_j^*}{\bf M}(k;x)=\lim_{k\rightarrow k_j^*}
{\bf M}(k;x)\begin{bmatrix}0 & -c_j(x)^*\\0 & 0\end{bmatrix}\,,
\end{equation}
and
\begin{equation}
\mathop{\text{Res}}_{k=-k_j^*}{\bf M}(k;x)=\lim_{k\rightarrow -k_j^*}
{\bf M}(k;x)\begin{bmatrix}0 & -c_j(x)^*\\0 & 0\end{bmatrix}\,.
\end{equation}
\item[]{\bf Normalization:}  The matrix ${\bf M}(k;x)$ is normalized in the
sense that
\begin{equation}
\lim_{k\rightarrow 0}{\bf M}(k;x)=\mathbb{I}\,,
\end{equation}
with the limit being taken in any direction.
\end{itemize}

Reversing the point of view by taking the function $r(k)$, the
fourfold symmetric set $D$, and the numbers $c_j^0$ (which taken
together form the set of \emph{scattering data} associated with
$\phi$) to be given, the above properties are said to constitute a
\emph{Riemann-Hilbert problem} for the unknown matrix ${\bf M}(k;x)$.
If the matrix ${\bf M}(k;x)$ can be determined from the scattering
data, then the function $\phi$ may also be recovered therefrom, since
from the asymptotics in \eqref{eq:Ydiagasymp},
\eqref{eq:Yoffdiagasymp}, and \eqref{eq:Sasymp} applied to the
definition \eqref{eq:Mdefine} we see that
\begin{multline}
{\bf M}(k;x)=\begin{bmatrix}1 + o(1) & \displaystyle
\frac{\alpha}{2k}\phi(x) + o(k^{-1})\\
\displaystyle -\frac{\alpha}{2k}\phi(x)^* + o(k^{-1}) & 1+o(1)\end{bmatrix}\\
{}\cdot\begin{bmatrix}
\displaystyle
\exp\left(-i\frac{\alpha}{\e}\int_{-\infty}^x|\phi(z)|^2\,dz\right) & 0\\
0 & 
\displaystyle\exp\left(i\frac{\alpha}{\e}\int_{-\infty}^x|\phi(z)|^2\,dz\right) 
\end{bmatrix}
\end{multline}
as $k\rightarrow\infty$ in any non-horizontal and non-vertical direction.
Consequently,  we have the reconstruction formula
\begin{equation}
\phi(x)=\lim_{k\rightarrow\infty} \frac{2k}{\alpha}
\frac{M_{12}(k;x)}{M_{22}(k;x)}\,.
\end{equation}
Note that this Riemann-Hilbert problem differs from others described
in the literature even for this specific problem (see, for example,
\cite{D}) in that the point of normalization is $k=0$, while the
potential $\phi$ is extracted via asymptotics of the solution for $k$
large.  Indeed, nowhere in the theory of matrix Riemann-Hilbert
problems is it essential that the point of normalization be any
specific value.  On the other hand, the fact that the potential is
extracted by asymptotics at $k=\infty$ is important because while
$\phi$ may also be obtained from the expansion near $k=0$, these
latter formulae involve derivatives of the elements of ${\bf M}(k;x)$
with respect to $x$, and controlling these derivatives in addition to
the matrix elements themselves in asymptotic problems adds an
unneeded layer of difficulty.
\subsection*{Time dependence of scattering data.}
Now we suppose that $\phi$ is evolving in time subject to the MNLS
equation \eqref{MNLScauchy}, and we seek to determine the way that the
scattering data corresponding to $\phi$ depend on $t$.  We need to
assume here that for all $t$ we have $\phi$ and $\phi_x$ in
$L^1(\mathbb{R})$ so that the notion of scattering data makes sense in
the way we have discussed above.  We will also assume $\phi_{xx}$ in
$L^1(\mathbb{R})$.

Since $\phi$ satisfies \eqref{MNLScauchy}, the zero-curvature
condition \eqref{eq:ZCC} is satisfied making the Lax pair
\eqref{eq:Laxx} and \eqref{eq:Laxt} compatible.  Therefore, there
exists a full basis of simultaneous solutions of \eqref{eq:Laxx} and
\eqref{eq:Laxt}.  As $\phi$ now depends on $t$ as well as $x$, we now
recall the explicit time dependence of the Jost matrices by writing
${\bf J}_\pm(x,t;k)$.  Although the columns of ${\bf J}_\pm(x,t;k)$
certainly solve \eqref{eq:Laxx} for each $t$ and although
\eqref{eq:Laxx} and \eqref{eq:Laxt} are compatible, the columns of
${\bf J}_\pm(x,t;k)$ do not satisfy \eqref{eq:Laxt} because they
satisfy time-independent boundary conditions as
$x\rightarrow\pm\infty$.  In other words, we can obtain simultaneous
solutions of \eqref{eq:Laxx} and \eqref{eq:Laxt} by multiplying the
columns of ${\bf J}_\pm(x,t;k)$ by scalar factors depending only on
$t$ and $k$ and then choosing these factors so that \eqref{eq:Laxt}
holds for the products.  To do choose the factors, it is enough to
examine the limiting form of the matrix ${\bf B}$ at $\pm\infty$
(${\bf B}$ has a limit as $x\rightarrow\pm\infty$ because
$\phi\rightarrow 0$ and $\phi_x\rightarrow 0$ in this limit by virtue
of the assumption that $\phi$, $\phi_x$, and $\phi_{xx}$ are all in
$L^1(\mathbb{R})$ for all $t$).  In this way it can be shown that
\begin{equation}
\epsilon\frac{\partial}{\partial t}
\left({\bf J}_\pm(x,t;k)e^{i\Lambda^2t\sigma_3/\e}\right)
={\bf B}\left({\bf J}_\pm(x,t;k)e^{i\Lambda^2t\sigma_3/\e}\right)\,.
\label{eq:Jtimedep}
\end{equation}
These equations hold columnwise throughout the closed sectors of
convergence in the complex $k$-plane for the corresponding Neumann
series.

In particular, both columns of \eqref{eq:Jtimedep} are meaningful for
$\text{Im}\{k^2\}=0$.  In this case we have the relation \eqref{eq:Smatrix}
which we rewrite here in the form
\begin{equation}
\left({\bf J}_+(x,t;k)e^{i\Lambda^2t\sigma_3/\e}\right)=
\left({\bf J}_-(x,t;k)e^{i\Lambda^2t\sigma_3/\e}\right)
\left(e^{-i\Lambda^2t\sigma_3/\e}{\bf S}(k;t)
e^{i\Lambda^2t\sigma_3/\e}\right)\,,
\end{equation}
where we have allowed time dependence in the scattering matrix ${\bf S}$.
Differentiating with respect to $t$ and applying \eqref{eq:Jtimedep}
gives
\begin{multline}
{\bf B}
\left({\bf J}_+(x,t;k)e^{i\Lambda^2t\sigma_3/\e}\right)=
{\bf B}\left({\bf J}_-(x,t;k)e^{i\Lambda^2t\sigma_3/\e}\right)
\left(e^{-i\Lambda^2t\sigma_3/\e}{\bf S}(k;t)
e^{i\Lambda^2t\sigma_3/\e}\right)\\
{}+\left({\bf J}_-(x,t;k)e^{i\Lambda^2t\sigma_3/\e}\right)\cdot
\e\frac{\partial}{\partial t}
\left(e^{-i\Lambda^2t\sigma_3/\e}{\bf S}(k;t)
e^{i\Lambda^2t\sigma_3/\e}\right)\,.
\end{multline}
Using \eqref{eq:Jtimedep} again and noting that ${\bf J}_-(x,t;k)$ and
$e^{i\Lambda^2t\sigma_3/\e}$ are invertible we see that
\begin{equation}
{\bf S}(k;t)=e^{i\Lambda^2t\sigma_3/\e}{\bf S}(k;0)e^{-i\Lambda^2t\sigma_3/\e}\,.
\label{eq:Sevolve}
\end{equation}
From this relation we learn the following facts:
\begin{itemize}
\item $S_{11}(k;t)$ and $S_{22}(k;t)$ are independent of $t$.  Therefore
so are the locations of the poles of ${\bf M}(k;x,t)$ in the complex $k$-plane,
as well as the derivatives $S_{11}'(k_j;t)$.
\item The function $r(k;t)$ satisfies
\begin{equation}
r(k;t)=r(k;0)e^{2i\Lambda^2 t/\e}\,.
\end{equation}
\end{itemize}

The proportionality constants $\gamma_j$ will also depend on $t$, and
to deduce the time dependence we first rewrite the defining relation
\eqref{eq:proportionality} in the form
\begin{equation}
\left(e^{i\Lambda_j^2 t/\e}{\bf j}_+^{(1)}(x,t;k_j)\right)=
\left(
e^{-i\Lambda_j^2 t/\e}{\bf j}_-^{(2)}(x,t;k_j)\right)
\left(e^{2i\Lambda_j^2 t/\e}\gamma_j\right)\,.
\end{equation}
Differentiating with respect to $t$ and applying \eqref{eq:Jtimedep} gives
\begin{multline}
{\bf B}\left(e^{i\Lambda_j^2 t/\e}{\bf j}_+^{(1)}(x,t;k_j)\right)=
{\bf B}\left(
e^{-i\Lambda_j^2 t/\e}{\bf j}_-^{(2)}(x,t;k_j)\right)
\left(e^{2i\Lambda_j^2 t/\e}\gamma_j\right)\\
{}+\left(
e^{-i\Lambda_j^2 t/\e}{\bf j}_-^{(2)}(x,t;k_j)\right)\cdot\e\frac{d}{dt}
\left(e^{2i\Lambda_j^2 t/\e}\gamma_j\right)\,.
\end{multline}
Again using \eqref{eq:Jtimedep} and noting that ${\bf j}_-^{(2)}(x,t;k_j)$
is not the zero vector, we learn that
\begin{equation}
\gamma_j(t)=e^{-2i\Lambda_j^2t/\e}\gamma_j(0)\,.
\end{equation}

Given this time evolution of the scattering data we are led to the
following algorithm for solving the Cauchy problem for the MNLS
equation \eqref{MNLScauchy}.  From the initial data $\phi(x,0)$ given
in a suitable space, calculate the scattering matrix ${\bf S}(k;0)$.
Check that the function $S_{11}(k;0)$ has a finite number of purely
simple zeros $k_j$ with $\text{Im}\{k_j^2\}<0$, and calculate the
constants $c_j^0:=\gamma_j(0)/S_{11}'(k_j;0)$.  Set
$r(k;0):=-S_{12}(k;0)/S_{22}(k;0)$ for $\text{Im}\{k^2\}=0$, and let $D$
denote the set of points
$\{k\in\mathbb{C}|S_{11}(k;0)=0\;\;\text{or}\;\;S_{22}(k;0)=0\}$.  The
scattering data are then used to formulate a Riemann-Hilbert problem:
find a $2\times 2$ matrix ${\bf M}(k;x,t)$, $x\in\mathbb{R}$, $t\ge 0$
such that the following properties are satisfied:
\begin{itemize}
\item[]{\bf Analyticity:} ${\bf M}(k;x,t)$ is analytic for
  $\text{Im}\{k^2\}\neq 0$ and $k\not\in D$ and takes continuous
  boundary values on the axes $\text{Im}\{k^2\}=0$ from each of the four
  sectors of its analyticity.  Moreover, ${\bf M}(k;x,t)$ is uniformly
  bounded for large $k$.
\item[]{\bf Jump Condition:} Letting ${\bf M}_\pm(k;x,t)$ denote the
  boundary value taken from the region where $\pm\text{Im}\{k^2\}<0$ as
  in \eqref{eq:Mboundaryvalues}, the boundary values are related by
  the formula
\begin{equation}
{\bf M}_+(k;x,t)={\bf M}_-(k;x,t)
e^{(\Lambda x+i\Lambda^2 t)\sigma_3/\e}\begin{bmatrix}1\pm |r(k;0)|^2 & r(k;0)\\
\pm r(k;0)^* & 1\end{bmatrix}e^{-(\Lambda x+i\Lambda^2 t)\sigma_3/\e}
\,,\quad \pm k^2>0\,.
\end{equation}
\item[]{\bf Singularities:}  The matrix ${\bf M}(k;x,t)$ has simple poles at
the points of the finite set $D$.  If $k_j\in D$ with $\text{Im}\{k_j\}>0$ and
$\text{Re}\{k_j\}<0$, then 
\begin{equation}
\mathop{\text{Res}}_{k=\pm k_j}{\bf M}(k;x,t)=\lim_{k\rightarrow \pm k_j}
{\bf M}(k;x,t)\begin{bmatrix}0 & 0\\c_j(x,t) & 0\end{bmatrix}
\label{eq:Msing}
\end{equation}
and
\begin{equation}
\mathop{\text{Res}}_{k=\pm k_j^*}{\bf M}(k;x,t)=\lim_{k\rightarrow \pm k_j^*}
{\bf M}(k;x,t)\begin{bmatrix}0 & -c_j(x,t)^*\\0 & 0\end{bmatrix}\,,
\label{eq:Msingstar}
\end{equation}
where 
\begin{equation}
c_j(x,t):=c_j^0 e^{-2(\Lambda_jx+i\Lambda_j^2t)/\e}\,,\quad\quad
\Lambda_j:=\Lambda\Big|_{k=k_j}\,.
\label{eq:cjdefine}
\end{equation}
\item[]{\bf Normalization:}  The matrix ${\bf M}(k;x,t)$ is normalized in the
sense that
\begin{equation}
\lim_{k\rightarrow 0}{\bf M}(k;x,t)=\mathbb{I}\,,
\end{equation}
with the limit being taken in any direction.
\end{itemize}

At a given $x\in\mathbb{R}$ and $t\ge 0$ we say that the Cauchy
problem for the MNLS equation \eqref{MNLScauchy} has a \emph{unique
  solution in the sense of inverse scattering} if this Riemann-Hilbert
problem has a unique solution for which the limit
\begin{equation}
\phi(x,t):=\lim_{k\rightarrow\infty}\frac{2k}{\alpha}\frac{M_{12}(k;x,t)}
{M_{22}(k;x,t)}
\label{eq:phireconstruction}
\end{equation}
exists at least in the sense of an approach to $k=\infty$ that is
nontangential to the axes $\text{Im}\{k^2\}=0$.  Lee \cite{Lee89} has
generalized the analysis of Beals and Coifman \cite{BealsC84} to study
the scattering and inverse-scattering maps for the derivative NLS
equation 
\begin{equation}
i\frac{\partial \psi}{\partial \tau} +
\frac{1}{2}\frac{\partial^2\psi}{\partial \xi^2} 
+ i\alpha\frac{\partial}{\partial \xi}(|\psi|^2\psi)=0
\label{eq:derivativeNLS}
\end{equation}
with initial data $\psi(\xi,0)=\psi_0(\xi)$.  This problem was shown
in the introduction to be related to the MNLS equation
\eqref{MNLScauchy} by a Galilean boost with velocity $c=\alpha^{-1}$.
Lee obtains global well-posedness of
the Cauchy problem in the Schwartz space $\mathscr{S}(\mathbb{R})$ for
a dense subset of initial data, but the theory is not entirely
complete because certain ``nongeneric'' initial conditions must be
excluded.  The nongeneric initial conditions include those for which
there are ``spectral singularities''; that is, the spectral functions
analogous to $S_{11}(k)$ and $S_{22}(k)$ have zeros on the axes
$\text{Im}\{k^2\}=0$.

The solution of the Cauchy problem in the sense of inverse scattering
would be expected to agree with solutions thereof that have been shown
to be unique using Fourier-based (nonlinear) methods.  For example, in
\cite{HayashiO92} Hayashi and Ozawa considered the Cauchy problem for
the derivative NLS equation and showed that if $\psi_0$ is a Schwartz
class ($\mathscr{S}(\mathbb{R})$) function, and if
\begin{equation}
\|\psi_0\|_2^2 = \int_{-\infty}^{+\infty}|\psi_0(\xi)|^2\,d\xi <\frac{\pi}{\alpha}
\label{eq:normbound}
\end{equation}
then there exists a unique global\footnote{It is conjectured in
  \cite{HayashiO92} that the inequality \eqref{eq:normbound} is sharp
  at least in the sense that there exist initial data in $\mathscr{S}$
  violating this bound leading to solutions that blow up in
  $L^\infty(\mathbb{R})$ in finite time.  This conjecture is based
  upon scaling arguments that suggest that the derivative NLS equation
  behaves similarly to the critical collapse case for the
  one-dimensional semilinear Schr\"odinger equation:
\[
i\frac{\partial\psi}{\partial\tau} +\frac{1}{2}\frac{\partial^2\psi}{\partial\xi^2} + |\psi|^4\psi=0\,.
\]
Despite the Hayashi-Ozawa conjecture, the literature is awash with
exact (global) solutions of the derivative NLS equation
\eqref{eq:derivativeNLS} that have finite $L^2(\mathbb{R})$ norm but
that violate \eqref{eq:normbound}.  For example, the soliton solutions of 
\eqref{eq:derivativeNLS} given by
\[
\psi=\frac{4\text{Im}\{z^2\}}{\alpha}\frac{ze^{-\zeta}+z^*e^\zeta}
{(ze^\zeta+z^*e^{-\zeta})^2}e^{-i\theta}\,,\;\;
\zeta:=\frac{4\text{Im}\{z^2\}}{\alpha}
\left(\xi-\xi_0+\frac{4}{\alpha}\text{Re}\{z^2\}\tau
\right)\,,\;\;
\theta:=\frac{4}{\alpha}\text{Re}\{z^2\}\xi+\frac{8}{\alpha^2}\text{Re}\{z^4\}\tau
+\theta_0\,,
\]
where $z\in\mathbb{C}$, $\xi_0\in\mathbb{R}$, and
$\theta_0\in\mathbb{R}$ are arbitrary parameters, are all Schwartz
class functions uniformly bounded in $L^2(\mathbb{R})$ norm for each
$\tau$, but the upper bound exceeds, exactly by a factor of two, the
restriction of \eqref{eq:normbound}:
\[
\sup_{z\in\mathbb{C}}\int_{-\infty}^{+\infty}|\psi(\xi,\tau;z)|^2\,d\xi = \frac{2\pi}{\alpha}\,.
\]
It follows also that there exist exact multisoliton solutions (that
is, solutions for which $r(k;0)\equiv 0$ in the Riemann-Hilbert
problem), necessarily in $\mathscr{S}$, with arbitrarily large
$L^2(\mathbb{R})$ norm.  To the degree that approximation by
multisoliton solutions is possible, there is therefore a dense set of
initial data for \eqref{eq:derivativeNLS} with arbitrary
$L^2(\mathbb{R})$ norm for which there exists a global Schwartz class
solution.  The validity of the Hayashi-Ozawa conjecture would then
seem to rest upon the contribution of nonzero $r(k;0)$ to the
Riemann-Hilbert problem.  Note that although Lee's analysis
\cite{Lee89} of the same Cauchy problem using inverse-scattering
techniques avoids entirely any assumptions about the size of the
$L^2(\mathbb{R})$-norm, it excludes nongeneric data leading to
spectral singularities and hence is not inconsistent with the
Hayashi-Ozawa conjecture.} solution $\psi(\xi,\tau)$ that is Schwartz
class for all $\tau\in\mathbb{R}$ and moreover
$\psi:\mathbb{R}\to\mathscr{S}$ is of class
$C^\infty(\mathbb{R};\mathscr{S})$.  It has also been shown that the
constraint \eqref{eq:normbound} on the $L^2(\mathbb{R})$ norm of the
initial data may be dispensed with at the expense of knowing existence
only for a finite time.  For example, in \cite{Pipolo99}, Pipolo
showed that for initial data $\psi_0$ of class
$H^{3,\infty}(\mathbb{R})$ (meaning that $\xi^k\psi_0(\xi)$,
$\xi^k\psi'_0(\xi)$, $\xi^k\psi''_0(\xi)$, and $\xi^k\psi'''_0(\xi)$
are all in $L^2(\mathbb{R})$ for all integer $k\ge 0$) then there is a
number $T>0$ such that a unique solution $\psi(\xi,\tau)$ of
\eqref{eq:derivativeNLS} exists of class $C^\infty([-T,T]\setminus
\{0\};C^\infty(\mathbb{R}))$.  There are many other results in the
literature for the Cauchy problem associated to the derivative NLS
equation \eqref{eq:derivativeNLS} (see in addition
\cite{CollianderKSTT01}, \cite{CollianderKSTT02}, \cite{Hayashi93},
\cite{HayashiO94}, \cite{Naumkin00}, and \cite{Ozawa96}) all differing
in detail but all granting existence of a unique solution under either
a sufficiently small $L^2(\mathbb{R})$ norm or a finite lifetime
condition.

All of these results apply equally well to the modified NLS equation
\eqref{MNLScauchy} by a Galilean change of coordinates followed by a
rescaling of $x$ and $t$ by $\e$.  It must be said, however, that
these results are unsatisfactory from the point of view of the
semiclassical ($\e\downarrow 0$) limit for the Cauchy problem for
\eqref{MNLScauchy} with $\phi(x,0)=A(x)e^{iS(x)/\e}$, because when $A$
and $S$ are fixed the $L^2(\mathbb{R})$ norm on the left-hand side of
\eqref{eq:normbound} becomes, after rescaling of $x$ by $\e$,
proportional to $1/\e$ which appears to rule out global solutions.  On
the other hand, upon rescaling of $t$ by $\e$, the fixed lifetime of a
solution guaranteed by Pipolo's result becomes arbitrarily small in
the limit $\e\downarrow 0$.  So it seems that there are no results in
the literature guaranteeing the existence of a unique solution to the
Cauchy problem for the MNLS equation \eqref{MNLScauchy} with initial
data of the form $\phi(x,0)=A(x)e^{iS(x)/\e}$ in the vicinity of fixed
$x$ and $t>0$ when $\e$ is sufficiently small.

Therefore, to deal with semiclassical problems we are forced to take
the approach of defining the solution of the Cauchy problem by the
Riemann-Hilbert problem.  Generally speaking, establishing the
existence of solutions for the Riemann-Hilbert problem is itself a
difficult problem requiring careful analysis and the correct
conditions on the function $r(k;0)$.  See \cite{BealsC84} or
\cite{Zhou98} for more details about the corresponding problem in a
simpler situation, and \cite{Lee89} for information relevant to the
equivalent derivative NLS problem.  However, when one is considering
semiclassical asymptotics the problem becomes somewhat simpler because
there is the possibility of the systematic construction of an explicit
parametrix by means of which the Riemann-Hilbert problem is reduced to
a ``small-norm'' problem that can be solved uniquely by iteration.
This is the essence of the steepest descent method of Deift and Zhou.
See \cite{KMM} and \cite{TovbisVZ04} for examples of this sort of
analysis for the case of the Riemann-Hilbert problem corresponding to
the focusing NLS equation. We intend to carry out such a steepest
descent analysis for the Riemann-Hilbert problem corresponding to the
MNLS equation in a subsequent work.  We may have confidence in the
success of such a program also because certain convergence results for
the semiclassical MNLS equation (or more correctly the derivative NLS
equation) have been obtained by Desjardins, Lin, and Tso using
``nonintegrable'' methods \cite{DesjardinsLT00}\footnote{Although as
  pointed out in Section~\ref{sec:Localconslaws} these authors do not
  treat the modulationally unstable case.}.  As the effective
solution size and lifetime are both large in this situation, these
results suggest that global solutions indeed may typically exist for
large data.  However, the approach of \cite{DesjardinsLT00} fails when
the solutions of the associated modulation equations become
nonclassical, and it is exactly at such junctures that the integrable
theory typically yields its best results, explaining and giving strong
approximations for the wild oscillations that are introduced by
dispersive regularization.

\subsection*{Even spectral symmetry.}
It turns out that our Riemann-Hilbert problem for ${\bf M}(k;x,t)$ has
``too much symmetry''.  As we will see below, there is a relation
between ${\bf M}(k;x,t)$ and ${\bf M}(-k;x,t)$ that will, in the
course of semiclassical analysis by the steepest descent method, lead
to Riemann surfaces with too many branch points.  For example, in
Section \ref{sec:stabilityRHP} it was shown that the solutions of the
spectral problem \eqref{eq:Laxx} corresponding to complex exponential
plane-wave solutions of the MNLS equation naturally live on a
hyperelliptic curve that is two-sheeted covering of the complex
$k$-plane with four distinct square-root type branch points.  This
surface has genus one, and functions on such a Riemann surface are
constructed from Riemann theta functions of genus one (that is,
elliptic functions).  One would therefore expect that the potential
$\phi$ would itself be constructed from elliptic functions, which
appears to contradict our starting point: that $\phi$ was of the
simple form $\rho^{1/2}e^{iux}$.

In this situation, judicious use of factorization identities of
Riemann theta functions may be required to see that the ``true genus''
is smaller than first thought.  To avoid these difficulties it is
desirable to formulate a new Riemann-Hilbert problem that effectively
quotients out the extra symmetry.  This approach was also used by
Deift, Venakides, and Zhou \cite{DVZ97} in their analysis of the
Korteweg-de Vries equation, and the specific reduction we use here can
be inferred also from the paper of Kaup and Newell \cite{KaupN78} on
the inverse-scattering transform for the derivative NLS equation.

It was pointed out earlier that the Jost functions enjoy a holomorphic
symmetry; see \eqref{eq:Jholo}.  This symmetry, taken together with
the implied relation \eqref{eq:Ssymholo} and used in the definition 
\eqref{eq:Mdefine} shows that for each $t$ the solution ${\bf M}(k;x,t)$
of the Riemann-Hilbert problem (assuming existence and uniqueness for
the moment) has the property that
\begin{equation}
{\bf M}(-k;x,t)=i^{\sigma_3}{\bf M}(k;x,t)i^{-\sigma_3}=\begin{bmatrix}
M_{11}(k;x,t) & -M_{12}(k;x,t)\\
-M_{21}(k;x,t) & M_{22}(k;x,t)\end{bmatrix}\,,\quad\quad \text{Im}\{k^2\}\neq 0\,.
\label{eq:Mholo}
\end{equation}
It follows that the matrix ${\bf N}(z;x,t)$ defined by
\begin{equation}
{\bf N}(z;x,t):=k^{\sigma_3/2}{\bf M}(k;x,t)k^{-\sigma_3/2} =
\begin{bmatrix} M_{11}(k;x,t) & kM_{12}(k;x,t)\\
k^{-1}M_{21}(k;x,t) & M_{22}(k;x,t)\end{bmatrix}
\label{eq:Ndefine}
\end{equation}
is well-defined as a function of $z=-k^2$ for $\text{Im}\{z\}\neq 0$.  From
the properties defining ${\bf M}(k;x,t)$ we may then deduce a
Riemann-Hilbert problem solved by the first row of ${\bf N}(z;x,t)$.

It is clear that ${\bf N}(z;x,t)$ is analytic for $\text{Im}\{z\}\neq 0$, and
that the boundary values taken for $z\in\mathbb{R}$ are continuous at least
for $z\neq 0$.  To analyze the behavior near $z=0$, we see immediately from
the normalization condition on ${\bf M}(k;x,t)$ that
\begin{equation}
\lim_{z\rightarrow 0}N_{11}(z;x,t)=\lim_{z\rightarrow 0}N_{22}(z;x,t)=1\quad\quad
\text{and}\quad\quad
\lim_{z\rightarrow 0}N_{12}(z;x,t)=0\,.
\end{equation}
The statement that $M_{21}(k;x,t)\rightarrow 0$ as $k\rightarrow
0$ alone is not enough to determine the limiting value of $N_{21}(k;x,t)$; 
however using the integral equations for
$\tilde{Y}_{21\pm}(k;x,t)$ it is possible to show that
\begin{equation}
\mathop{\lim_{z\rightarrow 0}}_{\pm\text{Im}\{z\}>0}
N_{21}(z;x,t)=\frac{2i}{\e}\int_{\pm\infty}^x\phi(y,t)e^{-i(y-x)/(\e\alpha)}
\,dy
\end{equation}
This limit cannot be used to define a normalization condition in the
usual way because it has a different value in different half-planes
for $z$, and more crucially, it depends on the (unknown) potential
$\phi(x,t)$.  This is the reason why we will only seek to determine
the first row of ${\bf N}(z;x,t)$ as the solution of a Riemann-Hilbert
problem (a similar situation occurs in \cite{DVZ97}).  Being able to
find only the first row of ${\bf N}(z;x,t)$ is not an obstruction to
determining the potential however, since from the reconstruction
formula \eqref{eq:phireconstruction} together with the antiholomorphic
symmetry relations \eqref{eq:Janti} and \eqref{eq:Ssymanti} we may
equally well find $\phi(x,t)$ from the limit
\begin{equation}
\phi(x,t)=\frac{2}{\alpha}\lim_{k\rightarrow\infty}\frac{kM_{12}(k;x,t)}
{M_{11}(k^*;x,t)^*} = \frac{2}{\alpha}\lim_{z\rightarrow\infty}
\frac{N_{12}(z;x,t)}{N_{11}(z^*;x,t)^*}\,.
\end{equation}

Next, we calculate the jump conditions satisfied by ${\bf N}(z;x,t)$
on the real $z$-axis.  This is straightforward, and the consistency of
the calculation (two values of $k$ with $\text{Im}\{k^2\}=0$ correspond
to each $z\in\mathbb{R}$) depends on the symmetry relations
\eqref{eq:Ssymholo} which imply that $k^{-1}r(k;0)$ is actually a function 
of $z=-k^2$ that is bounded as $z\rightarrow 0$.  
Defining the boundary values by
\begin{equation}
{\bf N}_\pm(z;x,t):=\mathop{\lim_{w\rightarrow z}}_{\pm\text{Im}\{w\}>0} 
{\bf N}(w;x,t)\,,\quad\quad z\in\mathbb{R}\,,
\label{eq:Npmdefine}
\end{equation}
and noting that in terms of $z=-k^2$ we have 
\begin{equation}
\Lambda=\frac{2i}{\alpha}\left(z+\frac{1}{4}\right)\,,
\end{equation}
one finds that 
\begin{equation}
{\bf N}_+(z;x,t)={\bf N}_-(z;x,t)e^{(\Lambda x+i\Lambda^2t)\sigma_3/\e}
\begin{bmatrix}1\pm |r(k;0)|^2 & kr(k;0)\\
\pm k^{-1}r(k;0)^* & 1\end{bmatrix}
e^{-(\Lambda x+i\Lambda^2t)\sigma_3/\e}\,,\quad \pm z<0\,.
\label{eq:Njump1}
\end{equation}
Therefore, defining
\begin{equation}
\rho(z):= k^{-1}r(k;0)  \,, \quad\quad z\in\mathbb{R}\,,
\label{eq:rhoreflectiondefine}
\end{equation}
(this is indeed well-defined because $r(k;0)$ is an odd function of
$k$) we resolve the dichotomy of signs in \eqref{eq:Njump1} and find
simply
\begin{equation}
{\bf N}_+(z;x,t)={\bf N}_-(z;x,t)e^{(\Lambda x+i\Lambda^2t)\sigma_3/\e}
\begin{bmatrix}1-z|\rho(z)|^2 & -z\rho(z)\\
\rho(z)^* & 1\end{bmatrix}e^{-(\Lambda x+i\Lambda^2t)\sigma_3/\e}\,,\quad
z\in\mathbb{R}\,.
\label{eq:Njump2}
\end{equation}

Finally, we examine the singularities of ${\bf N}(z;x,t)$.  Near a pole $k=k_j$
of ${\bf M}(k;x,t)$ with $\text{Im}\{k_j^2\}<0$ we know from \eqref{eq:Msing}
that 
\begin{equation}
{\bf M}(k;x,t)=\begin{bmatrix} a_jc_j(k-k_j)^{-1} + O(1) & a_j+O(k-k_j)\\
b_jc_j(k-k_j)^{-1} + O(1) & b_j+O(k-k_j)
\end{bmatrix}\,,\quad k\rightarrow k_j
\end{equation}
for some scalar functions $a_j=a_j(x,t)$ and $b_j=b_j(x,t)$ (and
$c_j=c_j(x,t)$ is defined by \eqref{eq:cjdefine}).  Applying the
symmetry \eqref{eq:Mholo} we see that
\begin{equation}
{\bf M}(k;x,t)=\begin{bmatrix} -a_jc_j(k+k_j)^{-1} + O(1) & -a_j+O(k+k_j)\\
b_jc_j(k+k_j)^{-1} + O(1) & b_j+O(k+k_j)
\end{bmatrix}\,,\quad k\rightarrow -k_j\,.
\end{equation}
From these and the definition \eqref{eq:Ndefine} it follows easily
that
\begin{equation}
{\bf N}(z;x,t)=\begin{bmatrix} a_jc_j(k-k_j)^{-1} + 
O(1) & k_ja_j+O(k-k_j)\\
k_j^{-1}b_jc_j(k-k_j)^{-1} + O(1) & b_j+O(k-k_j)
\end{bmatrix}\,,\quad k\rightarrow k_j
\end{equation}
and
\begin{equation}
{\bf N}(z;x,t)=\begin{bmatrix} -a_jc_j(k+k_j)^{-1} + O(1) & k_ja_j+O(k+k_j)\\
-k_j^{-1}b_jc_j(k+k_j)^{-1} + O(1) & b_j+O(k+k_j)
\end{bmatrix}\,,\quad k\rightarrow -k_j\,.
\end{equation}
In turn, these relations are simultaneously equivalent to
\begin{equation}
{\bf N}(z;x,t)=\begin{bmatrix} 
-2k_ja_jc_j(z-z_j)^{-1} + 
O(1) & k_ja_j+O(z-z_j)\\
-2b_jc_j(z-z_j)^{-1} + O(1) & b_j+O(z-z_j)
\end{bmatrix}\,,\quad z\rightarrow z_j:=-k_j^2\,.
\end{equation}
Therefore, we have
\begin{equation}
\mathop{\text{Res}}_{z=z_j}{\bf N}(z;x,t)=\lim_{z\rightarrow z_j}
{\bf N}(z;x,t)\begin{bmatrix}0 & 0\\-2c_j(x,t) & 0\end{bmatrix}\,,
\label{eq:Nsing}
\end{equation}
a relation equivalent to both relations in \eqref{eq:Msing}.  In a
similar way one derives the condition
\begin{equation}
\mathop{\text{Res}}_{z=z_j^*}{\bf N}(z;x,t)=\lim_{z\rightarrow z_j^*}
{\bf N}(z;x,t)\begin{bmatrix}0 & -2z_j^*c_j(x,t)^*\\0 & 0\end{bmatrix}\,,
\label{eq:Nsingstar}
\end{equation}
a relation equivalent to both relations in \eqref{eq:Msingstar}.

We are now in a position to formulate a ``less symmetric''
Riemann-Hilbert problem to determine $\phi(x,t)$.  Given $\rho(z)$ for
$z\in\mathbb{R}$ and a set of points $z_1,\dots,z_N$ in the upper
half-plane along with a corresponding set of nonzero complex numbers
$c_1^0,\dots,c_N^0$, seek a two-component row vector ${\bf
  n}^T(z;x,t)=[n_1(z;x,t),n_2(z;x,t)]$ with the following properties
\begin{itemize}
\item[]{\bf Analyticity:} The row vector ${\bf n}^T(z;x,t)$ is an
  analytic function of $z$ for $z\in\mathbb{C}\setminus
  (\mathbb{R}\cup\{z_1,\dots,z_N,z_1^*,\dots,z_N^*\})$ that takes
  continuous boundary values on the real line and is uniformly bounded
  as $z\rightarrow\infty$.
\item[]{\bf Jump Condition:} With the boundary values defined by
\begin{equation}
{\bf n}^T_\pm(z;x,t):=\mathop{\lim_{w\rightarrow z}}_{\pm\text{Im}\{w\}>0} 
{\bf n}^T(w;x,t)\,,\quad\quad z\in\mathbb{R}\,,
\end{equation}
the relation 
\begin{equation}
{\bf n}^T_+(z;x,t)={\bf n}^T_-(z;x,t)e^{(\Lambda x+i\Lambda^2t)\sigma_3/\e}
\begin{bmatrix}1-z|\rho(z)|^2 & -z\rho(z)\\
\rho(z)^* & 1\end{bmatrix}e^{-(\Lambda x+i\Lambda^2t)\sigma_3/\e}
\label{eq:reducedjump}
\end{equation}
holds for $z\in\mathbb{R}$.
\item[]{\bf Singularities:}  
The row vector ${\bf n}^T(z;x,t)$ has simple poles at the points $z_1,\dots,z_N$
and their complex conjugates.  The residues at these points satisfy
\begin{equation}
\mathop{\text{Res}}_{z=z_j}{\bf n}^T(z;x,t)=\lim_{z\rightarrow z_j}
{\bf n}^T(z;x,t)\begin{bmatrix}0 & 0\\-2c_j(x,t) & 0\end{bmatrix}\,,
\end{equation}
and 
\begin{equation}
\mathop{\text{Res}}_{z=z_j^*}{\bf n}^T(z;x,t)=\lim_{z\rightarrow z_j^*}
{\bf n}^T(z;x,t)\begin{bmatrix}0 & -2z_j^*c_j(x,t)^*\\0 & 0\end{bmatrix}\,,
\end{equation}
for $j=1,\dots,N$. 
\item[]{\bf Normalization:}  The row vector ${\bf n}^T(z;x,t)$ satisfies
\begin{equation}
\lim_{z\rightarrow 0}{\bf n}^T(z;x,t)=\begin{bmatrix}1 & 0\end{bmatrix}
\end{equation}
with the limit being taken in any direction.
\end{itemize}
From the solution of this Riemann-Hilbert problem, the corresponding solution
of the Cauchy problem for the MNLS equation \eqref{MNLScauchy} is then 
given by
\begin{equation}
\phi(x,t)=\frac{2}{\alpha}\lim_{z\rightarrow\infty}\frac{n_2(z;x,t)}
{n_1(z^*;x,t)^*}\,.
\end{equation}
We wish to emphasize that this Riemann-Hilbert problem involves
discontinuities only along the real $z$-axis and poles in
complex-conjugate pairs.  With the extra symmetry removed, this
problem is therefore quite similar to the Riemann-Hilbert problem
associated with the nonselfadjoint Zakharov-Shabat eigenvalue problem
and the focusing nonlinear Schr\"odinger equation.

\end{document}